\documentclass[preprint,3p,12pt]{elsarticle}
\usepackage{mathrsfs}
\usepackage{amsmath}
\usepackage{stmaryrd}
\usepackage{bbding}
\usepackage{dcolumn}
\usepackage{graphicx}
\usepackage{amsfonts}
\usepackage{amssymb}
\usepackage{psfrag}
\usepackage{wrapfig}
\usepackage{makeidx}
\usepackage{bm}
\usepackage{epsf}
\usepackage{epsfig}
\usepackage{setspace}
\usepackage{graphicx}
\usepackage{epstopdf}
\usepackage{psfrag}
\usepackage{subfigure}
\usepackage{color}
\usepackage{comment}

\newtheorem{rmk}{Remark}

\begin{document}
	
	\title{A HWENO Reconstruction Based High-order Compact Gas-kinetic Scheme on Unstructured Meshes}
	
	\author[HKUST1]{Xing Ji}
	\ead{xjiad@connect.ust.hk}
	
	\author[HKUST2]{Fengxiang Zhao}
	\ead{fzhaoac@connect.ust.hk}
	
	\author[HKUST2]{Wei Shyy}
	\ead{weishyy@ust.hk}
	
	\author[HKUST1,HKUST2,HKUST3]{Kun Xu\corref{cor1}}
	\ead{makxu@ust.hk}

	\address[HKUST1]{Department of Mathematics, Hong Kong University of Science and Technology, Clear Water Bay, Kowloon, Hong Kong}
	\address[HKUST2]{Department of Mechanical and Aerospace Engineering, Hong Kong University of Science and Technology, Clear Water Bay, Kowloon, Hong Kong}
\address[HKUST3]{Shenzhen Research Institute, Hong Kong University of Science and Technology, Shenzhen, China}
	\cortext[cor1]{Corresponding author}

	\begin{abstract}
		As an extension of previous fourth-order compact gas kinetic scheme (GKS) on structured meshes \cite{ji2018compact},
		this work is about the development of
		a third-order compact GKS on unstructured meshes for the compressible Euler and Navier-Stokes solutions.
		Based on the time accurate high-order gas-kinetic evolution solution,
		the time dependent gas distribution function at a cell interface in GKS provides not only
		the flux function and its time derivative, but also the time accurate flow variables there at next time level.
		As a result, besides updating the conservative flow variables inside each control volume through the interface fluxes,
		the cell averaged first-order spatial derivatives of flow variables in the cell can be also obtained using the
		updated flow variables at the cell interfaces around that cell through the divergence theorem.
		Therefore, with the flow variables and their first-order spatial derivatives inside each
		cell, the Hermite WENO (HWENO) techniques can be naturally implemented for the
		compact high-order reconstruction at the beginning of next time step.
		Following the reconstruction technique in \cite{zhu2018new}, 
        a new HWENO reconstruction on triangular meshes is designed in the current scheme.
		Combined with the two-stage temporal discretization and second-order time accurate flux function,
		a third-order compact scheme with a fourth-order accuracy for smooth flow on unstructured mesh has been constructed.
 		Accurate solutions can be obtained for both inviscid and viscous flows without sensitive dependence on the quality of triangular meshes.
		The robustness of the scheme has been validated as well through many cases, including strong shocks in the  hypersonic viscous flow simulations.

	\end{abstract}
	
	
	\maketitle

\section{Introduction}
In computational fluid dynamics (CFD) applications with complex geometry, the unstructured mesh
is widely used due to its flexible adaptability.
In such a mesh, many techniques used on structured meshes cannot be directly extended here.
For example, the third-order WENO method \cite{hu1999weighted} on a unstructured mesh needs many neighboring cells in the reconstruction,
and the number of cells used in the reconstruction may not be fixed.
In general, it's basically more complicated in unstructured mesh in terms of reconstruction, boundary treatment, and parallelization.
Theoretically, a large disparity between the numerical domain of dependence and the physical domain of dependence
indicates the intrinsic weakness in either physical model or the numerical discretization \cite{xu2017paradigm}.
Thus, it is preferred to design a compact high order scheme, which connects the target cell with its closest neighbors, and to use a
CFL number as large as possible.
Great efforts have been paid on the development of compact schemes in the past decades\cite{high-order-review}.
Two main representatives of compact schemes are the Discontinuous Galerkin (DG) method \cite{cockburn1989tvb} and correction procedure via reconstruction (CPR) \cite{yu2014accuracy}, with very restricted  CFL number in the determination of time step.
Most of those methods use Riemann solvers or approximate Riemann solvers for the flux evaluation, 
and adopt the Runge-Kutta time stepping for the
time accuracy.
Based on the time-dependent gas-kinetic flux function \cite{GKS-2001},
the corresponding schemes under the DG and CPR frameworks have been developed with triangular meshes \cite{luo2010bgk,zhang2018third}.
But, the advantages of GKS have not been fully utilized in the above approaches, at least the time step has not been enlarged in comparison with
Riemann solver based DG methods.

Higher than second-order gas kinetic schemes (HGKS) have been developed systematically in the past decades \cite{3rdGKS-Li}.
In comparison with traditional Riemann solver based high-order CFD
methods, the distinguishable points of HGKS include the following:
(i) The time-dependent gas distribution function at a cell interface provides a multiple
scale flow physics from the kinetic particle transport to the
hydrodynamic wave propagation, which bridges the evolution from the
kinetic flux vector splitting (KFVS) to the central difference Lax-Wendroff
type discretization.
(ii) Both inviscid and viscous fluxes are
obtained from the moments of a single gas distribution function.
(iii) The GKS is a multi-dimensional scheme
\cite{implicitGKS}, where both normal and tangential derivatives of
flow variables around a cell interface participate the time
evolution of the gas distribution function.
(iv) Besides fluxes, the time-dependent gas distribution function at a cell interface also provides
time evolving flow variables at the cell interface, which can be used to construct compact schemes.
The first high-order GKS is the one-step 3rd-order scheme with a third order time accurate flux function \cite{pan2016unstructuredcompact,3rdGKS-Li}.
In this scheme, both cell averaged and cell interface flow variables are directly implemented in the reconstruction. 
The one-step time-stepping formulation and the  rigid use of interface values makes the scheme complex and lack of robustness.

Recently, with the incorporation of multi-derivative multi-stage
technique, a family of HGKS has been developed \cite{ji2018family}.
Based on the 5th-order WENO reconstruction \cite{weno,wenoz}, the performance of
HGKS shows great advantages in terms of efficiency, accuracy, and
robustness in comparison with traditional high-order scheme with Riemann solver, especially in the
capturing of shear instabilities due to its multi-dimensionality in the GKS flux function.
Among the multi-derivative multi-stage GKS,
the two-stage fourth-order GKS (S2O4 GKS) \cite{Pan2016twostage} seems to be
an optimal choice in practical CFD computations. The S2O4 is both efficient and accurate, and
as robust as the 2nd-order GKS.
With the evaluation of cell averaged slopes from the cell interface values, 
and the adoption of two-stage time discretization and compact Hermite WENO (HWENO) reconstruction \cite{qiu2004hermite},
a class of compact GKS with the spatial accuracy from fifth-order to eighth-order  on two-dimensional structured mesh has been developed \cite{ji2018compact,zhao2019compact}.
The fifth-order compact scheme  \cite{ji2018compact} can take a CFL number around $0.5$, and
it shows better performance than the same order and the same stencils based DG method in all aspects of efficiency, robustness, and accuracy
 in the compressible viscous flow simulations with shocks.
 The sixth- and eighth-order compact scheme \cite{zhao2019compact} can achieve a spectral-like resolution at large wavenumber and give great advantages in both tracking the linear aero-acoustic wave and capturing shock-shock interactions.
As a continuation, here we further extend the compact two-stage GKS to the unstructured mesh.

In any high-order scheme, the reconstruction plays an important role.
The WENO-type reconstruction  achieves great success \cite{weno,wenoz,qiu2004hermite,levy1999central}, especially for the
high speed flow with discontinuities.
The coefficients for the reconstruction are solely geometric dependence, which could be pre-determined in the computation.
The limiting process depends on the non-linear weights.
The classical WENO techniques are based on the reconstructions from both 
 low-order sub-stencils to  high-order large stencils \cite{weno,wenoz,qiu2004hermite}, which are very effective on structured mesh.
The similar idea has been used in the construction of third-order compact HWENO on unstructured  triangular mesh \cite{hu1999weighted,zhu2009hermite}.
The direct applications of the reconstruction from the structured case to the unstructured one meet the following problems.
Firstly, the linear combinations of the point values from six sub-stencils in \cite{zhu2009hermite} could not exactly recover the corresponding values from large stencils in smooth cases, where at least eight sub-stencils are required \cite{hu1999weighted}.
Secondly, in general the linear weights are not all positive and some linear weights could take very large values,
which subsequently distort the numerical solutions.
Some techniques have been used to resolve these problems for non-compact WENO reconstruction \cite{zhao2017weighted},
but they cannot be directly extended to HWENO reconstruction.
To overcome these difficulties, Zhu et al. \cite{zhu2018new} designed a new type of WENO reconstruction on triangular mesh
for finite volume method.
The idea is to use the central WENO technique \cite{levy1999central}, where
the WENO procedure is performed on the whole polynomials rather than at each Gaussian point.
The linear weights can be chosen to be any positive number as long as the summation goes to one,
and the scheme keeps the expected order of accuracy in smooth region.
The smooth indicators are carefully designed to achieve such a goal.
In addition, the number of sub-stencils can be reduced in this method.
Therefore, in this paper we are going to design a new compact third-order Hermite WENO reconstruction by following the similar idea.
A quadratic polynomial is constructed first from a total of $10$ available cell averaged flow variables and their slopes within the compact stencils.
Each of the three sub-stencils is composed of three cells with averaged values.

In this paper, combining the second order gas kinetic flux and the two-stage temporal discretization,
a new compact third-order GKS will be proposed.
 The compact scheme inherits the advantages of original two-stage GKS \cite{Pan2016twostage,ji2018compact}.
It allows a larger CFL number than the same order DG method.
Compared with a 3rd-order Runge Kutta (RK) time stepping method, it could achieve a 3rd-order accuracy in time with one middle stage only.
At the same time, benefiting from the newly designed Hermite WENO reconstruction,
 the compact scheme has less number of sub-stencils than the previous method \cite{zhu2009hermite}, 
 resulting in an improvement in efficiency.
More importantly, the current scheme demonstrates excellent robustness in the test cases with strong shocks,
i.e., hypersonic flow passing a cylinder up to Mach number  $20$.
From the perspective of programming, the current algorithm could be easily developed from a finite volume code,
which has the same advantages as the reconstructed-DG (rDG)/PnPm methods \cite{luo2010reconstructed,dumbser2010arbitrary}.
But, different from the rDG method, the slopes within a cell in the current scheme are updated by the time accurate solutions
 as Gaussian points at cell interfaces through the Green-Gauss theorem, rather than by the evolution solution of the slopes directly.
The different slope update makes fundamental differences in the current compact GKS from all other DG methods, such as the use of the large
CFL time step here and the robustness in capturing discontinuous solutions.

This paper is organized as follows.
In Section 2, a brief review of finite volume  GKS on triangular mesh is presented.
The general formulation for the two-stage temporal discretization is introduced in Section 3.
In Section 4, the compact third-order Hermite WENO reconstruction on triangular mesh is presented.
Section 5 includes inviscid and viscous test cases. The last section is the conclusion.

\section{Finite volume gas-kinetic scheme}

\subsection{Finite volume scheme on unstructured mesh}
The two-dimensional gas-kinetic BGK equation \cite{BGK} can be
written as
\begin{equation}\label{bgk}
f_t+\textbf{u}\cdot\nabla f=\frac{g-f}{\tau},
\end{equation}
where $f$ is the gas distribution function, $g$ is the corresponding
equilibrium state, and $\tau$ is the collision time. The collision
term satisfies the following compatibility condition
\begin{equation}\label{compatibility}
\int \frac{g-f}{\tau}\psi \text{d}\Xi=0,
\end{equation}
where $\psi=(1,u,v,\displaystyle \frac{1}{2}(u^2+v^2+\xi^2))$,
$\text{d}\Xi=\text{d}u\text{d}v\text{d}\xi_1...\text{d}\xi_{K}$, $K$
is the number of internal degree of freedom, i.e.
$K=(4-2\gamma)/(\gamma-1)$ for two-dimensional flows, and $\gamma$
is the specific heat ratio.

Based on the Chapman-Enskog expansion for BGK equation
\cite{xu2014direct}, the gas distribution function in the continuum
regime can be expanded as
\begin{align*}
f=g-\tau D_{\textbf{u}}g+\tau D_{\textbf{u}}(\tau
D_{\textbf{u}})g-\tau D_{\textbf{u}}[\tau D_{\textbf{u}}(\tau
D_{\textbf{u}})g]+...,
\end{align*}
where $D_{\textbf{u}}={\partial}/{\partial t}+\textbf{u}\cdot
\nabla$.
By truncating on different orders of $\tau$, the
corresponding macroscopic equations can be derived.
For the Euler
equations, the zeroth order truncation is taken, i.e. $f=g$.
For the Navier-Stokes equations, the first order truncated distribution function is
\begin{align*}
f=g-\tau (ug_x+vg_y+g_t).
\end{align*}

For a polygon cell $\Omega_i$, the boundary can be expressed as
\begin{equation*}
\displaystyle \partial \Omega_i=\bigcup_{p=1}^m\Gamma_{ip},
\end{equation*}
where $m$ is the number of cell interfaces for cell $\Omega_i$, which has $m=3$
for triangular mesh.
Taking moments of the BGK equation Eq.\eqref{bgk} and integrating
over the cell $\Omega_i$, the
semi-discretized form of finite volume scheme can be written as
\begin{equation}\label{semidiscrete}
\frac{\text{d}W_{i}}{\text{d}t}=\mathcal{L}(W_i)=-\frac{1}{\left| \Omega_i \right|} \sum_{p=1}^m\oint_{\Gamma_{ip}}
\textbf{F}(W)\cdot\textbf{n}_p \text{d}s,
\end{equation}
where $W_{i}$ is the cell averaged value over cell $\Omega_i$, $\left|
\Omega_i \right|$ is the area of $\Omega_i$, $\mathcal{L}(W)$ is the spatial operator of flux, $\textbf{F}=(F,G)^T$
and $\textbf{n}_p$ is the outer normal direction of $\Gamma_{ip}$.

In this paper, a third-order spatial reconstruction will be
introduced, and the line integral over $\Gamma_{ip}$ is discretized
according to Gaussian quadrature as follows
\begin{equation}\label{quadrature}
\oint_{\Gamma_{ip}}\textbf{F}(W)\cdot\textbf{n}_p
\text{d}s=\left|l_p\right| \sum_{k=1}^{2} \omega_k
\textbf{F}(\textbf{x}_{p,k},t)\cdot \textbf{n}_p,
\end{equation}
where $\left|l_p\right|$ is the length of the cell interface
$\Gamma_{ip}$, $\displaystyle\omega_1=\omega_2=1/2$ are the Gaussian
quadrature weights, and the Gaussian points
$\textbf{x}_{p,k}, k=1,2$ for $\Gamma_{ip}$ are defined as
\begin{equation*}
\textbf{x}_{p1}=\frac{3+\sqrt{3}}{6}\textbf{X}_{p1}+\frac{3-\sqrt{3}}{6}\textbf{X}_{p2},~
\textbf{x}_{p2}=\frac{3+\sqrt{3}}{6}\textbf{X}_{p2}+\frac{3-\sqrt{3}}{6}\textbf{X}_{p1},
\end{equation*}
where $\textbf{X}_{p1}, \textbf{X}_{p2}$ are the endpoints of $\Gamma_{ip}$.

The fluxes in unit length across each Gaussian point for the updates of flow variables in a global coordinate can be written as follows
\begin{equation}\label{global-flux}
\textbf{F}(\textbf{x}_{p,k},t)\cdot \textbf{n}_p=\int\psi f(x_{p,k},y_{p,k},t,u,v,\xi)\textbf{u}\cdot \textbf{n} \text{d}u\text{d}v\text{d}\xi,
\end{equation}
where $f(x_{p,k},y_{p,k},t,u,v,\xi)$ is the gas distribution function at
the corresponding Gaussian point.
Here we can first evaluate the fluxes in a local coordinate
\begin{align*}
\tilde{F}_{p,k}(t)=\int\psi \tilde{u} f(\tilde{x}_{p,k},\tilde{y}_{p,k},t,\tilde{u},\tilde{v},\xi)\text{d}\tilde{u}\text{d}\tilde{v}\text{d}\xi,
\end{align*}
where the original point of the local coordinate is $(\tilde{x}_{p,k},\tilde{y}_{p,k})=(0,0)$ with x-direction in $\textbf{n}_p$.
Then the velocities in the local coordinate are given by
\begin{align*}
\begin{cases}
\tilde{u}=u \cos \theta + v \sin \theta , \\
\tilde{v}=-u \sin \theta + v \cos \theta.
\end{cases}
\end{align*}
In 2-D case, the global and local fluxes are related as \cite{pan2016unstructuredcompact}
\begin{align*}
\textbf{F}(W)\cdot\textbf{n}=
F(W)\cos\theta+G(W)\sin\theta=T^{-1}F(TW) ,
\end{align*}
where $T=T(\theta)$ is the rotation matrix
\begin{equation*}
T= \left(
\begin{array}{cccc}
1 & 0         & 0         & 0 \\
0 & \cos\theta & \sin\theta & 0 \\
0 &-\sin\theta & \cos\theta & 0 \\
0 & 0         & 0         & 1 \\
\end{array}
\right).
\end{equation*}

\subsection{Second-order gas-kinetic flux solver}

The formulation of gas kinetic flux will be presented in a local coordinate.
We omit the tilde on flow variables for simplicity.
In order to construct the numerical fluxes, the integral solution of
BGK equation Eq.\eqref{bgk} is used
\begin{equation}\label{integral1}
f(x_{p,k},y_{p,k},t,u,v,\xi)=\frac{1}{\tau}\int_0^t g(x',y',t',u,v,\xi)e^{-(t-t')/\tau}dt'\\
+e^{-t/\tau}f_0(-ut,-vt,u,v,\xi),
\end{equation}
where $(x_{p,k}, y_{p,k})=(0,0)$ is the quadrature point of a cell interface in the local coordinate,
and $x_{p,k}=x'+u(t-t')$ and $y_{p,k}=y'+v(t-t')$ are the trajectory of
particles. $f_0$ is the initial gas distribution function and $g$ is the corresponding
equilibrium state.
The flow dynamics at a cell interface depends on the ratio of time step
to the  local particle collision time $\Delta t/\tau$, which covers a process from the particle free transport in $f_0$ to the formation of
equilibrium state $g$.

To construct a time evolution solution of gas distribution function at a cell interface,
the following notations are introduced first
\begin{align*}
a_1=&(\partial g/\partial x)/g, a_2=(\partial g/\partial y)/g,
A=(\partial g/\partial t)/g,
\end{align*}
where $g$ is the equilibrium state.  The variables $(a_1,a_2, A)$, denoted by $\omega$,
depend on particle velocity in the form of
\cite{GKS-2001}
\begin{align*}
\omega=\omega_{1}+\omega_{2}u+\omega_{3}v+\omega_{4}\displaystyle
\frac{1}{2}(u^2+v^2+\xi^2).
\end{align*}
For the kinetic part of the integral solution Eq.\eqref{integral1},
the initial gas distribution function can be constructed as
\begin{equation*}
f_0=f_0^l(x,y,u,v)\mathbb{H} (x)+f_0^r(x,y,u,v)(1- \mathbb{H}(x)),
\end{equation*}
where $\mathbb{H}(x)$ is the Heaviside function. Here $f_0^l$ and $f_0^r$ are the
initial gas distribution functions on both sides of a cell
interface, which have one to one correspondence with the initially
reconstructed macroscopic variables. For the 2nd-order scheme, the
Taylor expansion for the gas distribution function in space around
$(x,y)=(0,0)$ is expressed as
\begin{align}\label{third-1}
f_0^k(x,y)=f_G^k(0,0)&+\frac{\partial f_G^k}{\partial
    x}x+\frac{\partial f_G^k}{\partial y}y,
\end{align}
for $k=l,r$. According to the Chapman-Enskog expansion, $f_{G}^k$ has the form
\begin{align}\label{third-2}
f_{G}^k=g_k-\tau(a_{1k}u+a_{2k}v+A_k)g_k,
\end{align}
where $g_l,g_r$ are the equilibrium states which can be fully determined from  the
reconstructed macroscopic variables $W_l, W_r$ at the left and right sides of a cell interface.
Substituting Eq.\eqref{third-1} and Eq.\eqref{third-2} into Eq.\eqref{integral1},
the kinetic part for the integral solution can be written as
\begin{equation}\label{dis1}
\begin{aligned}
e^{-t/\tau}f_0^k(-ut,-vt,u,v,\xi)
=e^{-t/\tau}g_k[1-\tau(a_{1k}u+a_{2k}v+A_k)-t(a_{1k}u+a_{2k}v)],
\end{aligned}
\end{equation}
where the coefficients $a_{1k},...,A_k, k=l,r$ are defined according
to the expansion of $g_{k}$. After determining the kinetic part
$f_0$, the equilibrium state $g$ in the integral solution
Eq.\eqref{integral1} can be expanded in space and time as follows
\begin{align}\label{equli}
g=g_0+\frac{\partial g_0}{\partial x}x+&\frac{\partial g_0}{\partial
    y}y+\frac{\partial g_0}{\partial t}t,
\end{align}
where $g_{0}$ is the equilibrium state located at an interface, which
can be determined through the compatibility condition
Eq.\eqref{compatibility},
\begin{align}\label{compatibility2}
\int\psi g_{0}\text{d}\Xi=W_0=\int_{u>0}\psi
g_{l}\text{d}\Xi+\int_{u<0}\psi g_{r}\text{d}\Xi,
\end{align}
where $W_0$ are the macroscopic flow variables for the determination of the
equilibrium state $g_{0}$. Substituting Eq.\eqref{equli} into Eq.\eqref{integral1}, the hydrodynamic part for the integral solution
can be written as
\begin{equation}\label{dis2}
\begin{aligned}
\frac{1}{\tau}\int_0^t
g&(x',y',t',u,v,\xi)e^{-(t-t')/\tau}dt'
=C_1g_0+C_2g_0(\overline{a}_1u+\overline{a}_2v)+C_3g_0\overline{A},
\end{aligned}
\end{equation}
where the coefficients
$\overline{a}_1,\overline{a}_2,...,\overline{A},\overline{B}$ are
defined from the expansion of the equilibrium state $g_0$. The
coefficients $C_i, i=1,2,3$ in Eq.\eqref{dis2}
are given by
\begin{align*}
C_1=1-&e^{-t/\tau}, C_2=(t+\tau)e^{-t/\tau}-\tau, C_3=t-\tau+\tau e^{-t/\tau}.
\end{align*}
The coefficients in Eq.\eqref{dis1} and Eq.\eqref{dis2}
can be determined by the spatial derivatives of macroscopic flow
variables and the compatibility condition as follows
\begin{align}\label{co}
\langle a_1\rangle =\frac{\partial W }{\partial x},
\langle a_2\rangle =\frac{\partial W }{\partial y},
\langle A+a_1u+a_2v\rangle=0,\\
\end{align}
where $$ \langle (...) \rangle  = \int \psi (...) g d \Xi .$$
Finally, the second-order time dependent gas  distribution function at a cell interface is \cite{GKS-2001}
\begin{align}\label{flux}
f(x_{p,k},y_{p,k},t,u,v,\xi)=&(1-e^{-t/\tau})g_0+((t+\tau)e^{-t/\tau}-\tau)(\overline{a}_1u+\overline{a}_2v)g_0\nonumber\\
+&(t-\tau+\tau e^{-t/\tau}){\bar{A}} g_0\nonumber\\
+&e^{-t/\tau}g_r[1-(\tau+t)(a_{1r}u+a_{2r}v)-\tau A_r)]H(u)\nonumber\\
+&e^{-t/\tau}g_l[1-(\tau+t)(a_{1l}u+a_{2l}v)-\tau A_l)](1-H(u)).
\end{align}

\section{Two-stage temporal discretization}

The two-stage fourth-order temporal discretization which was
adopted in the previous compact fourth order scheme on structured mesh  is implemented here \cite{ji2018compact}.
Following the definition of Eq.\eqref{semidiscrete}, a
fourth-order time accurate solution for cell-averaged conservative flow variables $W_i$ is updated by
    \begin{equation}\label{step-hyper-1}
    \begin{aligned}
    W_i^*=W_i^n+\frac{1}{2}\Delta t\mathcal
    {L}(W_i^n)+\frac{1}{8}\Delta t^2\frac{\partial}{\partial
        t}\mathcal{L}(W_i^n),
    \end{aligned}
    \end{equation}
    \begin{equation}\label{step-hyper-2}
    \begin{aligned}
    W_i^{n+1}=W_i^n+\Delta t\mathcal
    {L}(W_i^n)+\frac{1}{6}\Delta t^2\big(\frac{\partial}{\partial
        t}\mathcal{L}(W_i^n)+2\frac{\partial}{\partial
        t}\mathcal{L}(W_i^*)\big),
    \end{aligned}
    \end{equation}
where $ \frac{\partial}{\partial t}\mathcal{L}(W)$ are the time  derivatives of the summation of flux transport over closed interfaces of the cell.
    The proof for fourth-order accuracy can be found in \cite{li2016twostage}.

For the gas-kinetic scheme, the gas evolution is a time-dependent
relaxation process from kinetic to hydrodynamic scale through the exponential function, and the corresponding flux becomes
 a complicated function of time.
In order to obtain the time derivatives of the flux function at $t_n$ and $t_*=t_n + \Delta t/2$,
the flux function can be approximated as a linear function of time within a time interval.
Let's first
introduce the following notation,
\begin{align*}
\mathbb{F}_{p,k}(W^n,\delta)=\int_{t_n}^{t_n+\delta} F_{p,k}(W^n,t)\text{d}t.
\end{align*}
For convenience, assume $t_n=0$,
the flux in the time interval $[t_n, t_n+\Delta t]$ is expanded as
the following linear form
\begin{align*}
F_{p,k}(W^n,t)=F_{p,k}^n+ t \partial_t F_{p,k}^n  .
\end{align*}
The coefficients $F_{p,k}^n$ and $\partial_tF_{p,k}^n$ can be
fully determined by
\begin{align*}
F_{p,k}(W^n,t_n)\Delta t&+\frac{1}{2}\partial_t
F_{p,k}(W^n,t_n)\Delta t^2 =\mathbb{F}_{p,k}(W^n,\Delta t) , \\
\frac{1}{2}F_{p,k}(W^n,t_n)\Delta t&+\frac{1}{8}\partial_t
F_{p,k}(W^n,t_n)\Delta t^2 =\mathbb{F}_{p,k}(W^n,\Delta t/2).
\end{align*}
By solving the linear system, we have
\begin{equation}\label{second}
\begin{aligned}
F_{p,k}(W^n,t_n)&=(4\mathbb{F}_{p,k}(W^n,\Delta t/2)-\mathbb{F}_{p,k}(W^n,\Delta t))/\Delta t,\\
\partial_t F_{p,k}(W^n,t_n)&=4(\mathbb{F}_{p,k}(W^n,\Delta t)-2\mathbb{F}_{p,k}(W^n,\Delta t/2))/\Delta
t^2.
\end{aligned}
\end{equation}
After determining the numerical fluxes and their time derivatives in the above equations,
$\mathcal{L}(W_i^n)$ and $\frac{\partial}{\partial t}\mathcal{L}(W_i^n)$ can be obtained
\begin{align}
\mathcal{L}(W_i^n)&= -\frac{1}{\left| \Omega_i \right|} \sum_{p=1}^m
\sum_{k=1}^{2} \omega_k \textbf{F}(\textbf{x}_{p,k},t_n)\cdot \textbf{n}_p,\label{operator-1}\\
\frac{\partial}{\partial t}\mathcal{L}(W_i^n)&= -\frac{1}{\left| \Omega_i \right|} \sum_{p=1}^m\sum_{k=1}^{2} \omega_k
\partial_t \textbf{F}(\textbf{x}_{p,k},t_n)\cdot \textbf{n}_p \label{operator-2}.
\end{align}
According to Eq.\eqref{step-hyper-1}, $W_{i}^*$ at $t_*$ can be updated. With the similar procedure, the numerical fluxes and temporal derivatives at the intermediate stage can be constructed and  $\frac{\partial}{\partial t}\mathcal{L}(W_{i}^*)$ is given by
\begin{align}
\frac{\partial}{\partial t}\mathcal{L}(W_{i}^*)&=-\frac{1}{\left| \Omega_i \right|} \sum_{p=1}^m\sum_{k=1}^{2} \omega_k
\partial_t \textbf{F}(\textbf{x}_{p,k},t_*)\cdot \textbf{n}_p\label{operator-3}.
\end{align}
Therefore, with the solutions Eq.\eqref{operator-1}, Eq.\eqref{operator-2}, and Eq.\eqref{operator-3},
$W_{i}^{n+1}$ at $\displaystyle t_{n+1}$ can be updated by Eq.\eqref{step-hyper-2}.

Different from the Riemann problem with a constant state at a cell interface,
the gas-kinetic scheme provides a time evolution solution.
Taking moments of the time-dependent distribution function in Eq.(\ref{flux}), the
pointwise values at a cell interface can be obtained
\begin{align}\label{point}
W_{p,k}(t)=\int\psi f(x_{p,k},t,u,v,\xi) d\Xi .
\end{align}
Similar to the two-stage temporal discretization for flux,
the time dependent gas distribution function at a cell interface is updated as

\begin{equation}\label{step-du}
\begin{split}
&f^*=f^n+\frac{1}{2}\Delta tf_t^n,\\
&f^{n+1}=f^n+\Delta tf_{t}^*,
\end{split}
\end{equation}
where a second-order evolution model is used for the update of gas distribution function on the cell interface and for the
evaluation of flow variables.

This temporal evolution for the interface value is similar to the one used in GRP solver \cite{du2018hermite}, which has a 4th-order
accuracy for a compact scheme in rectangular mesh, but may not give the same accuracy for the current scheme in unstructured triangular mesh.
However, numerical accuracy tests demonstrate that it is enough for a 3rd-order temporal accuracy.

In order to construct the first order time derivative of the gas distribution function, the distribution
function in Eq.(\ref{flux}) is approximated by the
linear function
\begin{align*}
f(t)=f(x_{p,k},y_{p,k},t,u, v,\xi)=f^n+
f_{t}^n(t-t^n).
\end{align*}
According to the gas-distribution function at
$t=0$ and $\Delta t$
\begin{align*}
f^n&=f(0),\\
f^n&+f_{t}^n\Delta t=f(\Delta t),
\end{align*}
the coefficients $f^n, f_{t}^n$ can be
determined by
\begin{align*}
f^n&=f(0),\\
f^n_t&=(f(\Delta t)-f(0))/\Delta t.\\
\end{align*}
Thus, $f^*$ and $f^{n+1}$ are fully determined at the cell interface for the evaluation of macroscopic flow variables.

The above time  accuracy could be kept for the simulation of inviscid smooth flow.
For dissipative terms, the theoretical accuracy can be only first order in time, and the details can  found in \cite{Pan2016twostage}.
By taking benefits of the above two-stage time stepping method, the current compact GKS with 2nd-order flux function could achieve the expected time accuracy, which reduces the computational cost significantly in comparison with the
early one step third-order compact scheme with a 3rd-order flux function \cite{pan2016unstructuredcompact}.


After obtaining $W^{n+1}_{p,k}$ at each Gauss point at cell interfaces,
the cell-averaged first-order derivatives within each triangle can be evaluated

\begin{align} \label{green-formula}
\bar{W}_x^{n+1}&= \frac{1}{ \Delta S}\oint_{\Gamma} W^{n+1} dy =\frac{1}{ \Delta S} \sum_{0}^{m}\sum_{0}^{2} \omega_kW^{n+1}_{p,k}cos \alpha _{p}|l|_p,
\\
\bar{W}_y^{n+1}&=-\frac{1}{ \Delta S}\oint_{\Gamma} W^{n+1}dx=
-\frac{1}{ \Delta S}\sum_{0}^{m}\sum_{0}^{2}\omega_k W^{n+1}_{p,k}cos \beta _{p}|l|_p,
\end{align}
where $\alpha_{p}$ is angle of tangential direction of each edges with  positive $y$ direction, $\beta_{p}$ is angle of tangential direction of each edge with  positive $x$ direction, and $|l|_{p}$ is the length of each edge.
The cell-averaged derivatives will be referred as cell-averaged first-order derivatives for simplicity in the following.

\begin{figure}[!h]
    \centering
    \includegraphics[width=0.4\textwidth]{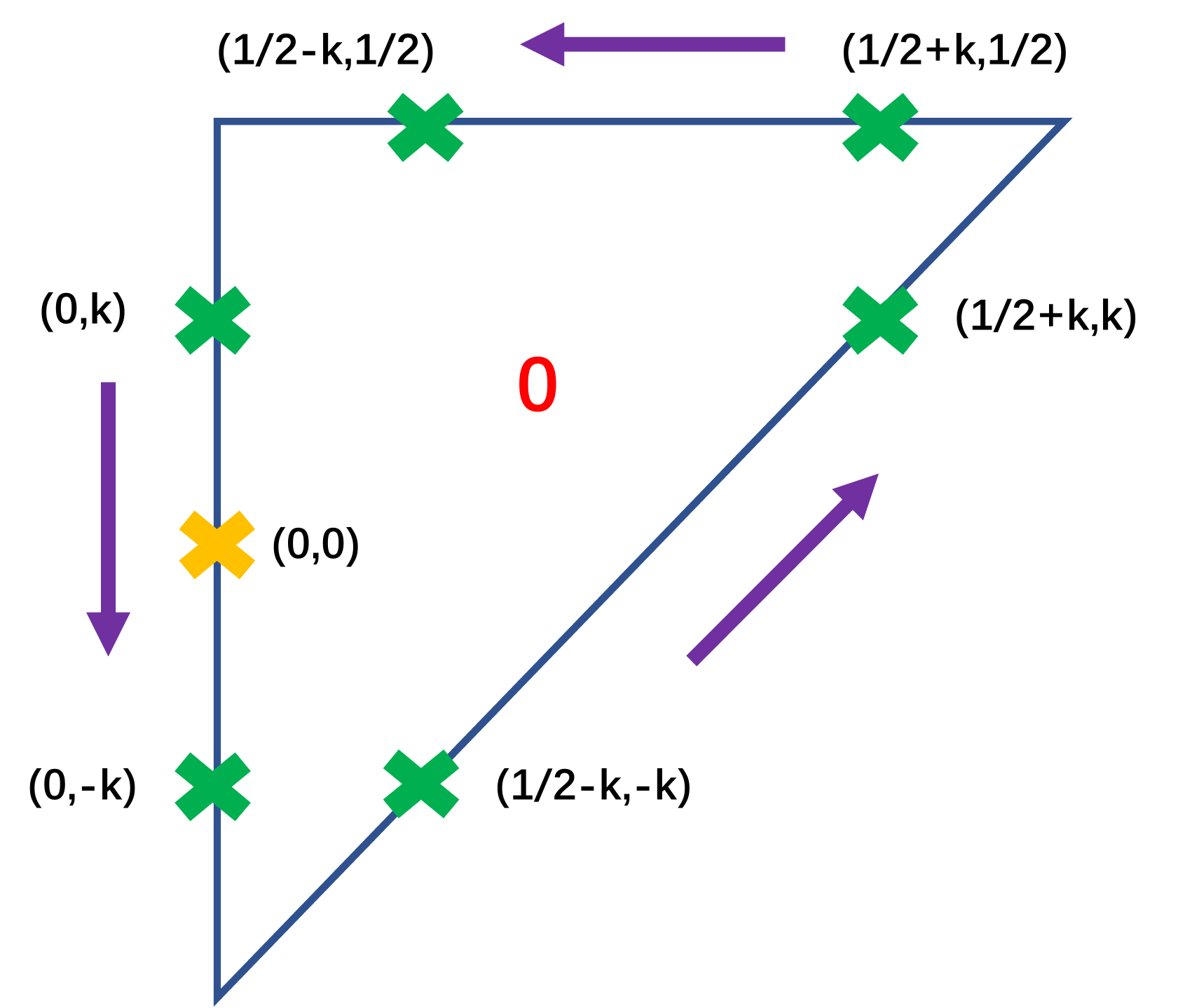}
    \caption{Tangential direction for an isosceles right cell 0 shown by purple arrows. The length of right-angle side is 1. The locations of Gauss points around it are labeled, where $k=\sqrt{3}/6$.}
    \label{green-gauss}
\end{figure}

The tangential direction is determined by "right hand rule", a sketch is shown in Fig.\ref{green-gauss}

\begin{rmk}
    For a better illustration of the cell-averaged derivatives evaluation, we investigate the accuracy of the numerical approximation of the average derivatives calculated by \eqref{green-formula} within an isosceles right cell 0 shown in Fig.\ref{green-gauss}.

    Assume a certain fluid variable is distributed as
    \begin{align}\label{w1}
    W(x,y)= a_0+a_1x+a_3 x^2 + a_4 y^2 + a_5 xy +a_6 x^3 +a_7 y^3 +a_8 x^2 y +a_9 xy^2,
    \end{align}
    taking x derivative of $W$, we have
    \begin{align}\label{w2}
    W_x= a_1+2 a_3 x + a_5 y +3 a_6 x^2 +2 a_8 x y +a_9 y^2.
    \end{align}
    Thus the exact averaged x derivative over cell 0 is
    \begin{align}\label{w3}
    \bar{W}_x=\frac{1}{ \Delta S} \iint W_x dxdy= a_1+ \frac{2}{3} a_3  + \frac{1}{6} a_5 +\frac{1}{2} a_6 + \frac{1}{6} a_8 +\frac{1}{12} a_9.
    \end{align}
    The numerical x derivative over cell 0 by Eq.\eqref{green-formula} is
    \begin{align*}\label{w5}
    \bar{W}_x^h&=\frac{1}{ 2 \Delta S} (W(1/2+k,k)+W(1/2-k,-k)-W(0,k)-W(0,-k)) \\
    &= a_1+ (\frac{1}{2}+2k^2) a_3 +2 k^2 a_5 + (\frac{1}{4}+3k^2) a_6 + 2 k^2 a_8 + k^2 a_9\\
    &= a_1+ \frac{2}{3} a_3  + \frac{1}{6} a_5 +\frac{1}{2} a_6 + \frac{1}{6} a_8 +\frac{1}{12} a_9.
    \end{align*}
    Compared with \eqref{w3}, the above numerical calculation from Gaussian points around the triangle gives a fourth-order
    accurate representation of the cell averaged gradients. Note that the above derivatives are obtained from the time accurate
    evolution solution of flow variables at cell interfaces, which are different from the directly updated derivatives in the DG methods.
    Here in the compact GKS there is no any direct numerical evolution equation for the cell averaged derivative.

\end{rmk}

\section{HWENO Reconstruction}
Our target in this section is to reconstruct reliable pointwise values and first-order derivatives at each Gaussian point on the cell interface.
Once the initial reconstruction is done, the corresponding time-dependent gas distribution function at that point is fully determined.
\subsection{Linear reconstruction}

As a starting point of WENO reconstruction, a linear reconstruction
will be presented. For
a piecewise smooth function $W(x,y)$ over cell $\Omega_{i}$, a
polynomial $P^r(x,y)$ with degrees $r$ can be constructed to
approximate $W(x,y)$ as follows
\begin{equation*}
P^r(x,y)=W(x,y)+O(\Delta x^{r+1},\Delta y^{r+1}).
\end{equation*}
In order to achieve a third-order accuracy and satisfy
conservative property, the following quadratic
polynomial over cell $\Omega_{i_0}$ is constructed
\begin{equation}\label{qua-def}
P^2(x,y)=W_{i_0}+\sum_{k=1}^5a_kp^k(x,y),
\end{equation}
where $W_{i_0}$ is the cell averaged value of $W(x,y)$ over cell $\Omega_{i_0}$ and
$p^k(x,y), k=1,...,5$ are basis functions, which are given by
\begin{align}\label{base}
\begin{cases}
\displaystyle p^1(x,y)=x-\frac{1}{\left| \Omega_{i_0} \right|}\displaystyle\iint_{\Omega_{i_0}}x dxdy, \\
\displaystyle p^2(x,y)=y-\frac{1}{\left| \Omega_{i_0} \right|}\displaystyle\iint_{\Omega_{i_0}}y dxdy, \\
\displaystyle p^3(x,y)=x^2-\frac{1}{\left| \Omega_{i_0} \right|}\displaystyle\iint_{\Omega_{i_0}}x^2 dxdy, \\
\displaystyle p^4(x,y)=y^2-\frac{1}{\left| \Omega_{i_0} \right|}\displaystyle\iint_{\Omega_{i_0}}y^2 dxdy, \\
\displaystyle p^5(x,y)=xy-\frac{1}{\left| \Omega_{i_0}
    \right|}\displaystyle\iint_{\Omega_{i_0}}xy dxdy.
\end{cases}
\end{align}

\subsubsection{Large stencil and sub-stencils}

\begin{figure}[!h]
	\centering
	\subfigure[]{
		\includegraphics[height=0.25\textwidth]{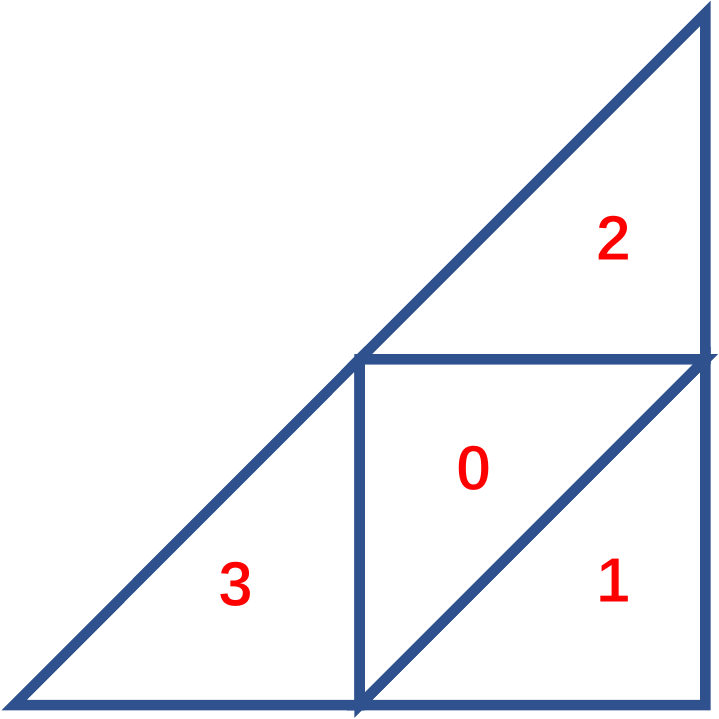}}
	\qquad
	\quad
	\subfigure[]{	
		\includegraphics[height=0.25\textwidth]{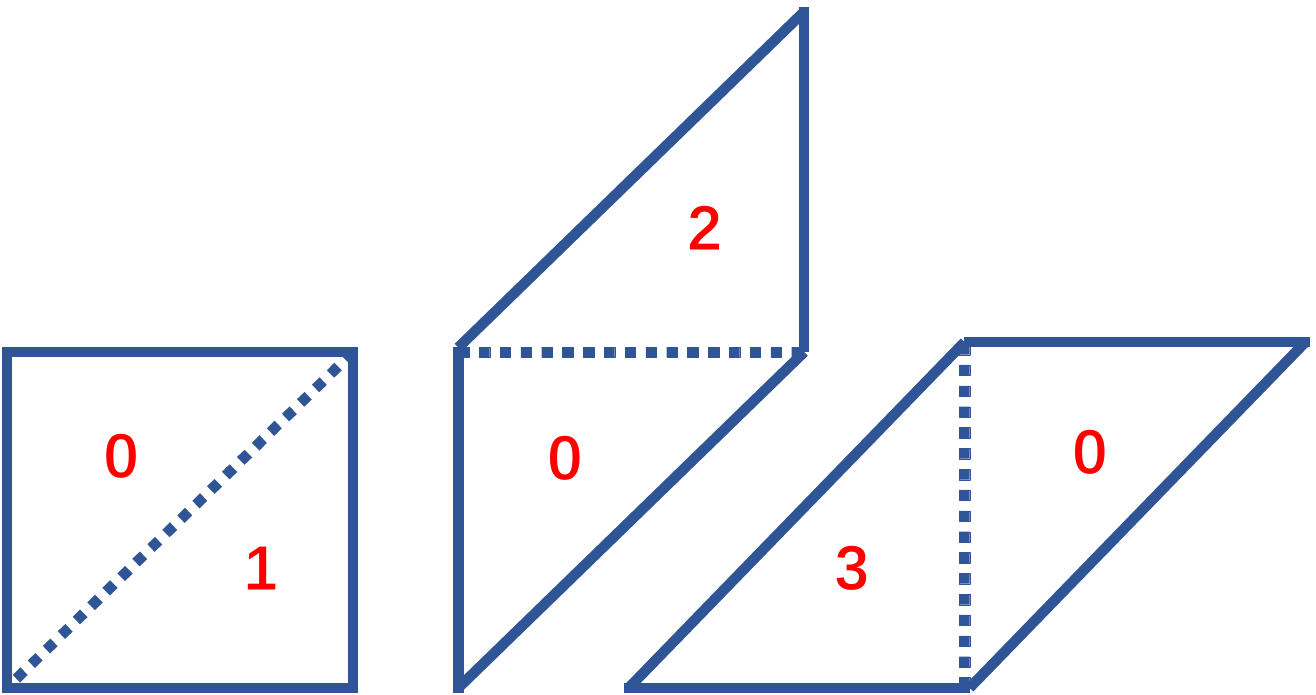}}
	\caption{The large stencils with the inclusion of (a) cell averaged values $\bar{W}$ for cell 0, 1, 2, 3
		(b) cell averaged derivatives $\bar{W}_x$ and $\bar{W}_y$ for cell "0\&1", "0\&2", "0\&3".}
	\label{stencil-a}
\end{figure}

In order to reconstruct the quadratic polynomial
$P^2(x,y)$, the stencil for reconstruction is selected in Fig. \ref{stencil-a}, where the averages of $W(x,y)$ and averaged derivatives
of $W(x,y)$ over each cell are known. For the current
3rd-order scheme, the following values are used to obtain
$P^2(x,y)$.

\begin{itemize}
    \item Cell averages $\bar{W}$ for cell 0, 1, 2, 3
    \item Cell averages of the x partial derivative $\bar{W}_x$ for cell "0\&1", "0\&2", "0\&3"
    \item Cell averages of the y partial derivative $\bar{W}_y$ for cell "0\&1", "0\&2", "0\&3"
\end{itemize}

To determine the polynomial $P^2(x,y)$, the following conditions can be used
\begin{align*}
\iint_{\Omega_{i_j}}P^2(x,y)\text{d}x\text{d}y=W_{i_j}\left| \Omega_{i_j}\right|,\\
\iint_{\Omega_{i_0}+\Omega_{i_j}}
\frac{\partial}{\partial x} P^2(x,y)\text{d}x\text{d}y=W_{x,i_0}|\Omega_{i_0}|+W_{x,i_j}|\Omega_{i_j}|,\\
\iint_{\Omega_{i_0}+\Omega_{i_j}}
\frac{\partial}{\partial y} P^2(x,y)\text{d}x\text{d}y=W_{y,i_0}|\Omega_{i_0}|+W_{y,i_j}|\Omega_{i_j}|,\\
\end{align*}
 where $W_{i_j}$ is the cell averaged value over
$\Omega_{i_j}$, $W_{x, i_j}$ and $W_{y, i_j}$ are the cell averaged $x$ and $y$
derivatives over $\Omega_{i_j}$ in a global coordinate,
respectively. On a regular mesh, the system has $10$ independent
equations. To solve the system uniquely and avoid the singularity
caused by the irregularity in the mesh, the technique in \cite{zhao2017weighted}
has been adopted.

\begin{figure}[!h]
	\centering
		\includegraphics[height=0.25\textwidth]{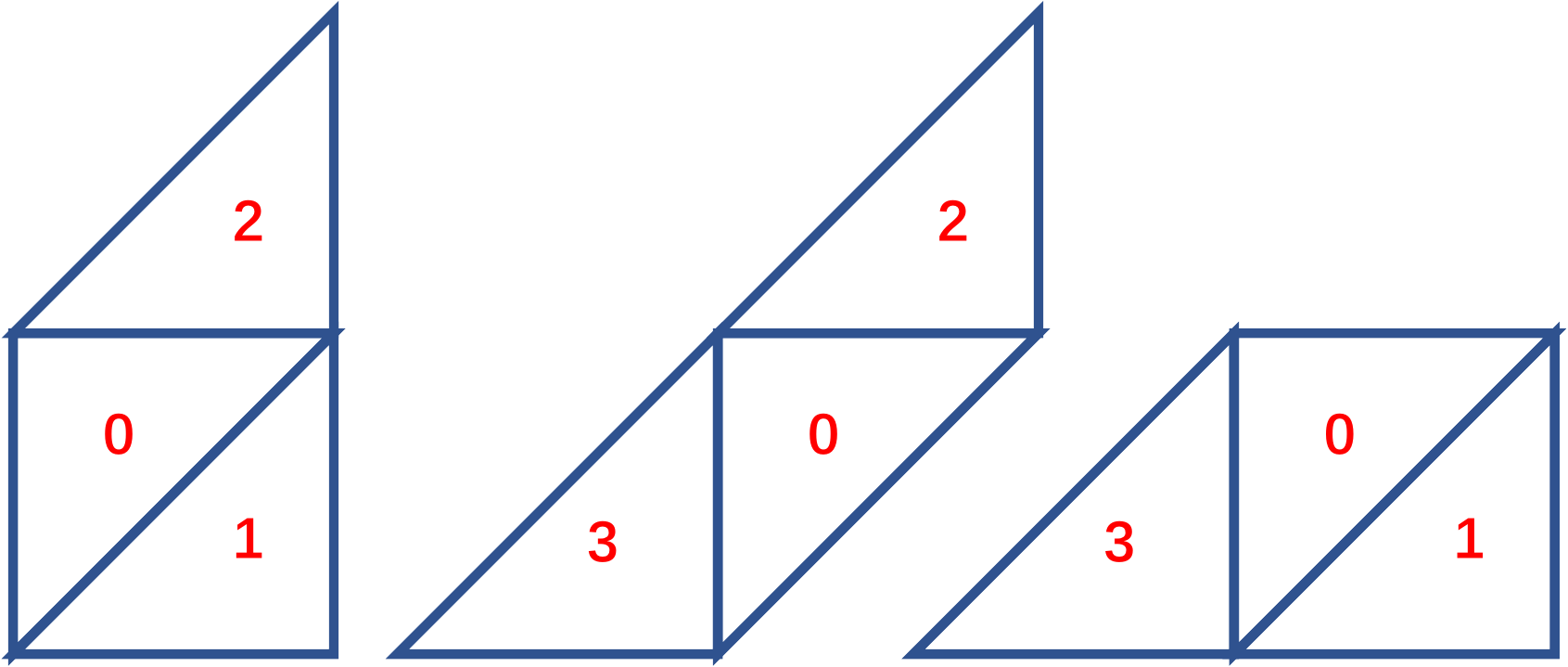}
	\caption{The sub-stencils for compact HWENO reconstruction. From left to right, sub-stencils 1, 2, 3.}
	\label{stencil-b}
\end{figure}

In order to deal with discontinuity, inspired by the existing WENO
reconstruction \cite{zhu2018new}, three sub-stencils $S_{j},
j=1,2,3$ are selected from the large stencil given in
Fig. \ref{stencil-b}. And the following cell averaged values for each sub-stencil are
used to get the linear polynomial $P^1_j(x,y)$,
\begin{align*}
P_{1}^1 ~&\text{on}~ S_1=\{\bar{W}_0,\bar{W}_1,\bar{W}_2\},
~~~P_{2}^1 ~\text{on}~ S_2=\{\bar{W}_0,\bar{W}_2,\bar{W}_3\},
~~~P_{3}^1 ~\text{on}~ S_3=\{\bar{W}_0,\bar{W}_3,\bar{W}_1\} .\\
\end{align*}
Through this process, for a targeting cell, there is always one sub-stencil in
smooth region even with the appearance of discontinuity near any one of the
cell interfaces. For $j=1,2,3$, the technique
\cite{zhao2017weighted} can be used to obtain $P^1_j(x,y)$, and the
linear polynomial can be expressed as
\begin{equation}\label{linear-def}
P^1_j(x,y)=W_{i_0}+\sum_{k=1}^2a_{j,k}p^k(x,y).
\end{equation}

\subsubsection{Define the values of linear weights}
We rewrite $P^2(x,y)$ as
\begin{align*}
P^2(x,y)=\gamma_0[\frac{1}{\gamma_0} P^2(x,y) - \sum_{j=1}^{3} \frac{\gamma_{j}}{\gamma_0} P_{j}^1(x,y)]+\sum_{j=1}^{3} \gamma_{j} P_{j}^1(x,y) ,
\end{align*}
where the linear weights are chosen as $\gamma_0=0.97, \gamma_1 =\gamma_2=\gamma_3= 0.01$ \cite{zhu2018new} without special statement.

 \subsubsection{Compute the nonlinear weights}

 The smoothness indicators $\beta_{j}, j=0,1,2,3$ are defined as
\begin{equation*}
\beta_j=\sum_{|\alpha|=1}^{K}|\Omega|^{|\alpha|-1}\iint_{\Omega}\big(D^{\alpha}P_j(x,y)\big)^2dxdy,
\end{equation*}
where $\alpha$ is a multi-index and $D$ is the derivative operator. It has be proved in \cite{zhu2018new} that the smoothness indicators in Taylor series at $(x_0,y_0)$ have the order
\begin{align*}
\beta_0&=O\{|\Delta_0|[1+O(|\Delta_0|)]\}=O(|\Delta_0|),\\
\beta_j&=O\{|\Delta_0|[1+O(|\Delta_0|^{\frac{1}{2}})]\}=O(|\Delta_0|),j=1,2,3.
\end{align*}
By using a similar technique \cite{zhu2018new}, we can define a global smoothness indicator $\sigma$ as
\begin{equation*}
\sigma = (|2\beta_0-\beta_1-\beta_2|+|2\beta_0-\beta_2-\beta_3|+|2\beta_0-\beta_1-\beta_3|)^2 = O(|\Delta_0|^3),
\end{equation*}
then the corresponding non-linear weights are defined by
\begin{equation}\label{non-linear-weight}
\delta_{j}=\frac{\omega_{j}}{\sum_{l=0}^{3}\omega_{l}},
~~\omega_{j}=\gamma_{j}(1+\frac{\sigma}{\epsilon+\beta_{j}}), j = 0,1,2,3,
\end{equation}
where $\epsilon$ takes $10^{-8}$ to avoid zero in the denominator.

The final reconstruction polynomial for the approximation of $W(x,y)$ yields
\begin{equation} \label{final_hweno_expression}
R(x,y)=\delta_0[\frac{1}{\gamma_0} P^2(x,y) - \sum_{j=1}^{3} \frac{\gamma_{j}}{\gamma_0} P_{j}^1(x,y)]+\sum_{j=1}^{3} \delta_{j} P_{j}^1(x,y).
\end{equation}

From Eq.\eqref{non-linear-weight}, the  non-linear weights approximate linear weights with  $\delta_j=\gamma_j+O(|\Delta_0|^2)$, which satisfy the required accuracy condition $\delta_j=\gamma_j+O(|\Delta_0|)$ \cite{wenoz}. As a result, the nonlinear reconstruction $R(x,y)$ achieves a 3rd-order accuracy  $R(x,y)=W(x,y)+O(|\Delta|^3)$.

\subsection{Derivative reconstruction for non-equilibrium and equilibrium parts }

Once the conservative variables at each Gaussian points are constructed, a quadratic polynomial could be reconstructed using the cell average values $W_i$ on cell i and all the values
of the Gaussian points along the three edges of cell 0, $Q(x_{G_j},y_{G_j}), j=1,2...,6$, in a least square sense.
The derivatives for non-equilibrium parts could be obtained by the reconstructed quadratic polynomial.

Then the equilibrium state $g_0$ corresponding to the conservative variables $W_0$ is obtained by Eq.\eqref{compatibility2}.
The derivatives for equilibrium part are
constructed by simple averages of the derivatives of flow variables on both sides of the interface for the
 construction of non-equilibrium distribution functions, which are  effective in all test cases in the current paper.

\subsection{Additional limiting technique in exceptional cases}
In most cases, the above HWENO reconstruction could give physically reliable values in the determination of gas distribution functions in \eqref{flux}. However in extreme cases, i.e., the Mach number $20$ hypersonic flow passing through a cylinder under irregular meshes,
the above procedure gives unreasonable large deviations in the smooth indicators. A simple limiter is added in the above reconstruction scheme, such as
\begin{align*}
\mbox{when} \quad \beta_0 > \max (100 \beta_j), \quad j=1,2,3 \quad \mbox{then} \quad u(x,y)=W_{i_0}.
\end{align*}
It will be indicated explicitly once the above limiter is used in the test cases.
In fact, this criteria is so strong that it could hardly be triggered in most simulation cases.

\section{Numerical examples}
In this section, numerical tests will be presented to validate the compact high-order GKS. For the
inviscid flow, the collision time $\tau$ is
\begin{align*}
\tau=\epsilon \Delta t+C\displaystyle|\frac{p_l-p_r}{p_l+p_r}|\Delta
t,
\end{align*}
where $\varepsilon=0.01$ and $C=1$. For the viscous flow, the collision time is related to the viscosity coefficient,
\begin{align*}
\tau=\frac{\mu}{p}+C \displaystyle|\frac{p_l-p_r}{p_l+p_r}|\Delta t,
\end{align*}
where $p_l$ and $p_r$ denote the pressure on the left and right
sides of the cell interface, $\mu$ is the dynamic viscous coefficient, and
$p$ is the pressure at the cell interface. In  smooth flow regions,
it will reduce to $\tau=\mu/p$. The ratio of specific heats takes
$\gamma=1.4$. The reason for including pressure jump term
in the particle collision time is to add artificial dissipation
in the discontinuous region, where the numerical cell size is not enough to resolve the shock structure,
and the enlargement of collision time is to keep the non-equilibrium in the kinetic flux function
to mimic the real physical mechanism in the shock layer.

Same as many other high-order schemes, all reconstructions will be done on the
characteristic variables. Denote $F(W)=(\rho U, \rho U^2+p, \rho
UV,U(\rho E+p))$ in the local coordinate. The Jacobian matrix
$\partial F/\partial W$ can be diagonalized by the right eigenmatrix
$R$. For a specific cell interface, $R_*$ is the right eigenmatrix
of $\partial F/\partial W^*$, and $W^*$ are the Roe-averaged
conservative flow variables from both sided of the cell interface. The characteristic variables
for reconstruction are defined as $U=R_*^{-1}W$.
Generally, the CFL number could be safely taken around $0.5$ in the cases without extreme strong shocks.

\subsection{2-D sinusoidal wave propagation }
The advection of density perturbation is commonly used to test the order of accuracy of a numerical scheme
for the smooth Euler solution.
The initial condition is given as a sinusoidal wave propagating in the diagonal direction
\begin{align*}
&\rho(x,y)=1+0.2\sin(\pi (x + y)),
\\& U(x,y)=1,V(x,y)=1,p(x,y)=1,
\end{align*}
with the exact solution
\begin{align*}
&\rho(x,y,t)=1+0.2\sin(\pi (x+y-2t)),
\\&U(x,y,t)=1,V(x,y,t)=1, p(x,y,t)=1.
\end{align*}

For the current compact reconstruction, we also need the derivative information. For this sin-wave test case, the initial derivative distributions of primary variables are given by
\begin{align*}
&\partial_x{\rho(x,y)}=\partial_y{\rho(x,y)}=0.2 \pi \cos(\pi (x + y)),
\\& \partial_x{U(x,y)}=\partial_y{U(x,y)}=0,
\\&\partial_x{V(x,y)}=\partial_y{V(x,y)}=0,
\\& \partial_x{P(x,y)}=\partial_y{p(x,y)}=0,
\end{align*}
and the initial derivatives of conservative variables can be calculated by the chain rule.

The computational domain is $[0,2]\times[0,2]$ and $2 \times N{\times}N$ uniform triangular meshes are used with periodic boundary conditions in both directions. CFL number is 0.1 in this test.
First the linear weights are chosen as $\gamma_0=1.0, \gamma_{j}=0.0, j=1,2,3$, where a smooth polynomial will be constructed solely on the big stencil by a least square sense.
The $L^1, L^2$ and $L^\infty$ errors and convergence orders are presented in Table \ref{2d_accuracy1}.
The expected third order of accuracy is obtained.
Next the linear weights are chosen as $\gamma_0=0.97, \gamma_{j}=0.01, j=1,2,3$. In this case, the non-linear weights will take effects on coarse meshes due to the numerical discontinuities. It can be observed as the mesh size is refined from 1/5 to 1/10, as shown in Table \ref{2d_accuracy2}. With a continuous mesh refinement, the convergence orders tend to be uniform and the absolute errors are more close to the ones shown in Table  \ref{2d_accuracy2}. This demonstrates that the current reconstruction strategy with non-linear weights could keep third order accuracy even with possible existing extremum \cite{wenoz}, which is consistent with the proof in \cite{zhu2018new}.

\begin{table}[!h]
    \small
    \begin{center}
        \def\temptablewidth{1\textwidth}
        {\rule{\temptablewidth}{1pt}}
        \begin{tabular*}{\temptablewidth}{@{\extracolsep{\fill}}c|cc|cc|cc}

            mesh length & $L^1$ error & Order & $L^2$ error & Order& $L^{\infty}$ error & Order  \\
            \hline
            1/5&1.06878e-1&~ & 1.18091e-1 &~ &1.65136e-1&~\\
            1/10&6.82928e-3&3.97 & 7.46663e-3 &3.98 &1.05518e-2&3.97\\
            1/20&4.07219e-4&4.07 & 4.50552e-4 &4.05 &6.31324e-4&4.06\\
            1/40&2.37196e-5&4.10 & 2.63661e-5 & 4.09& 3.7281e-5&4.08\\
            1/80&1.21062e-6&4.29 & 1.34495e-6 & 4.29& 1.90213e-6&4.29\\
        \end{tabular*}
        {\rule{\temptablewidth}{0.1pt}}
    \end{center}
    \vspace{-4mm} \caption{\label{2d_accuracy1} Accuracy test for the 2-D sin-wave
        propagation: the linear weights are chosen as $\gamma_0=1.0, \gamma_{j}=0.0, j=1,2,3$. }
\end{table}

\begin{table}[!h]
    \small
    \begin{center}
        \def\temptablewidth{1\textwidth}
        {\rule{\temptablewidth}{1pt}}
        \begin{tabular*}{\temptablewidth}{@{\extracolsep{\fill}}c|cc|cc|cc}

            mesh length & $L^1$ error & Order & $L^2$ error & Order& $L^{\infty}$ error & Order  \\
            \hline
            1/5&1.23029e-1&~ & 1.40881e-1 &~ &2.17303e-1&~\\
            1/10&1.26119e-2&3.28 & 1.80215e-2 &2.97 &3.37488e-2&2.69\\
            1/20&6.17707e-4&4.35 & 7.36097e-4 &4.61 &1.48205e-3&4.50\\
            1/40&2.63653e-5&4.55 & 2.99133e-5 & 4.62& 5.22265e-5&4.82\\
            1/80&1.21406e-6&4.44 & 1.35057e-6 & 4.47& 2.08196e-6&4.65\\
        \end{tabular*}
        {\rule{\temptablewidth}{0.1pt}}
    \end{center}
    \vspace{-4mm} \caption{\label{2d_accuracy2} Accuracy test for the 2-D sin-wave
        propagation: the linear weights are chosen as $\gamma_0=0.97, \gamma_{j}=0.01, j=1,2,3$. }
\end{table}

\subsection{One-dimensional Riemann problem}
One-dimensional Riemann problems are well-designed and commonly-used to test the performance of a numerical scheme for compressible flow. It is important that the current scheme under irregular unstructured meshes could keep good performance in simulating these problems.
Several test cases will be used to evaluate the mesh adaptability and the computational efficiency of the current method.

\begin{rmk}
    There are three type of meshes are used to test the mesh adaptability in this paper, which is shown on the left side of Fig. \ref{sod-3d}. The first type contains only isosceles right triangles, referred as uniform mesh. The second type is generated through the "Frontal" algorithm, and the third one through the "Delaunay" algorithm by using the Gmsh \cite{geuzaine2009gmsh}. They are referred as regular and irregular meshes respectively.
\end{rmk}
\subsubsection{Sod problem}
The initial condition for the Sod test case is given as follows
\begin{equation*}
(\rho,U,p) = \begin{cases}
(1,0,1),  0\leq x<0.5,\\
(0.125, 0, 0.1), 0.5\leq x\leq1.
\end{cases}
\end{equation*}

\begin{figure}[!h]
    \centering
    \includegraphics[width=0.35\textwidth]{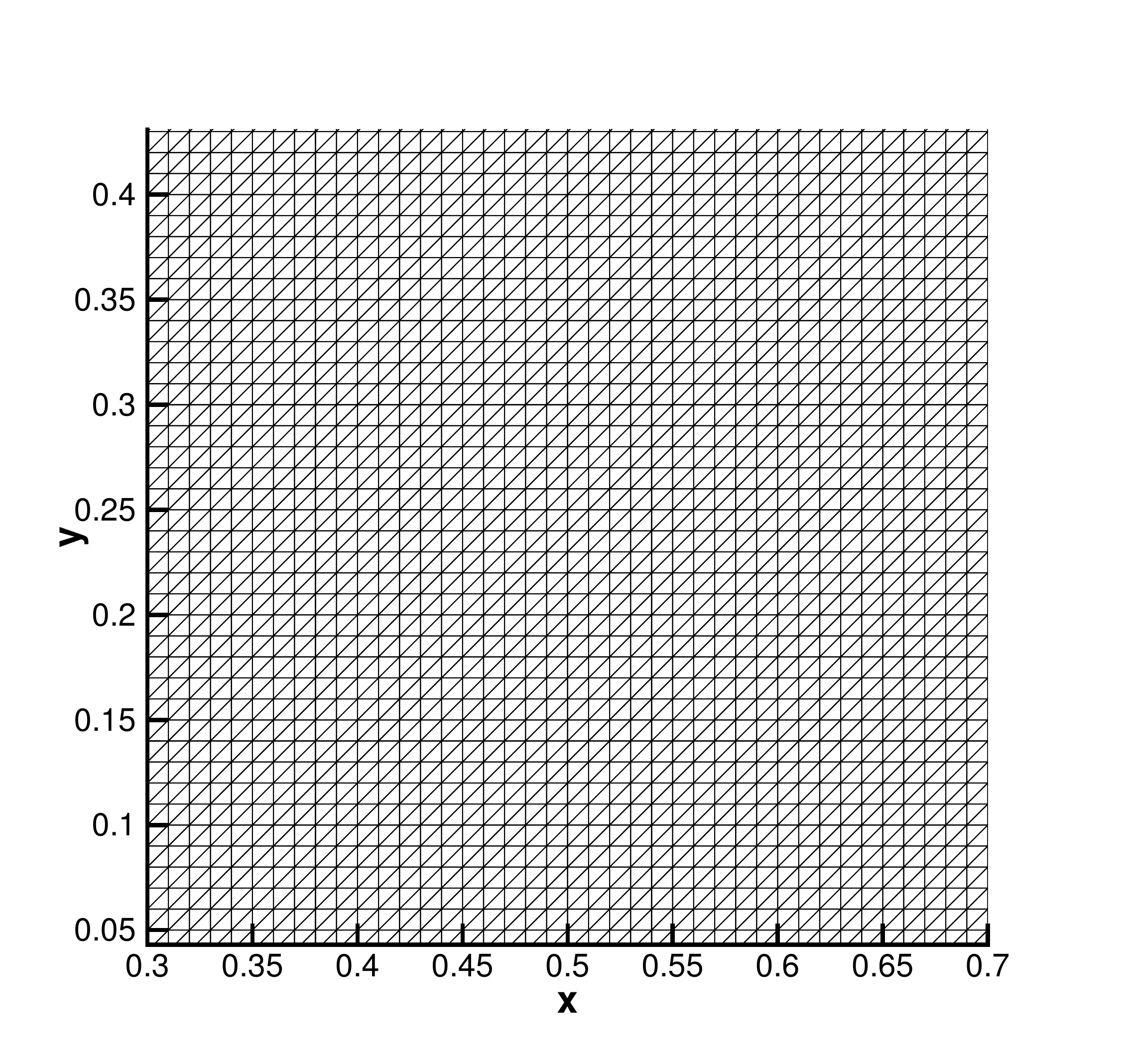}
    \includegraphics[width=0.35\textwidth]{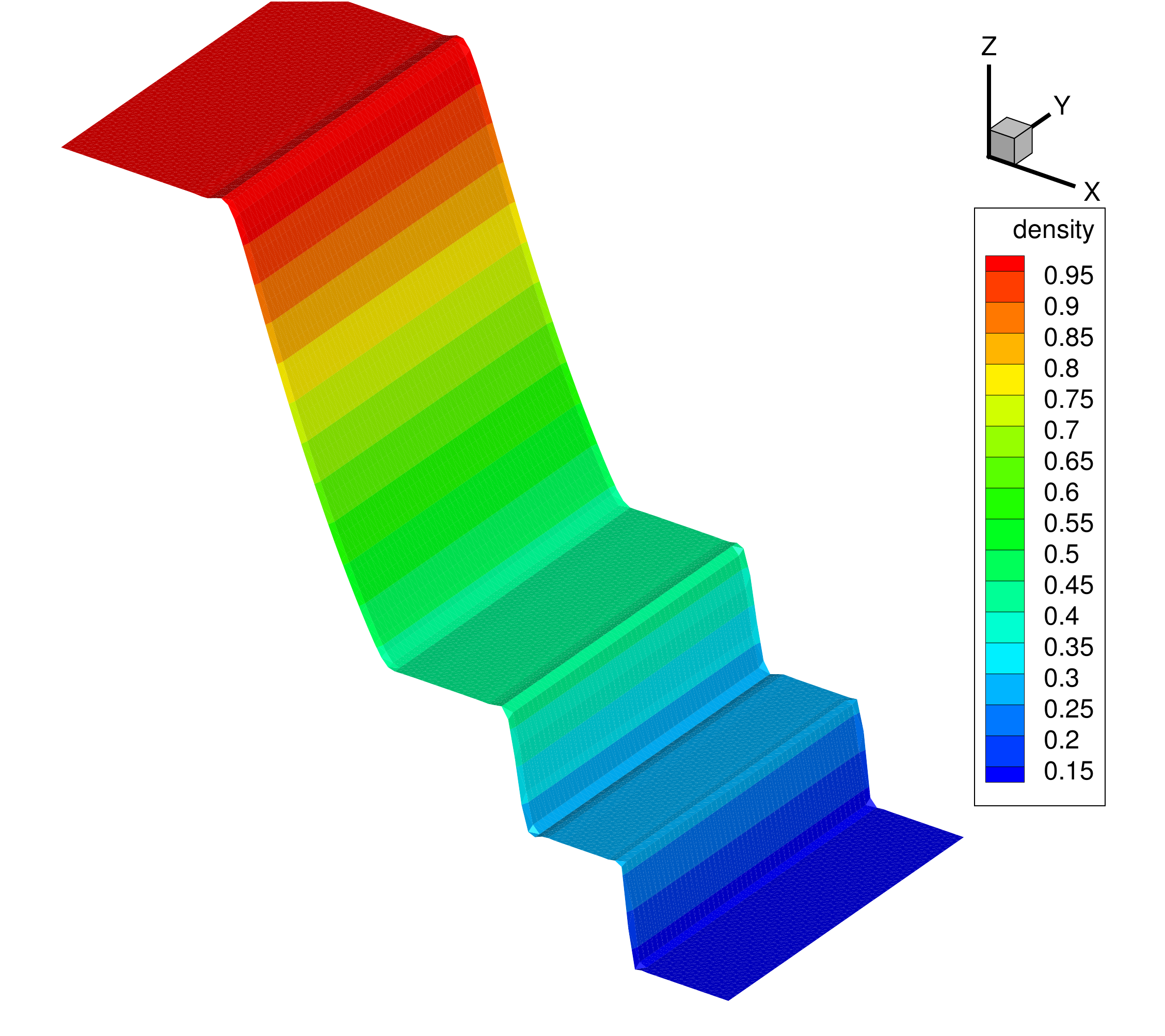}
    \includegraphics[width=0.35\textwidth]{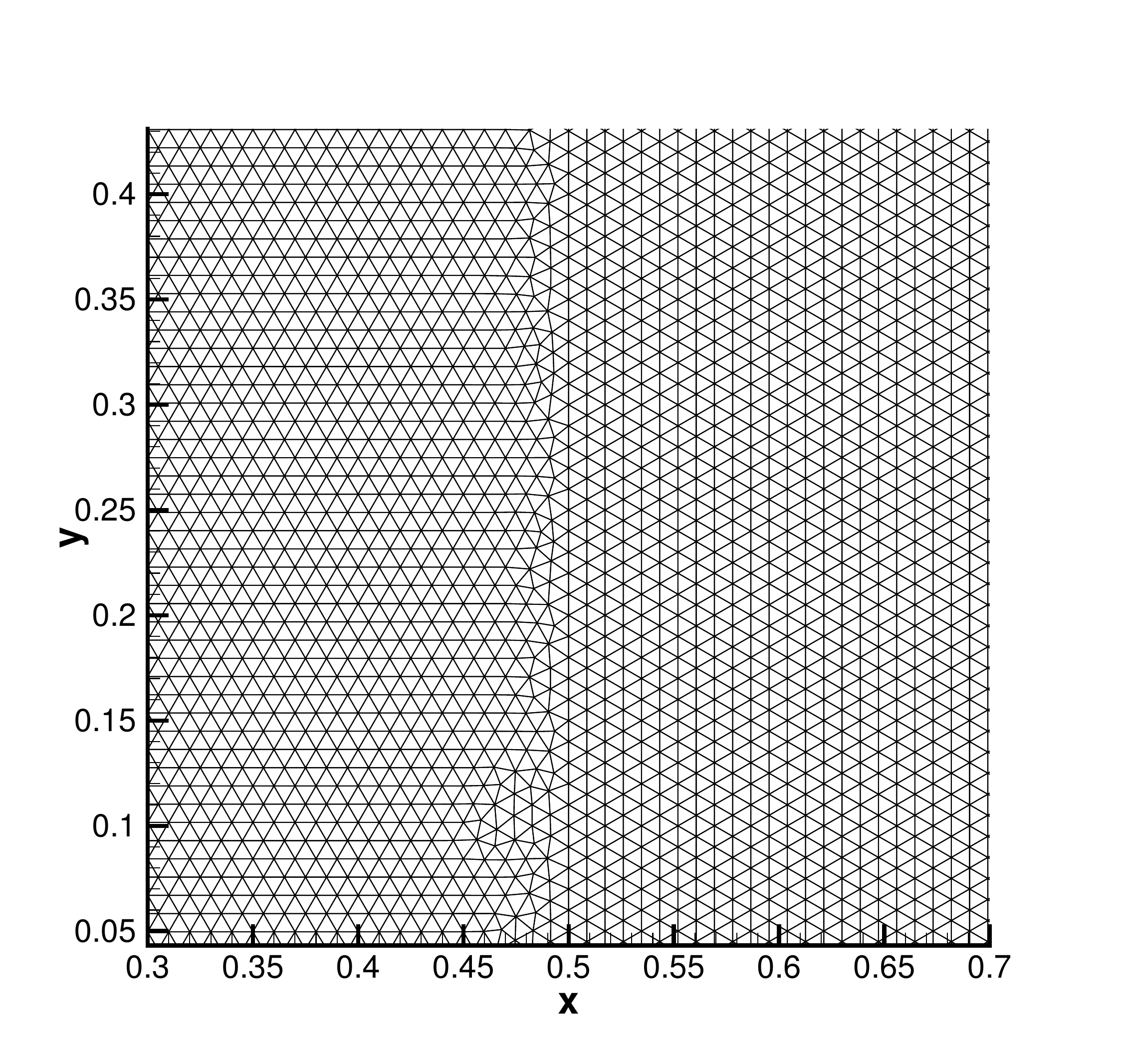}
    \includegraphics[width=0.35\textwidth]{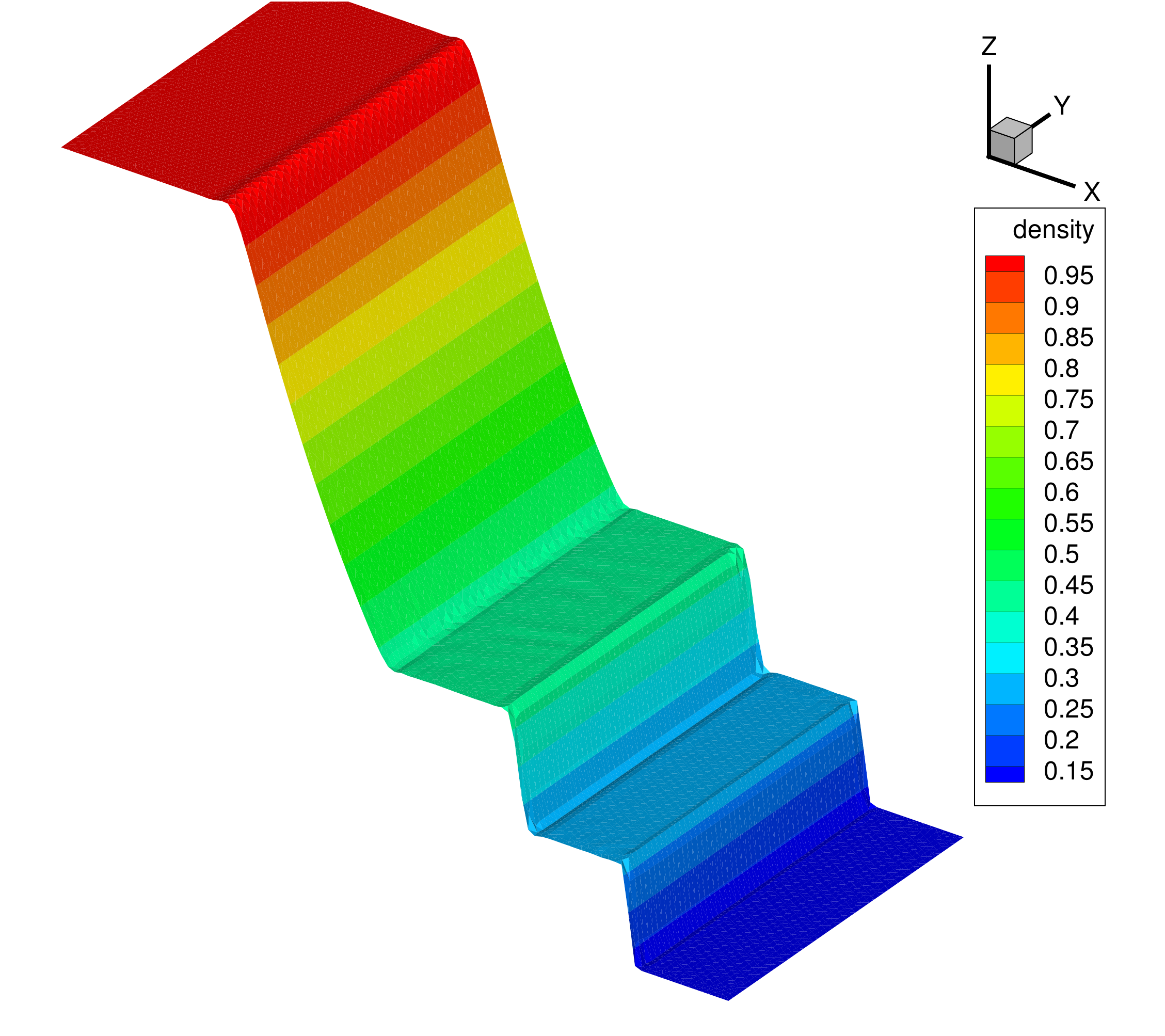}
    \includegraphics[width=0.35\textwidth]{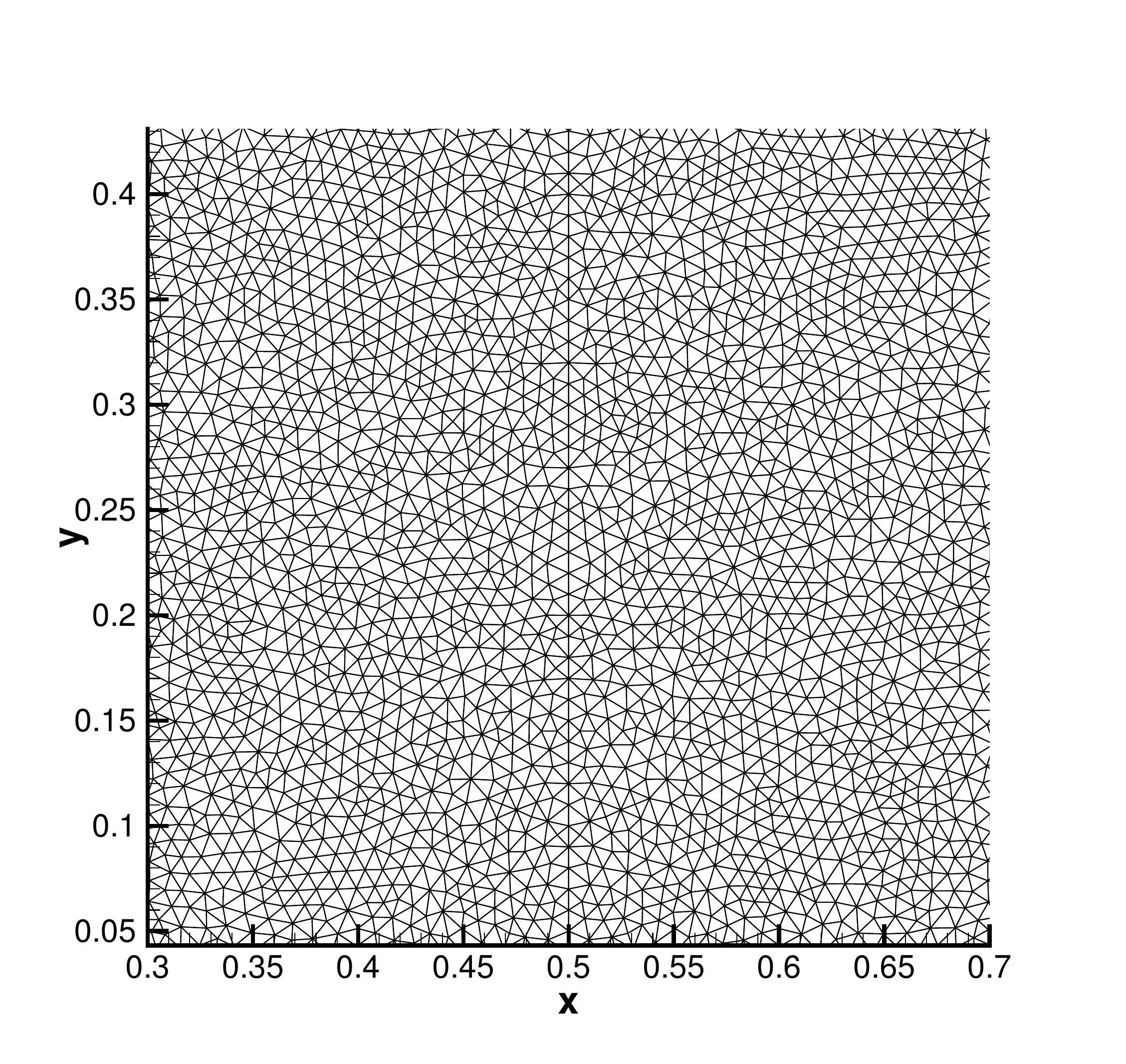}
    \includegraphics[width=0.35\textwidth]{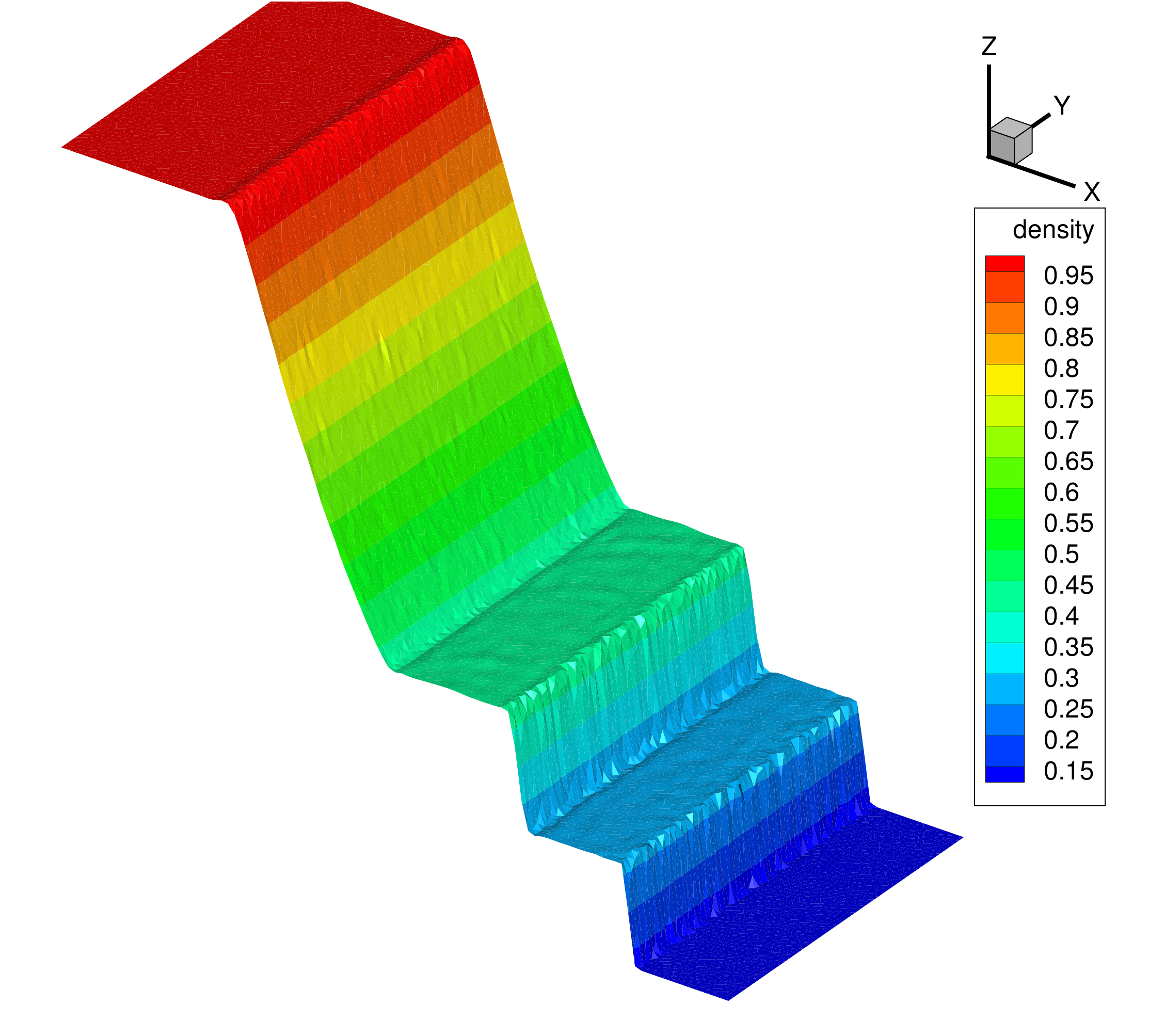}
    \caption{\label{sod-3d} Sod problem: the meshes and 3-D view of density distributions at $t=0.2$ with cell size $1/100$. The meshes are referred as uniform mesh , regular mesh  and irregular mesh  from the top to bottom. }
\end{figure}

\begin{figure}[!h]
    \centering
    \includegraphics[width=0.32\textwidth]{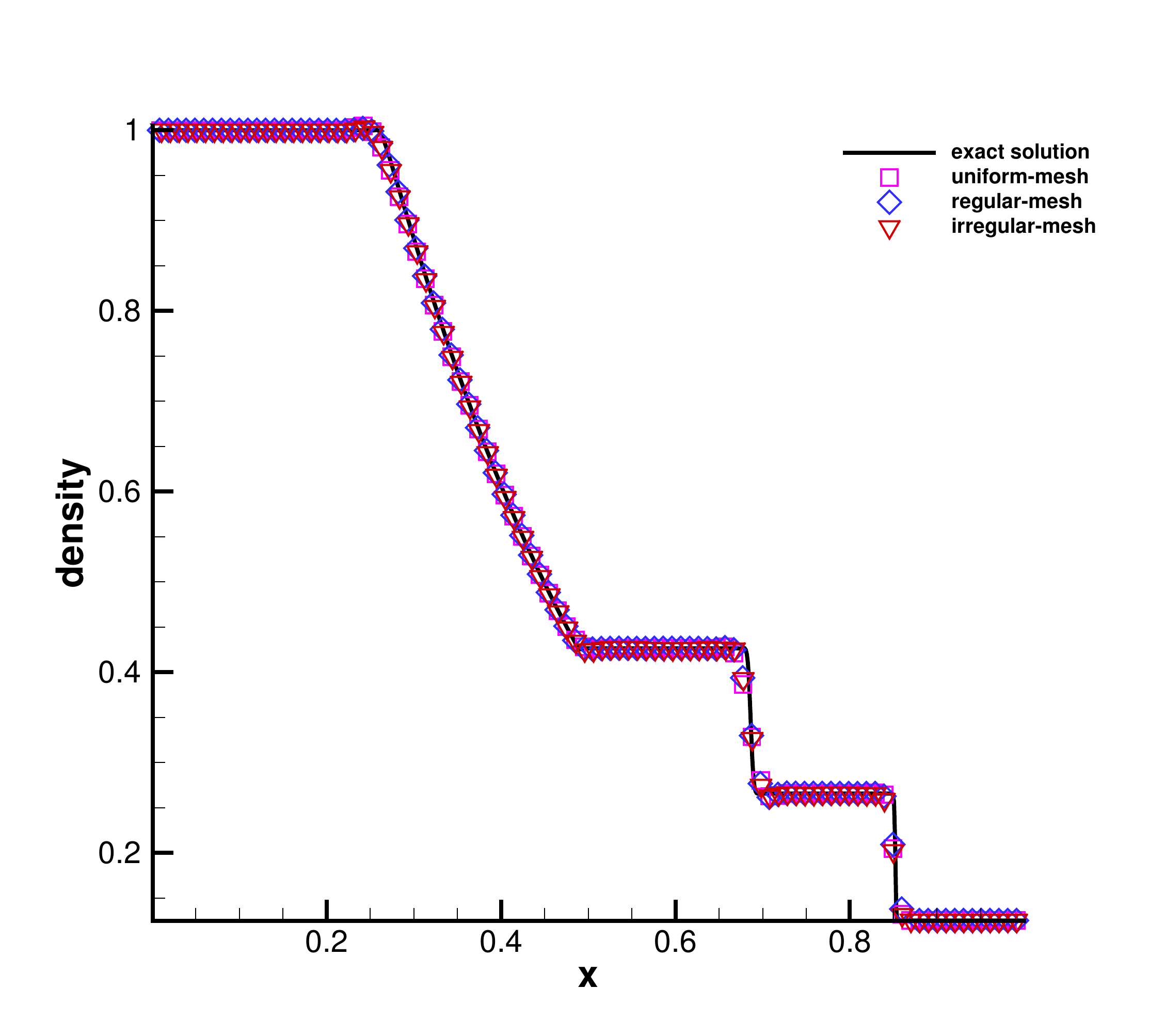}
    \includegraphics[width=0.32\textwidth]{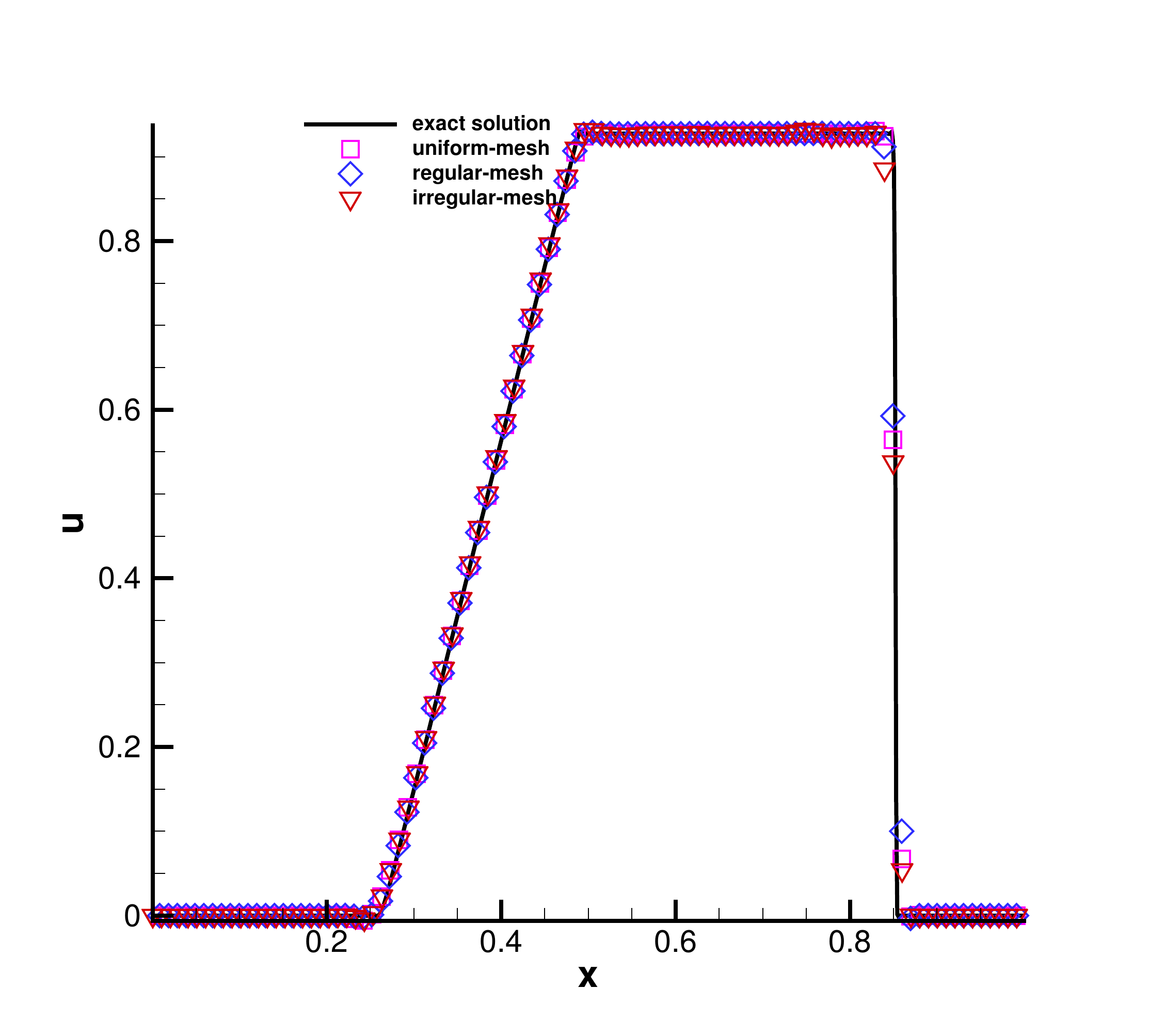}
    \includegraphics[width=0.32\textwidth]{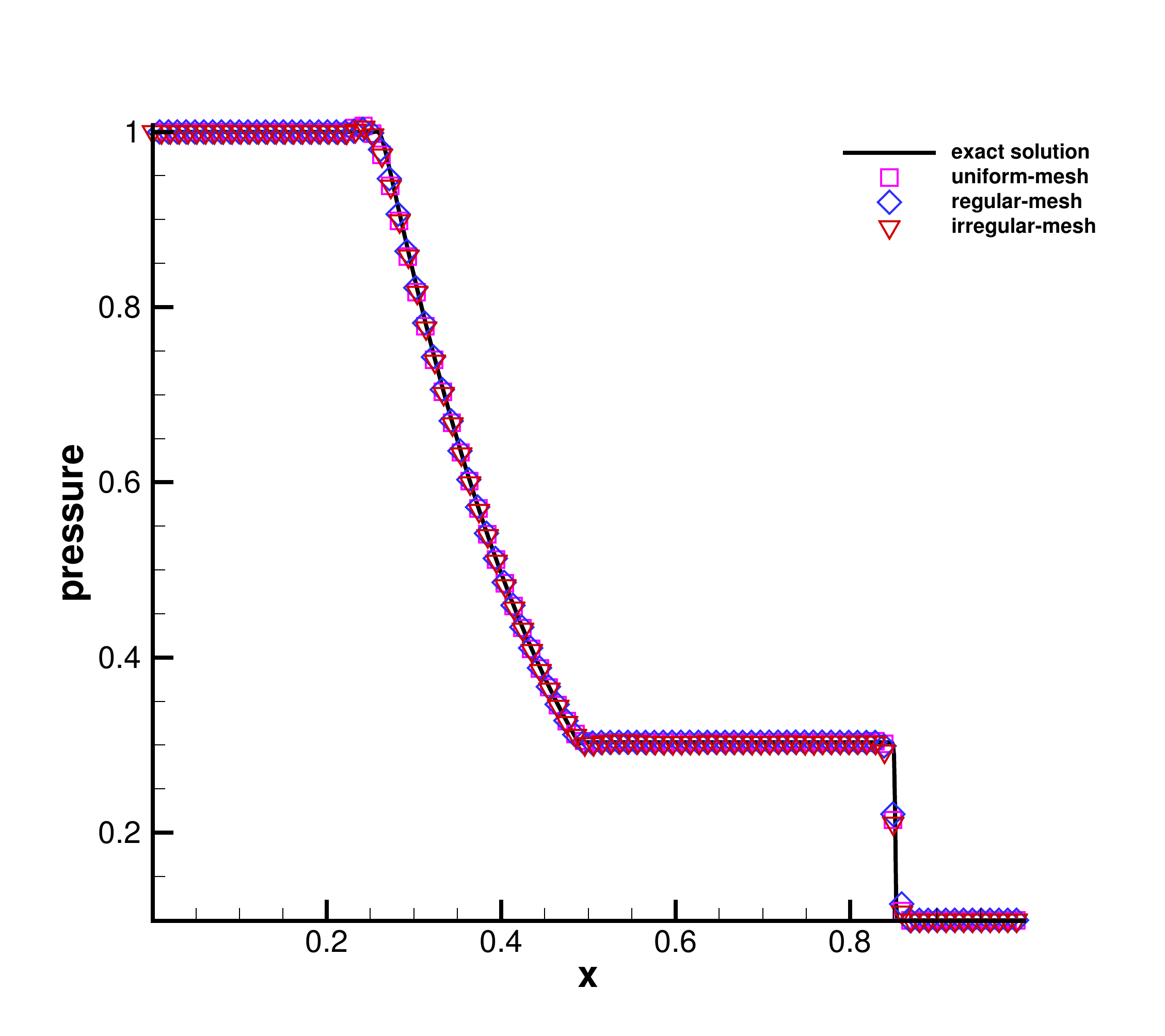}
    \caption{\label{sod-line} Sod problem: density, velocity, and pressure distributions at $t=0.2$ with cell size $1/100$ along the center horizontal lines under different meshes. }
\end{figure}

The computational domain is
$[0,1]\times[0,0.5]$, and the mesh size is $h=1/100$. Non-reflection
boundary condition is adopted at the left and right boundaries of the
computational domain, and periodic boundary condition is adopted at
the bottom and top boundaries of the computational domain.
The computations are preformed under uniform, regular, and irregular meshes.
The 3-D plot of density distributions in Fig. \ref{sod-3d} shows the uniformity in the flow distributions along $y$ direction
even with irregular mesh.
The density, velocity, and pressure distributions at the center horizontal line on
	different meshes are also extracted, shown in Fig. \ref{sod-line}.

\subsubsection{Computational efficiency}

Based on the above Sod test case, the computational efficiency from current method will be evaluated.
To make an assessment of the efficiency of the current scheme,
the comparisons with other schemes are conducted.
The CPU times are recorded after running $10$ explicit time steps for each scheme
with a single processor of Intel Xeon E5 2630v4 $@$2.10GHz in both cases.

First, the current method is compared with two different two-stage schemes: the original S2O4 GKS based on non-compact WENO5-Z reconstruction \cite{Pan2016twostage} and the compact S2O4 GKS based on HWENO reconstruction \cite{ji2018compact}.
Characteristic variables are used for reconstruction in all three schemes and two Gaussian points are used at each cell interface.
With the same number of cells, the new method is slightly slower than the other two schemes, as shown in Table \ref{cpu-time-1}.

Next, the schemes with different reconstruction methods and different time marching approaches are compared.
These methods are developed from the same in-house GKS codes.
The computational time is obtained in Table \ref{cpu-time-2}.
The second-order GKS on triangular mesh uses a min-mod limiter reconstruction, which is the same as that in \cite{pan2016unstructuredcompact}.
From the simulation results, the new compact-GKS is about $5$ times slower than the typical 2nd-order GKS and only about
$1/3$ times slower than the 2nd-order GKS if the same time discretization and Gaussian points are used.
The high efficiency of reconstruction used in the current scheme is mainly due to three reasons.
 Firstly, the coefficients for linear reconstruction could be all pre-stored in memory.
 Secondly, the linear weights are arbitrarily chosen and independent from geometry \cite{zhu2018new}.
 Thirdly, all the reconstruction procedure could be performed under a global coordinate.

\begin{table}[!h]
    \small
    \begin{center}
        \def\temptablewidth{1\textwidth}
        {\rule{\temptablewidth}{1pt}}
        \begin{tabular*}{\temptablewidth}{@{\extracolsep{\fill}}c|c|c|c}

            Scheme & No. of cell & No. of interface & CPU time  \\
            \hline
            Structured S2O4-WENO-GKS \cite{Pan2016twostage} &10000 &20200 & 3.51s\\
            Structured Compact S2O4-HWENO-GKS \cite{ji2018compact} &10000 &20200 & 4.00s \\
            Triangular Compact-GKS&10000&15150& 4.80s \\
        \end{tabular*}
        {\rule{\temptablewidth}{0.1pt}}
    \end{center}
    \vspace{-4mm} \caption{\label{cpu-time-1} Computational time (in seconds) of different schemes for the 1D Sod problem.  The results are obtained after 10 explicit time steps by an in-house C++ code with a single core of Intel Xeon 2630v4 @ 2.10 GHz. Characteristic reconstruction are used for all three schemes.  }
\end{table}

\begin{table}[!h]
    \small
    \begin{center}
        \def\temptablewidth{1\textwidth}
        {\rule{\temptablewidth}{1pt}}
        \begin{tabular*}{\temptablewidth}{@{\extracolsep{\fill}}c|c|c|c}

            Scheme & No. of stage & No. of Gauss point & CPU time  \\
            \hline
            Triangular 2nd-order GKS Case I  &1 &1 & 0.81s\\
            Triangular 2nd-order GKS Case II  &2 &2 & 3.68s\\
            Triangular Compact-GKS&2&2& 4.80s \\
        \end{tabular*}
        {\rule{\temptablewidth}{0.1pt}}
    \end{center}
    \vspace{-4mm} \caption{\label{cpu-time-2} Computational time (in seconds) of different schemes for the 1D Sod problem.  The results are obtained after 10 explicit time steps by an in-house C++ code with a single core of Intel Xeon 2630v4 @ 2.10 GHz. The same uniform triangular mesh with 10000 cells and 15150 interfaces are used for all three schemes.  }
\end{table}

\subsubsection{Shu-Osher problem}
\begin{figure}[!h]
    \centering
    \includegraphics[width=0.485\textwidth]{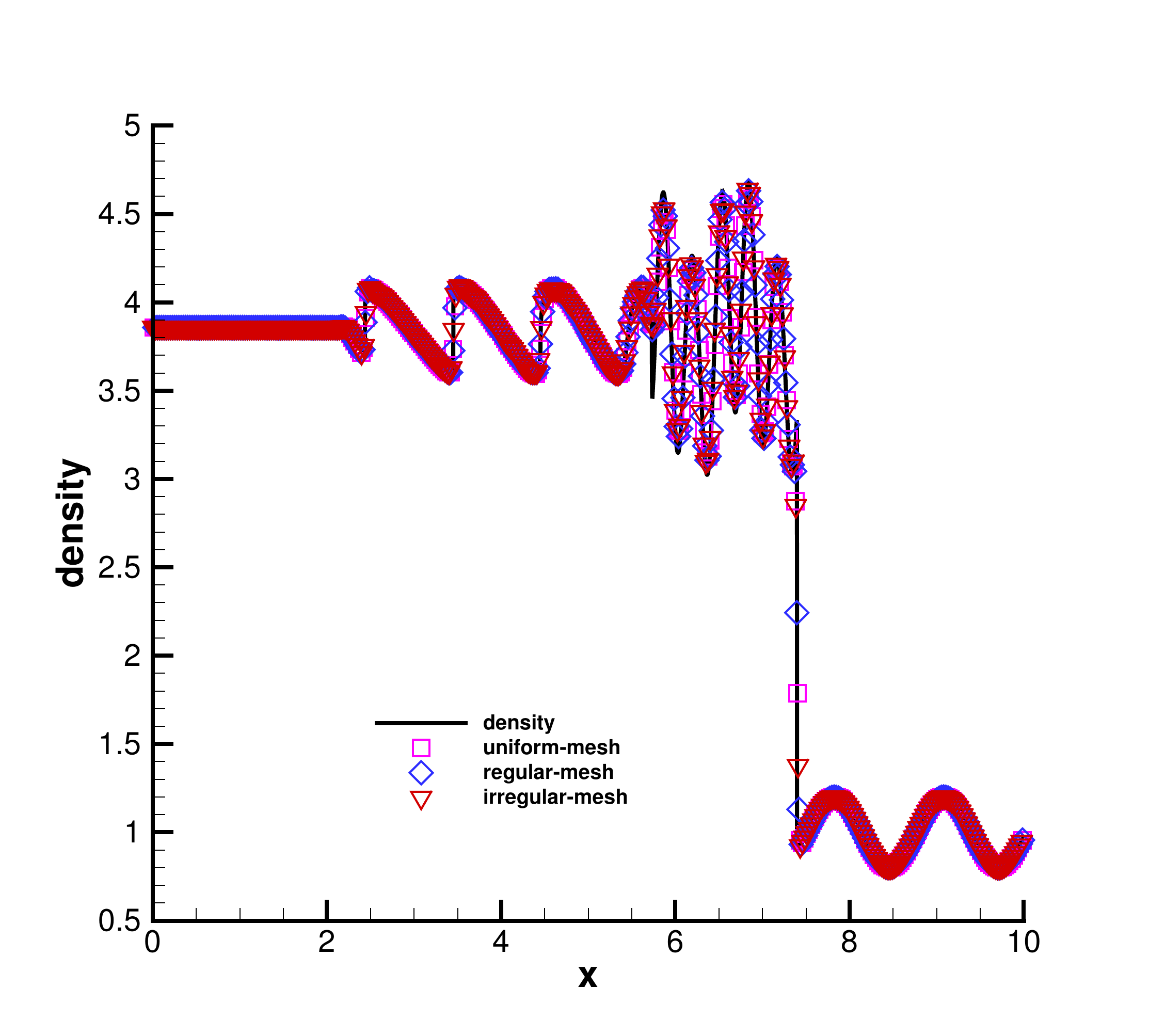}
    \includegraphics[width=0.485\textwidth]{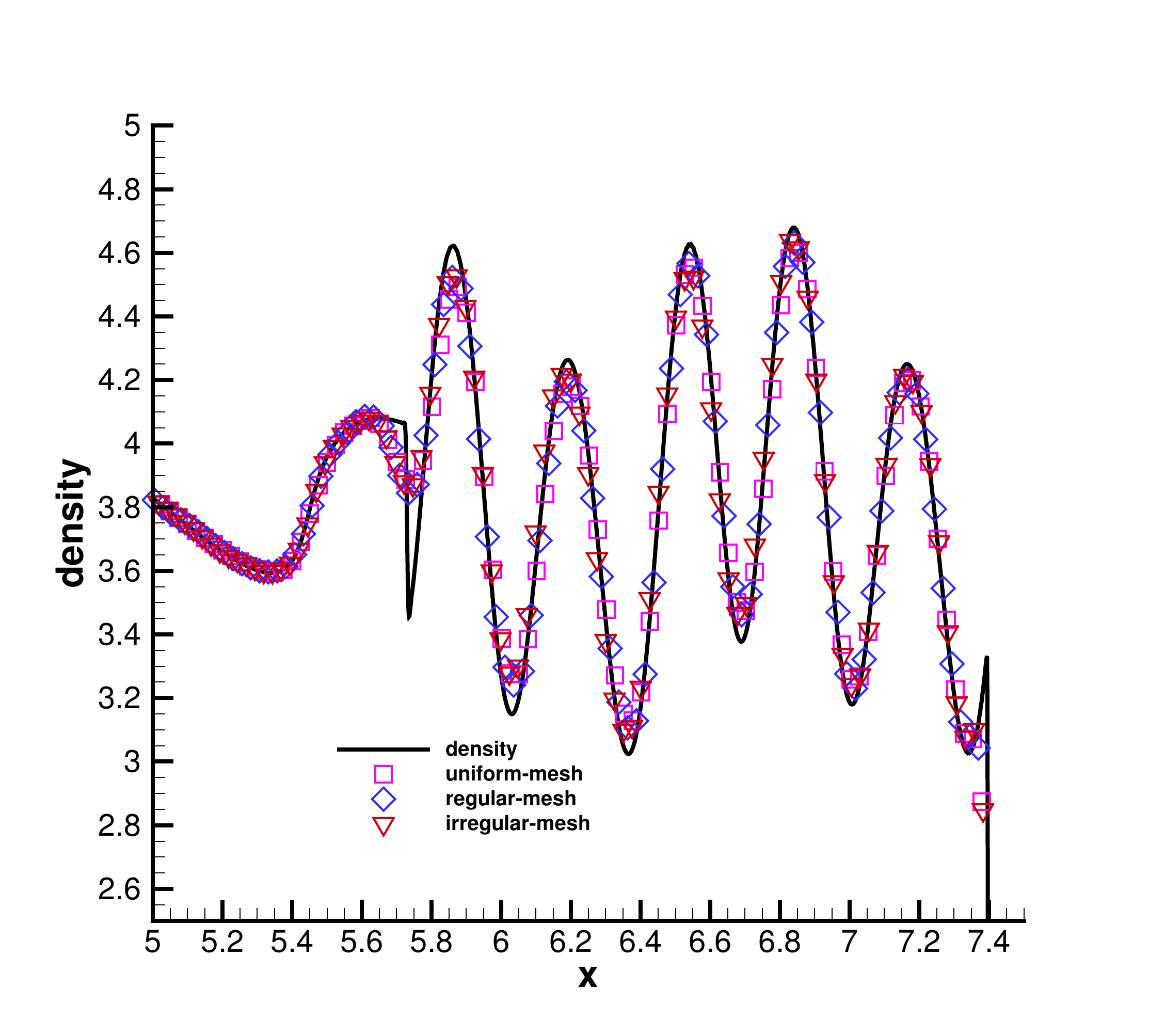}
    \caption{\label{shu-osher} Shu-Osher problem: the density
        distributions and local enlargement at $t=1.8$ with cell scale $1/40$ along the center horizontal line under different meshes.}
\end{figure}

\begin{figure}[!h]
    \centering
    \includegraphics[width=0.32\textwidth]{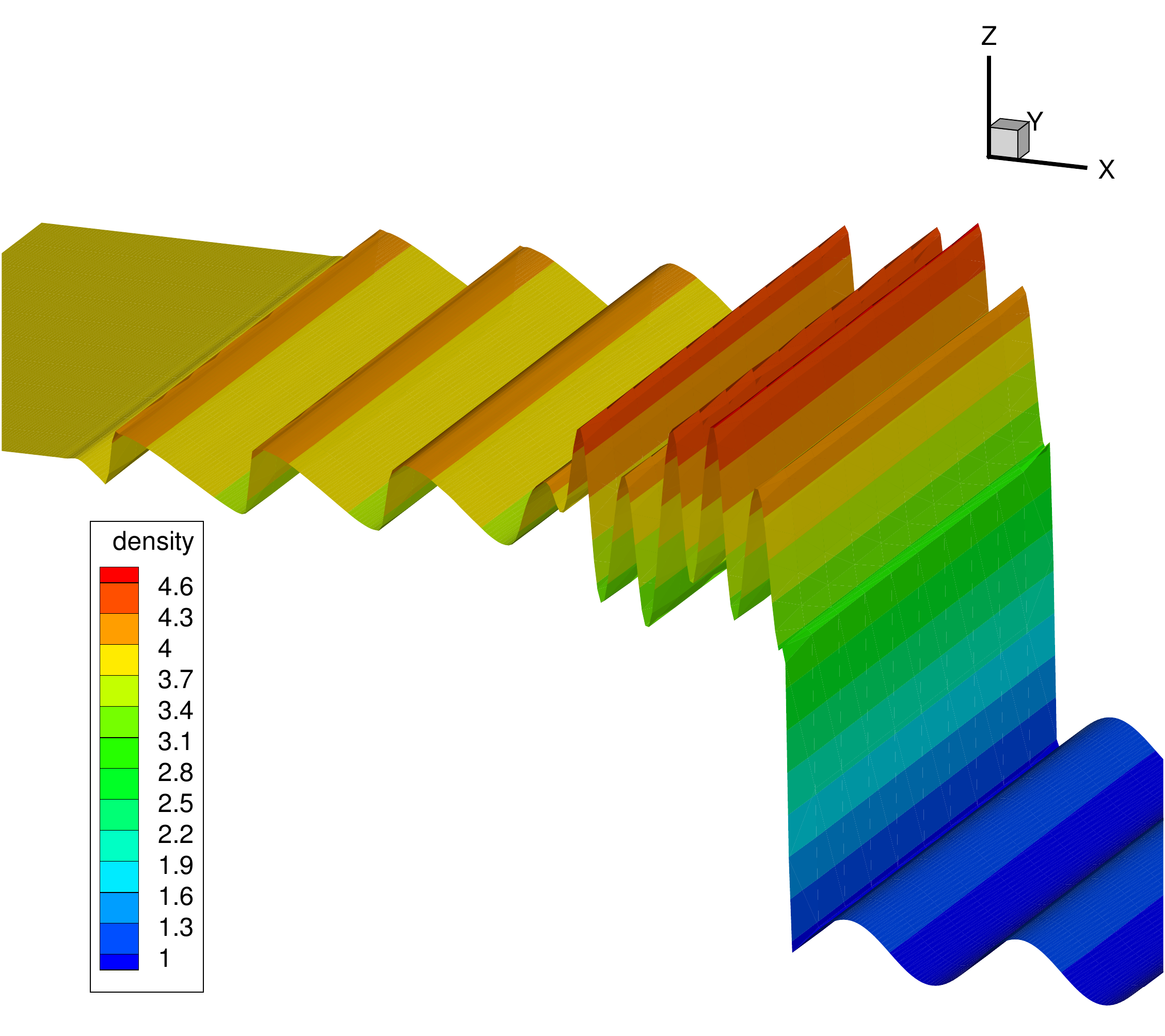}
    \includegraphics[width=0.32\textwidth]{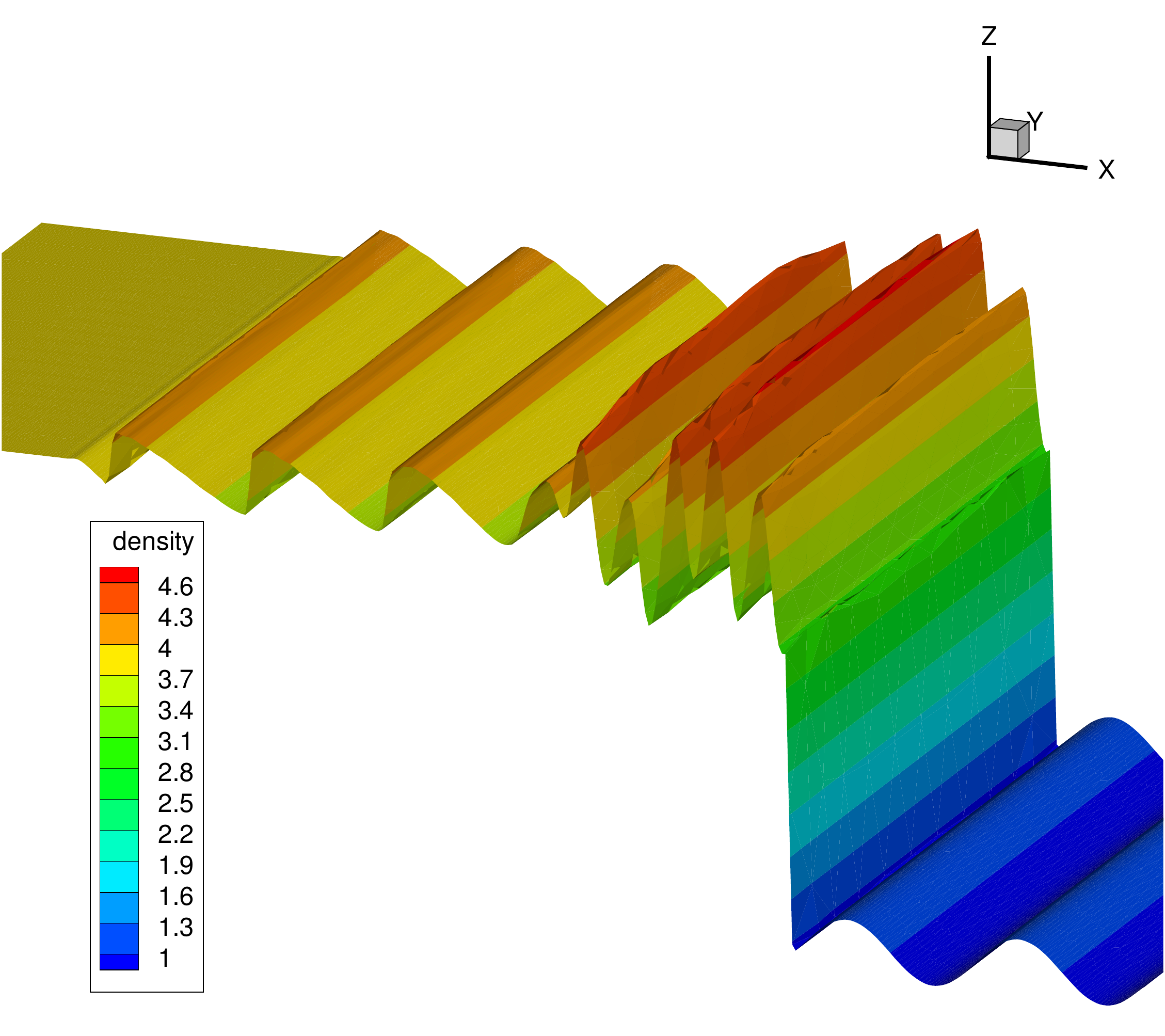}
    \includegraphics[width=0.32\textwidth]{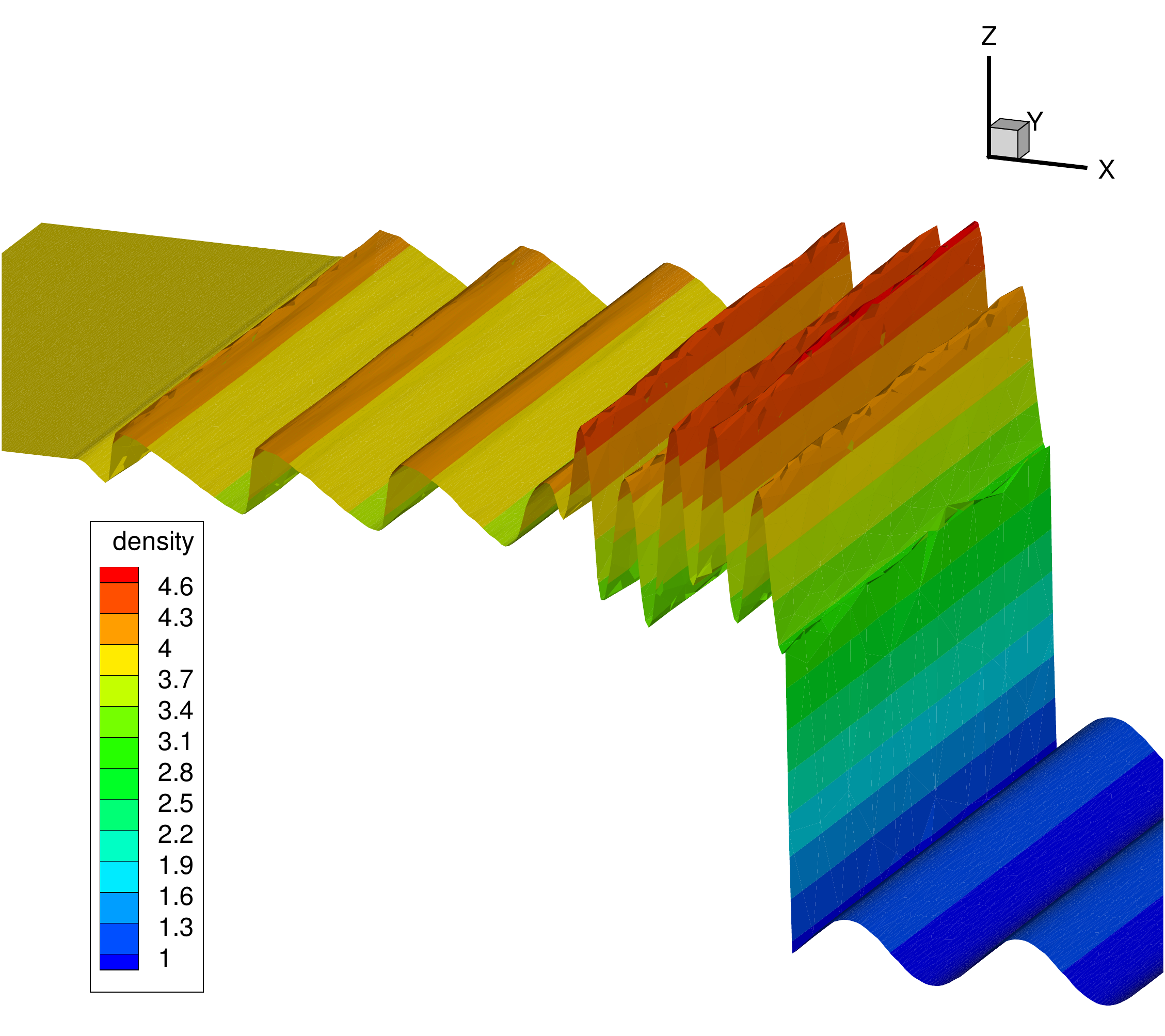}
    \caption{\label{shu-osher-3d} Shu-Osher problem: the 3-D view of density distributions at $t=1.8$ with cell size $1/40$ under different meshes. From left to right: uniform mesh, regular mesh, and irregular mesh.}
\end{figure}

To test the performance of capturing high frequency waves, the
Shu-Osher problem \cite{shu1989efficient} is tested, which is a case
with the density wave interacting with shock. The initial condition
is given as follows
\begin{equation*}
(\rho,U,p)=\left\{\begin{array}{ll}
(3.857134, 2.629369, 10.33333),  \ \ \ \ &  x \leq 1,\\
(1 + 0.2\sin (5x), 0, 1),  &  1 <x.
\end{array} \right.
\end{equation*}
The computational domain is $[0, 10]\times[0, 0.25]$ and $h=1/40$
triangular mesh is used.
The periodic boundary condition is applied in the $y$ direction.
Again numerical results under three different kinds of meshes are presented in Fig. \ref{shu-osher}.
The current third-order results are even compatible with the traditional S2O4 GKS with WENO5-Z reconstruction \cite{Pan2016twostage}.
For the same mesh size, due to more cells used in the non-uniform mesh case than the uniform
one, the results from regular and irregular meshes capture extremes slightly better than the case of uniform one.
The 3-D density distributions are  presented in
Fig. \ref{shu-osher-3d}. The numerical results preserve the uniformity along $y$-direction nicely even with the existence of acoustic waves in a large scale.

\subsubsection{Blast wave}

\begin{figure}[!h]
	\centering
	\includegraphics[width=0.48\textwidth]{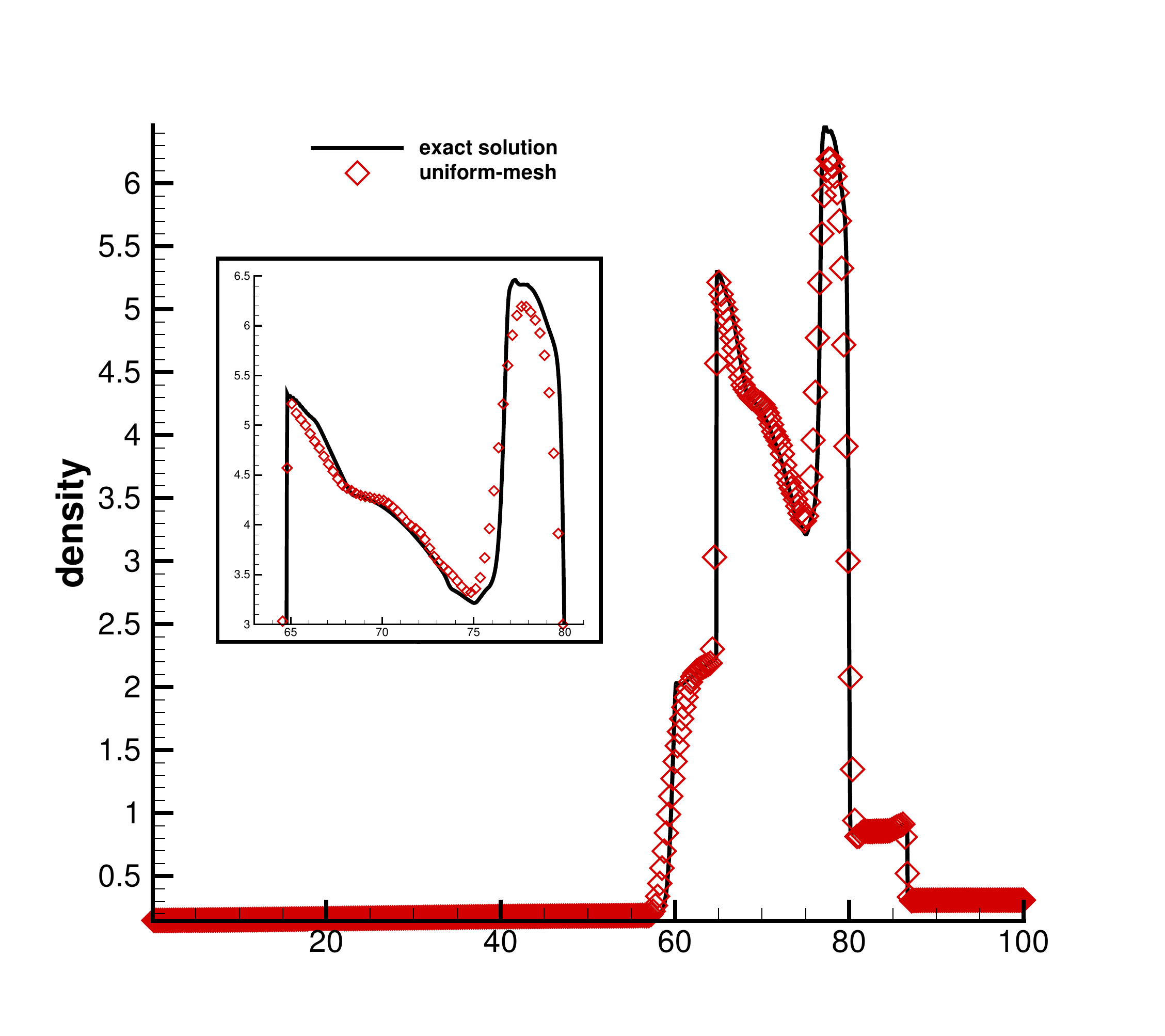}
	\includegraphics[width=0.48\textwidth]{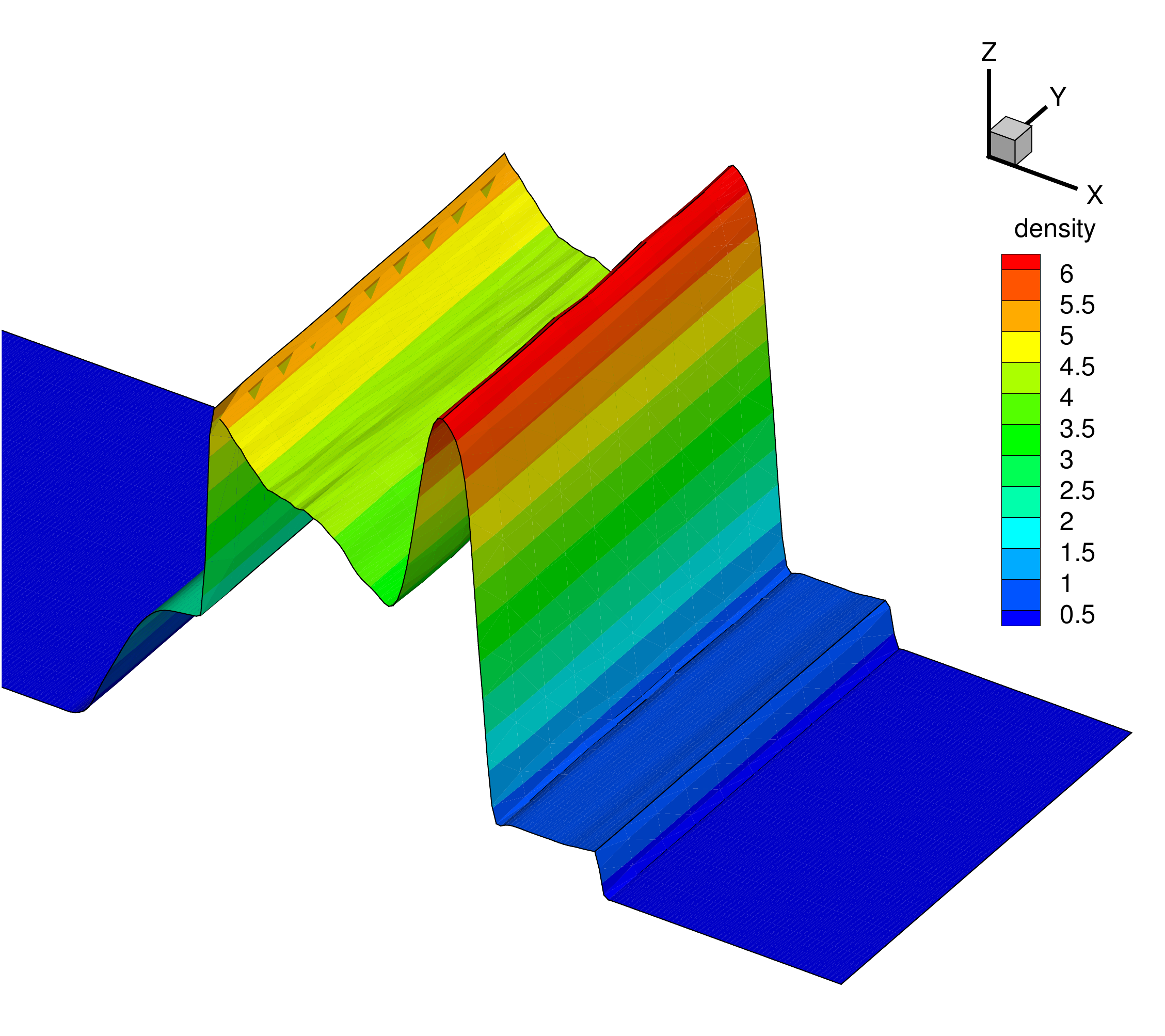}
	\caption{\label{blastwave-u} Blast wave problem: the density distribution along the center horizontal line and its 3-D view under uniform mesh at $t=1.8$ with cell size $h=1/400$. CFL=0.4. }
\end{figure}

The Woodward-Colella blast wave problem \cite{woodward1984numerical} is considered, and
the initial condition is given as follows
\begin{align*}
(\rho,U,p) =\begin{cases}
(1, 0, 1000), & 0\leq x<10,\\
(1, 0, 0.01), & 10\leq x<90,\\
(1, 0, 100),  &  90\leq x\leq 100.
\end{cases}
\end{align*}
The computational domain is $[0, 100]\times[0, 2.5]$ and the uniform mesh with $400\times10\times2$ is used.
The periodic boundary condition is applied in the $y$ direction.
The extracted density profile along the horizontal line  and the "exact" solution,
which is obtained by the 1-D fifth-order WENO GKS \cite{ji2018family} with 10,000 grid points  at $t =3.8$ are shown in Fig.\ref{blastwave-u}.
Again the current third-order results are as good as the ones from the traditional GKS with the WENO5-Z reconstruction and the same mesh size.
It seems that the current compact scheme can resolve the wave profiles clearly better than the non-compact scheme \cite{zhu2018new}, particularly for the local extreme values.
Moreover, the uniformity in the flow distributions along $y$ direction are well kept in the existence of the strong shocks, seen in Fig. \ref{blastwave-u}.

\subsection{Shock-vortex interaction}

\begin{figure}[!htb]
	\centering
	\includegraphics[width=0.475\textwidth]{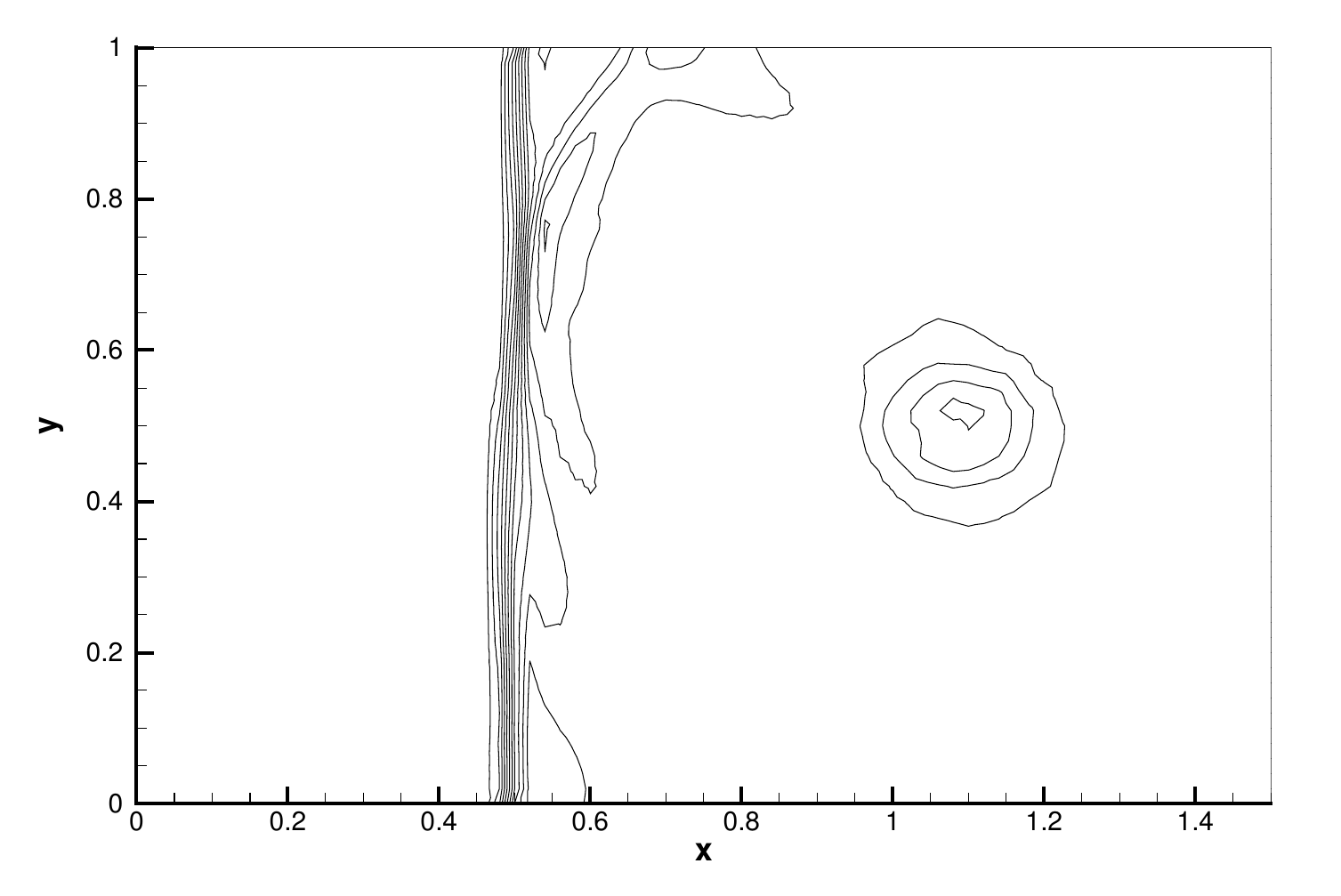}{a}
	\includegraphics[width=0.475\textwidth]{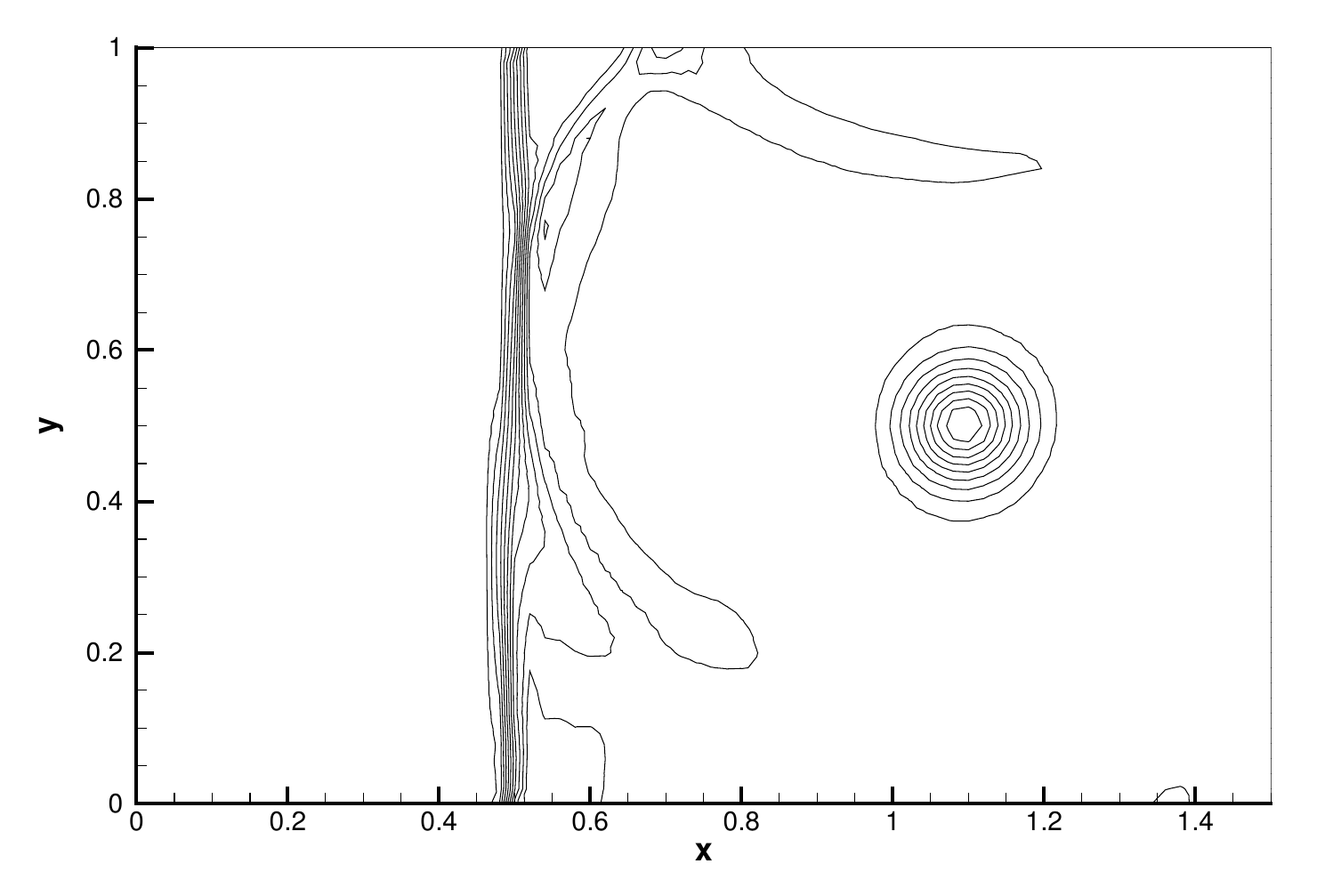}{b}
	\includegraphics[width=0.475\textwidth]{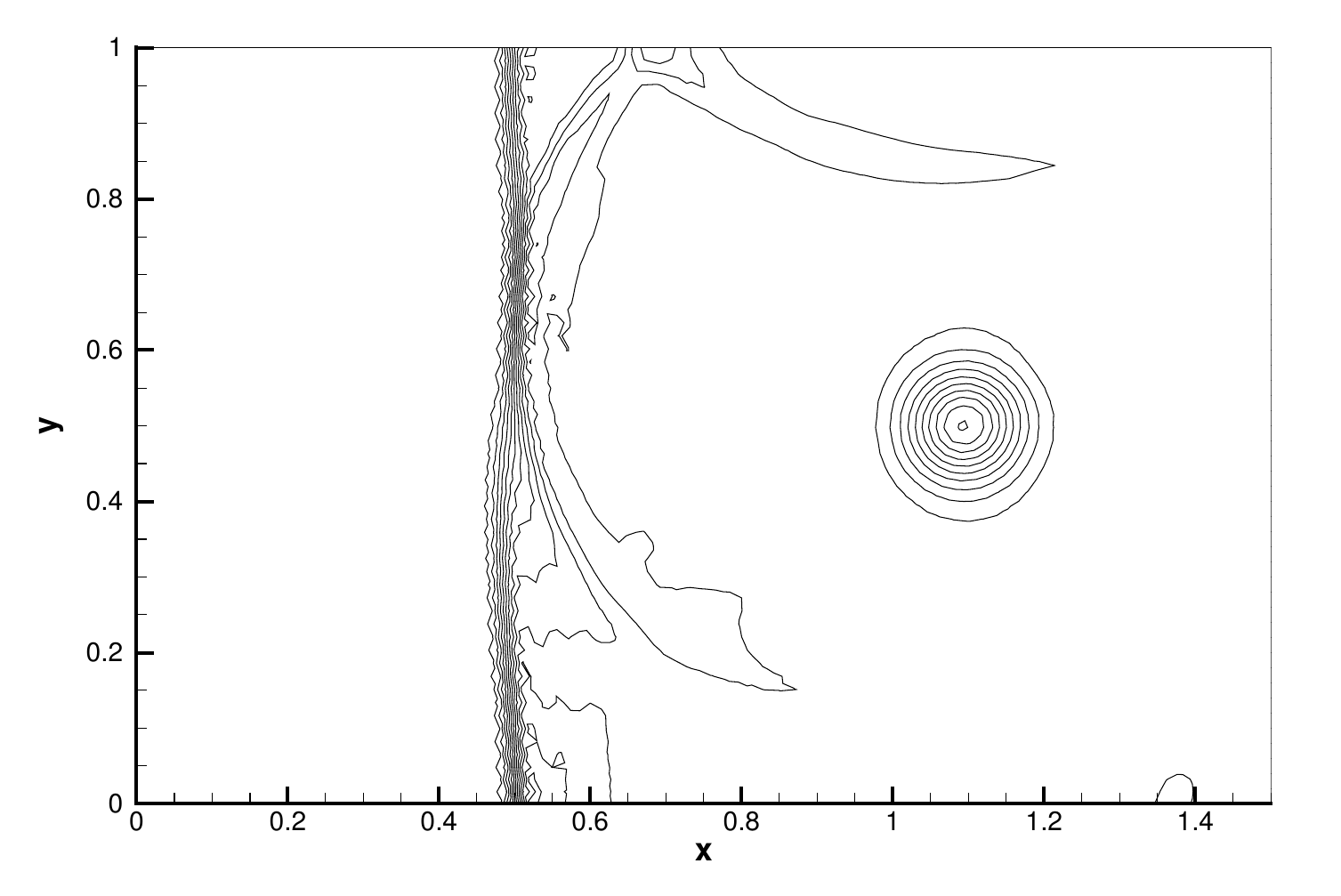}{c}
	\includegraphics[width=0.475\textwidth]{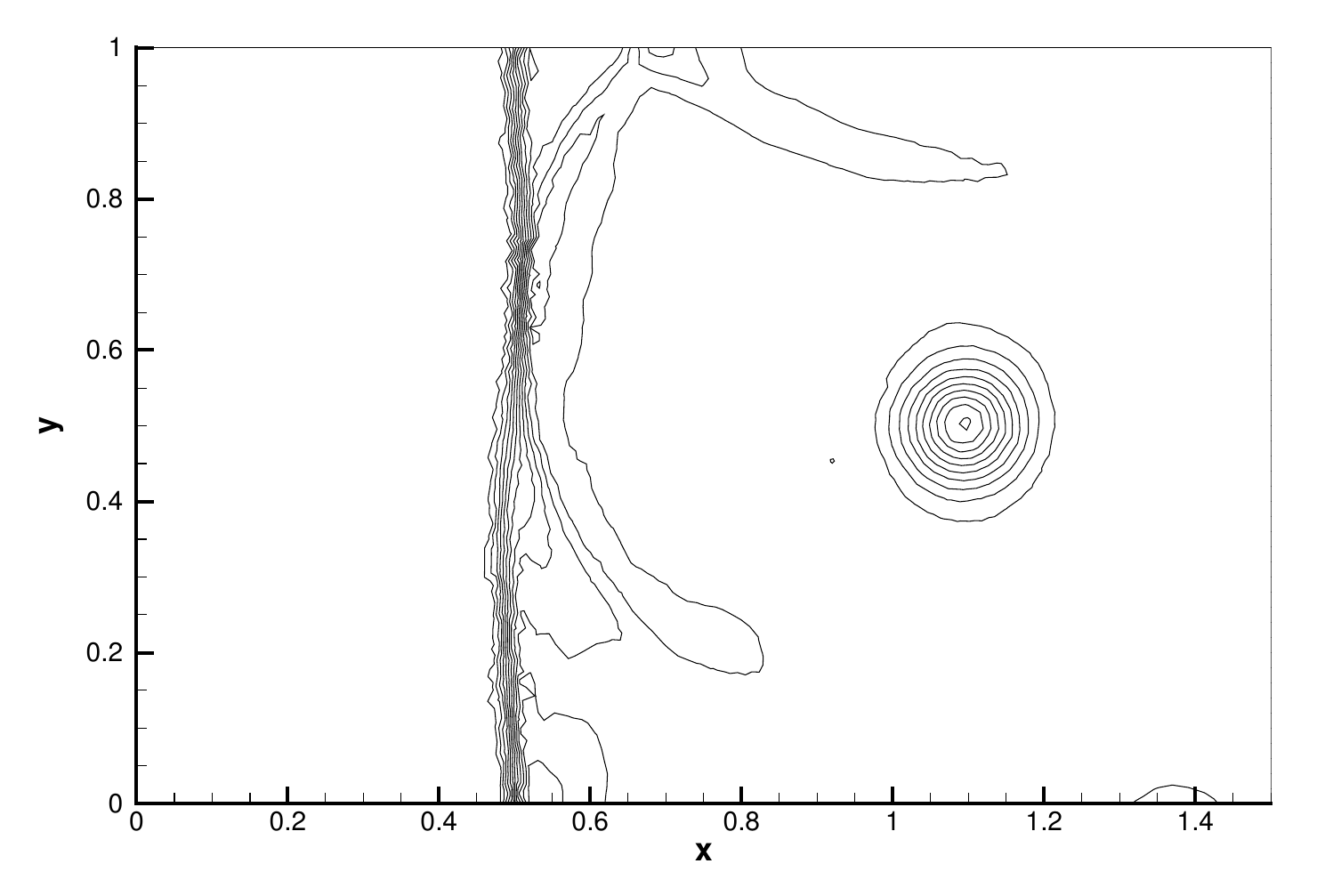}{d}
	\caption{\label{2d-sv-2}Shock-vortex interaction: density contours obtained
		under uniform mesh by a 2nd-order GKS (a), and uniform mesh, regular mesh and irregular mesh
		by current compact scheme (b,c,d) at $t=0.8$ with $h=1/50$.}
\end{figure}

\begin{figure}[!htb]
    \centering
    \includegraphics[width=0.475\textwidth]{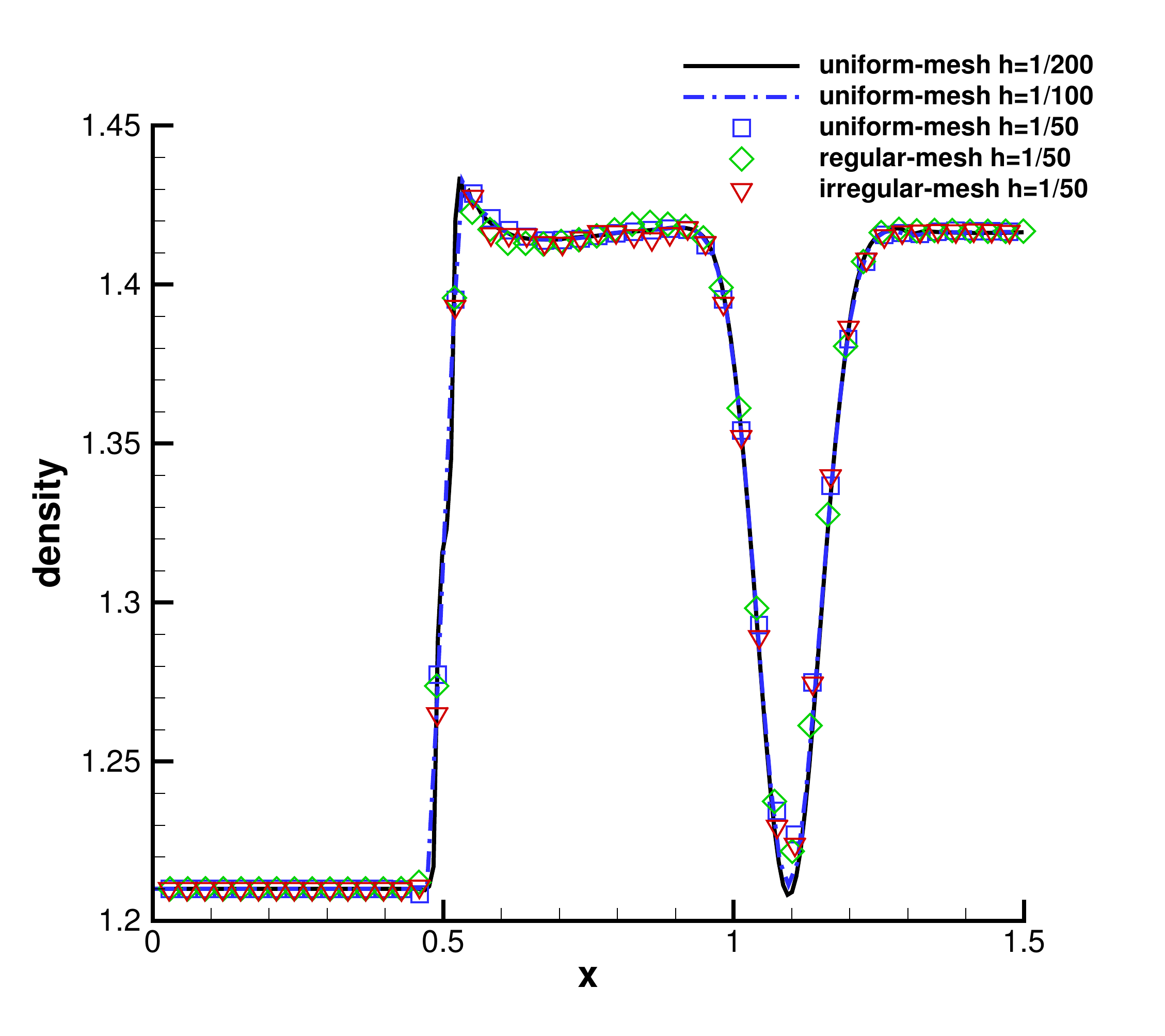}
    \includegraphics[width=0.475\textwidth]{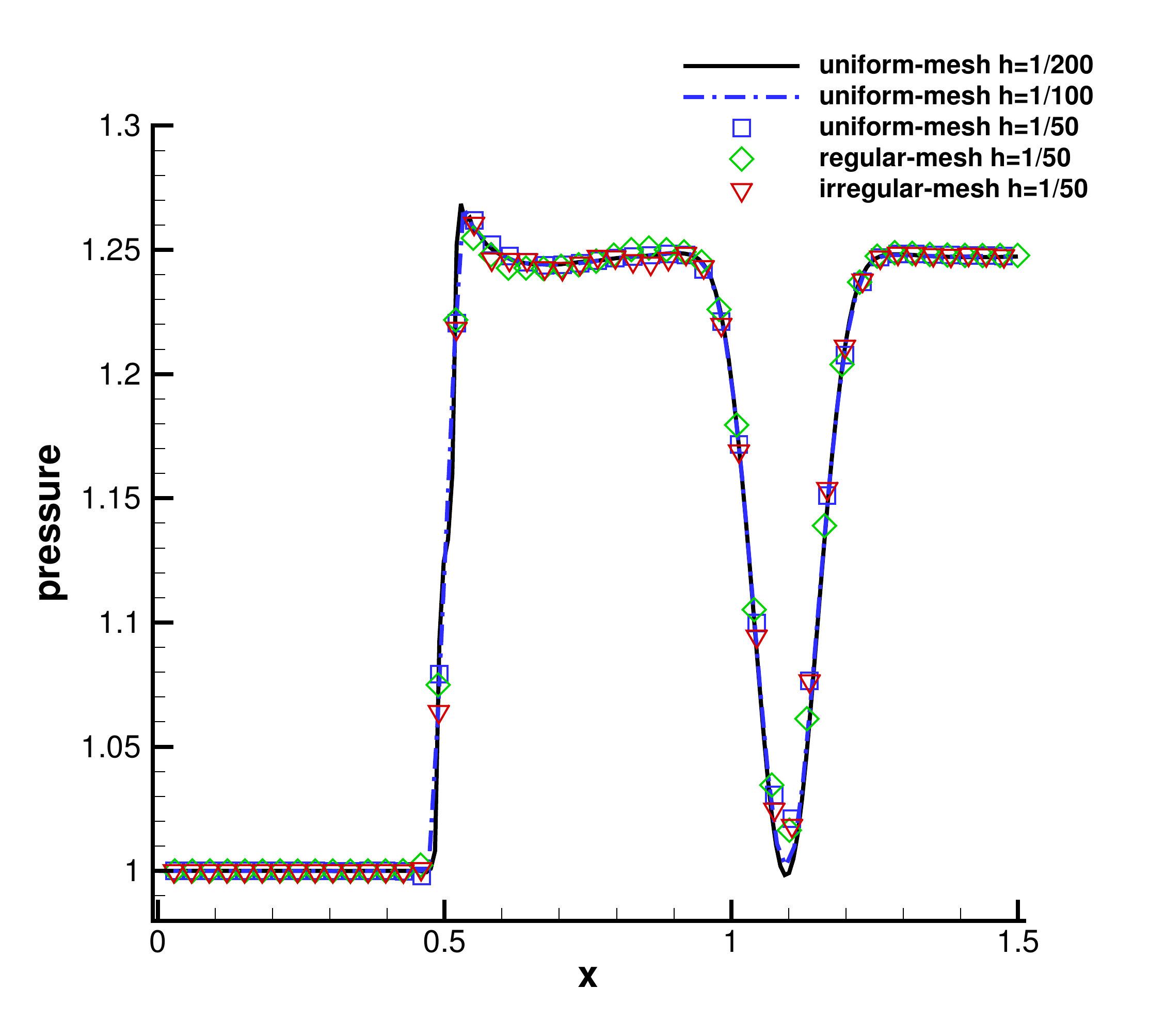}
    \caption{\label{2d-sv-line}Shock-vortex interaction: density and pressure
        distributions along the center horizontal line under different meshes at $t=0.8$.}
\end{figure}

The interaction between a vortex and a stationary shock for the
inviscid flow \cite{weno} is presented. The computational domain is
$[0, 1.5]\times[0, 1]$. A stationary Mach $1.1$ shock
is positioned at $x=0.5$ and normal to the $x$-axis. The
mean flow on the left is $(\rho, U, V, p) = (Ma^2,\sqrt{\gamma}, 0, 1)$,
where $Ma$ is the Mach number. A circular vortex is designed by a
perturbation on the mean flow with the velocity $(U, V)$,
temperature $T=p/\rho$, and entropy $S=\ln(p/\rho^\gamma)$.  The
perturbation is expressed as
\begin{align*}
&(\delta U,\delta V)=\kappa\eta e^{\mu(1-\eta^2)}(\sin\theta,-\cos\theta),\\
&\delta
T=-\frac{(\gamma-1)\kappa^2}{4\mu\gamma}e^{2\mu(1-\eta^2)},\delta
S=0,
\end{align*}
where $\eta=r/r_c$, $r=\sqrt{(x-x_c)^2+(y-y_c)^2}$, and $(x_c,
y_c)=(0.25, 0.5)$ is the center of the vortex. Here $\kappa$
represents the vortex strength, $\mu$ controls the decay rate
of the vortex, and $r_c$ is the critical radius where the vortex
reaches the maximum strength. In current computation, $\kappa=0.3$,
$\mu=0.204$, and $r_c=0.05$. The reflected boundary conditions are
used on the top and bottom boundaries.
Inflow and outflow boundary conditions are applied along the entrance and exit.
The numerical results by current scheme under a
coarse mesh are compared with traditional 2nd-order GKS in
Fig. \ref{2d-sv-2}. The density and pressure distributions along the
center horizontal line with meshes $h=1/50, 1/100, 1/200$ at $t=0.8$
are compared in Fig. \ref{2d-sv-line}. The peak values around $x=1.1$
are well resolved under all three types of triangular meshes with
$h=1/50$.

\subsection{Forward step problem}

\begin{figure}[!htb]
    \centering
    \includegraphics[width=1.0\textwidth]{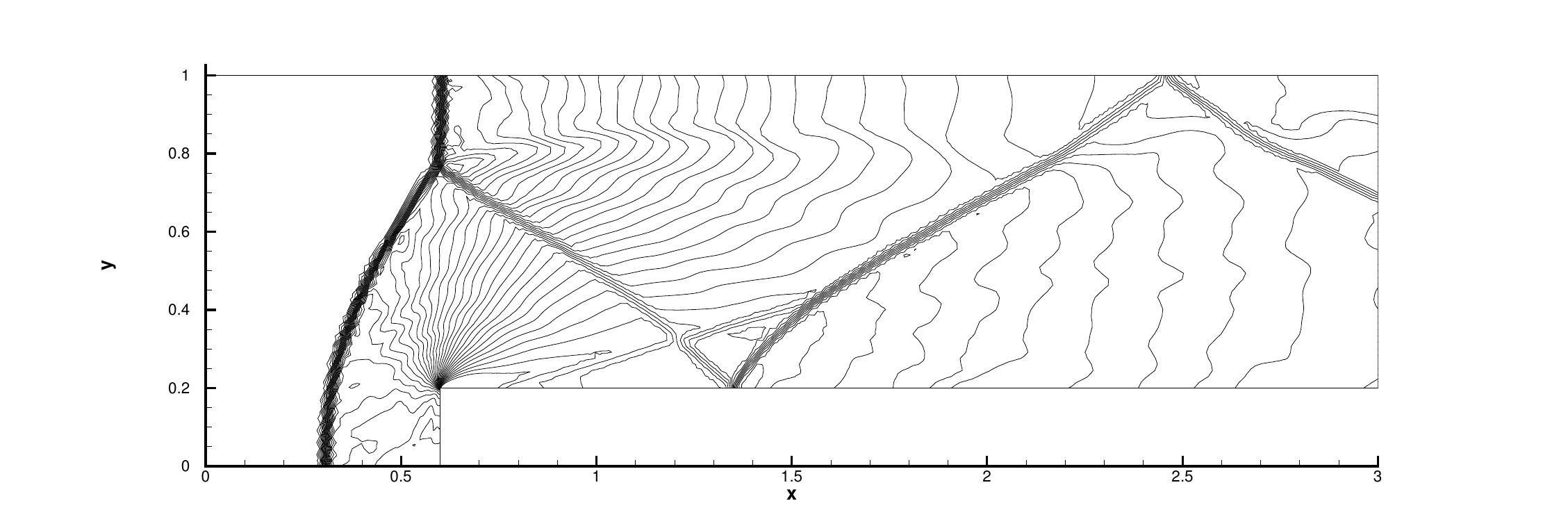}
    \includegraphics[width=1.0\textwidth]{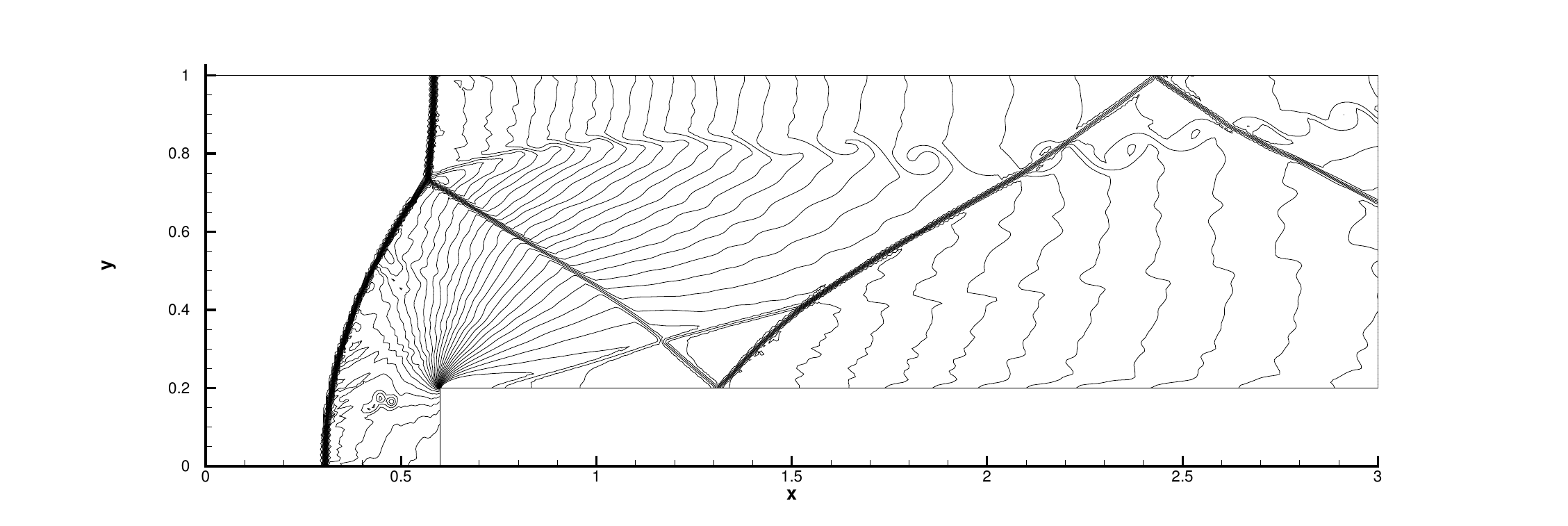}
    \includegraphics[width=1.0\textwidth]{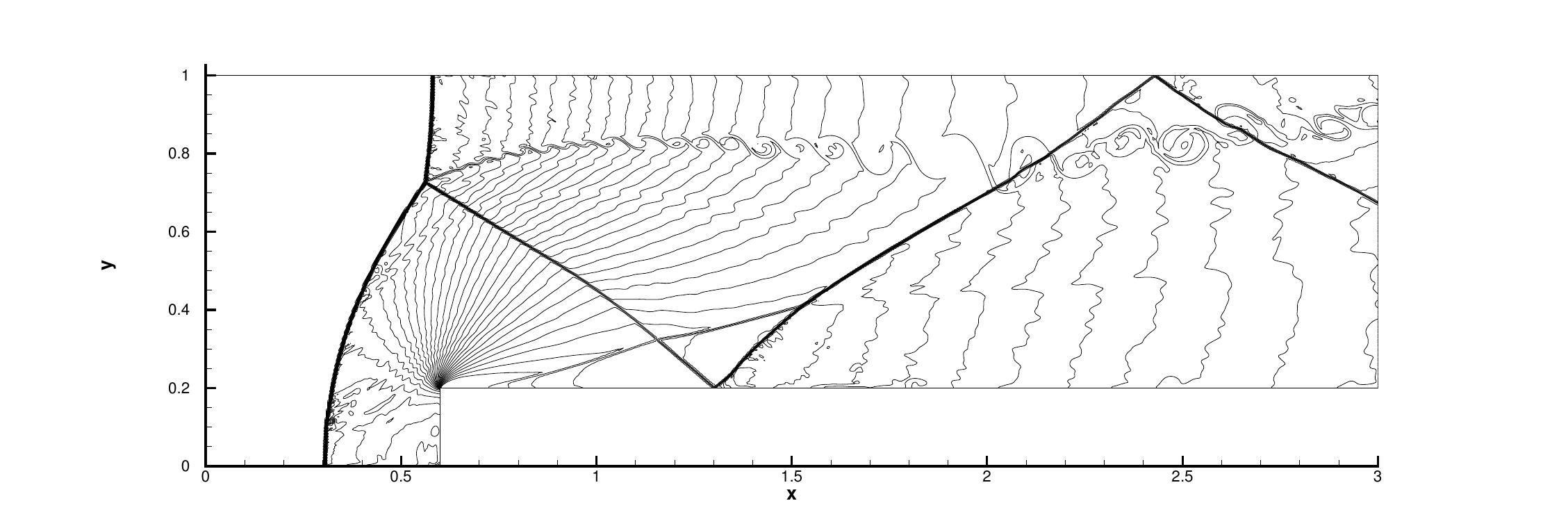}
    \caption{\label{forward-step-contour} Mach 3 forward step problem: density distributions on different meshes with $h=1/60$, $1/120$, $1/240$ at $t=4$. $30$ contours are used.}
\end{figure}

This standard test case is originally from \cite{woodward1984numerical} for inviscid flow. Initially, a Mach 3 flow
is moving from left to right in a wind tunnel.
The computational domain is a rectangle with $3$ unit long and $1$ unit wide.
The mesh sample is shown in Fig. \ref{forward-step-contour}. The walls of the tunnel are set as reflective boundary conditions. The inflow boundary condition is set at the left entrance while the outflow boundary condition is set at the right exit.
As a practice, the meshes near the corner are refined to $h=1/240$ to minimize flow separation from this singular corner.
The results with $h=1/60, 1/120, 1/240 $ at $t=4$ are presented in Fig. \ref{forward-step-contour}. The instability along the slip line starting from the triple point can be clearly observed with a rather coarse mesh of $h=1/120$.
\begin{figure}[!htb]
    \centering
    \includegraphics[width=1.0\textwidth]{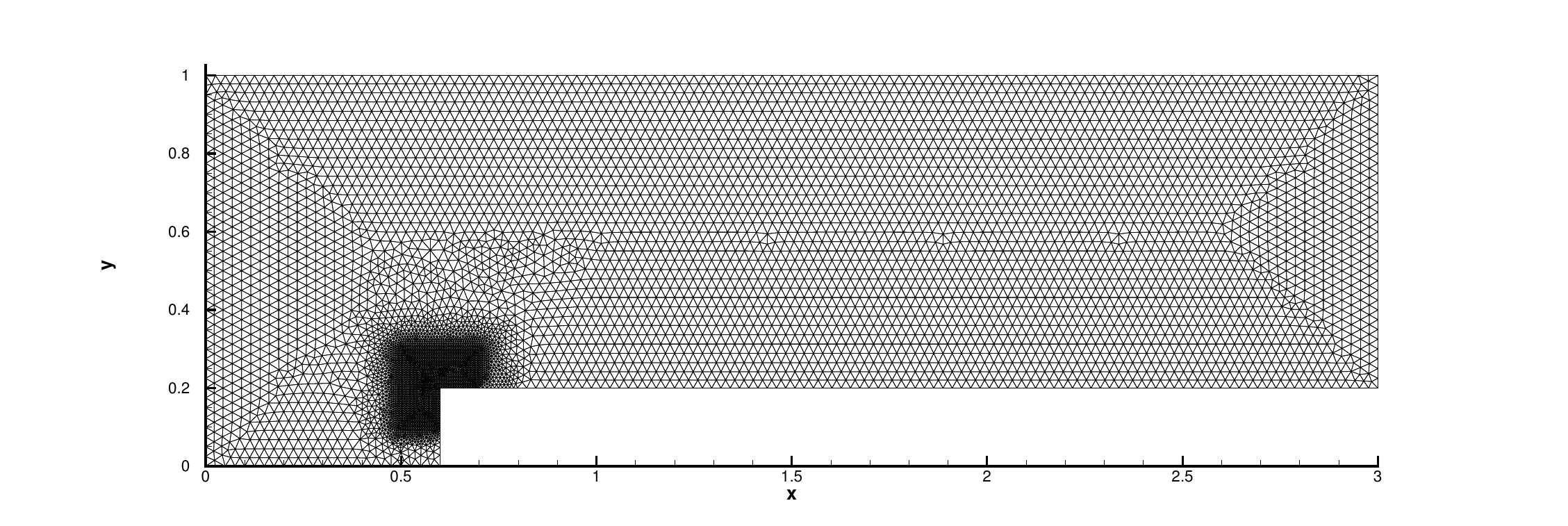}
    \caption{\label{forward-step-mesh}Mach 3 forward step problem: mesh sample with $h=1/40$.}
\end{figure}

\subsection{Double Mach reflection problem}
\begin{figure}[!htb]
	\centering
	\includegraphics[width=0.5\textwidth]{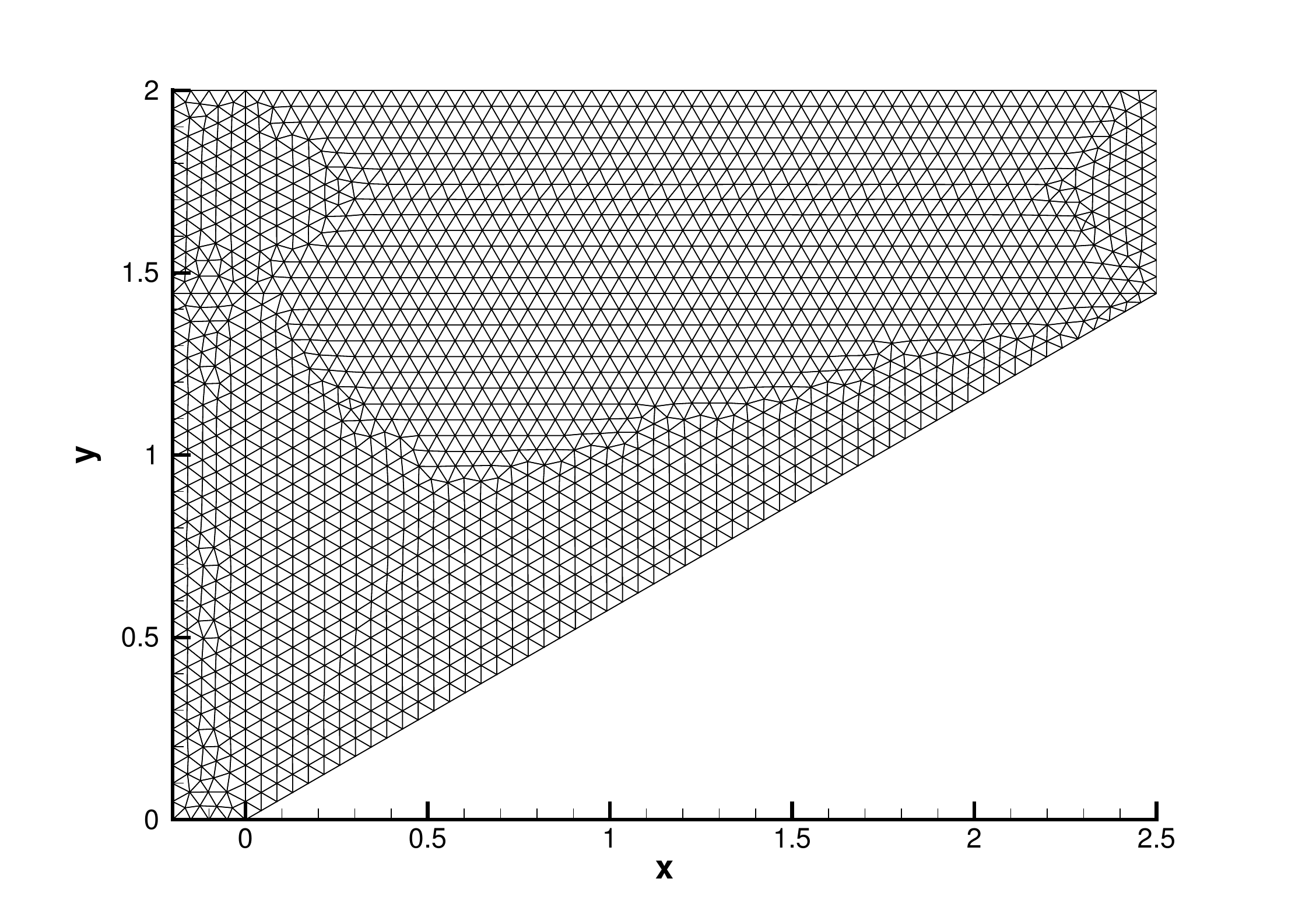}
	\caption{\label{double-mach-1}Double Mach reflection problem: Mesh distribution with $h=1/20$.}
\end{figure}

\begin{figure}[!htb]
	\centering
	\includegraphics[height=0.37\textwidth]{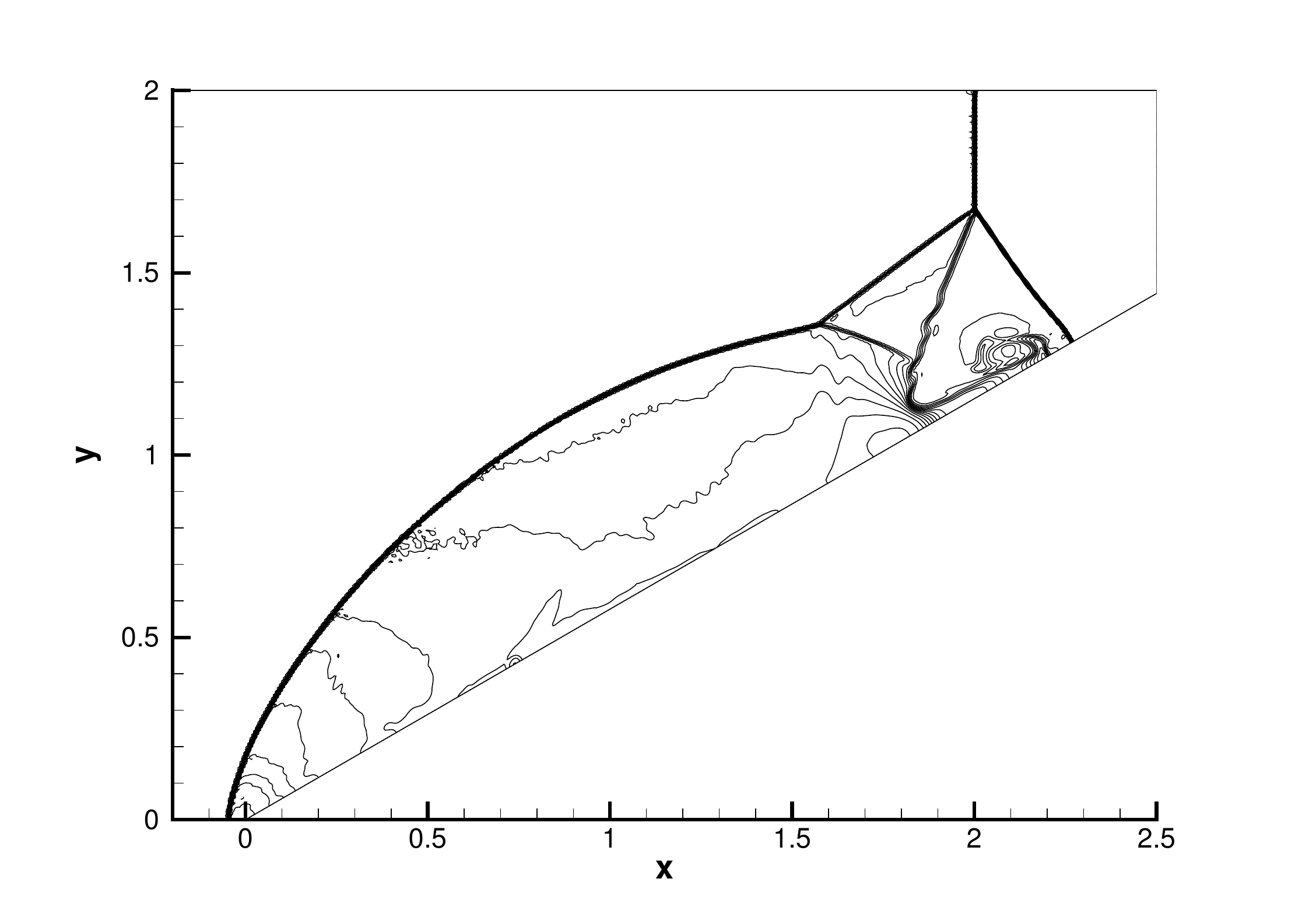}
	\includegraphics[height=0.37\textwidth]{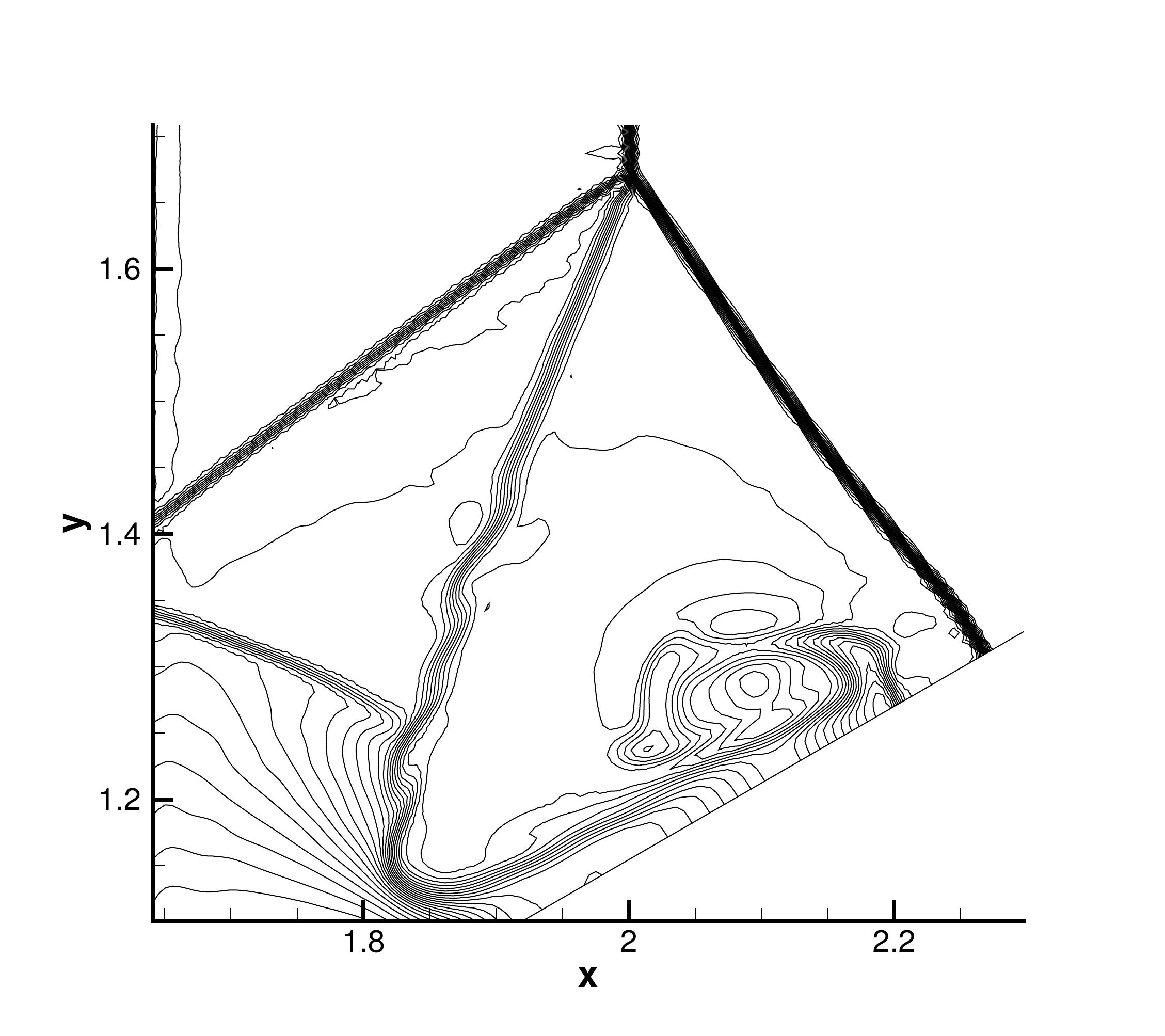}
	\includegraphics[height=0.37\textwidth]{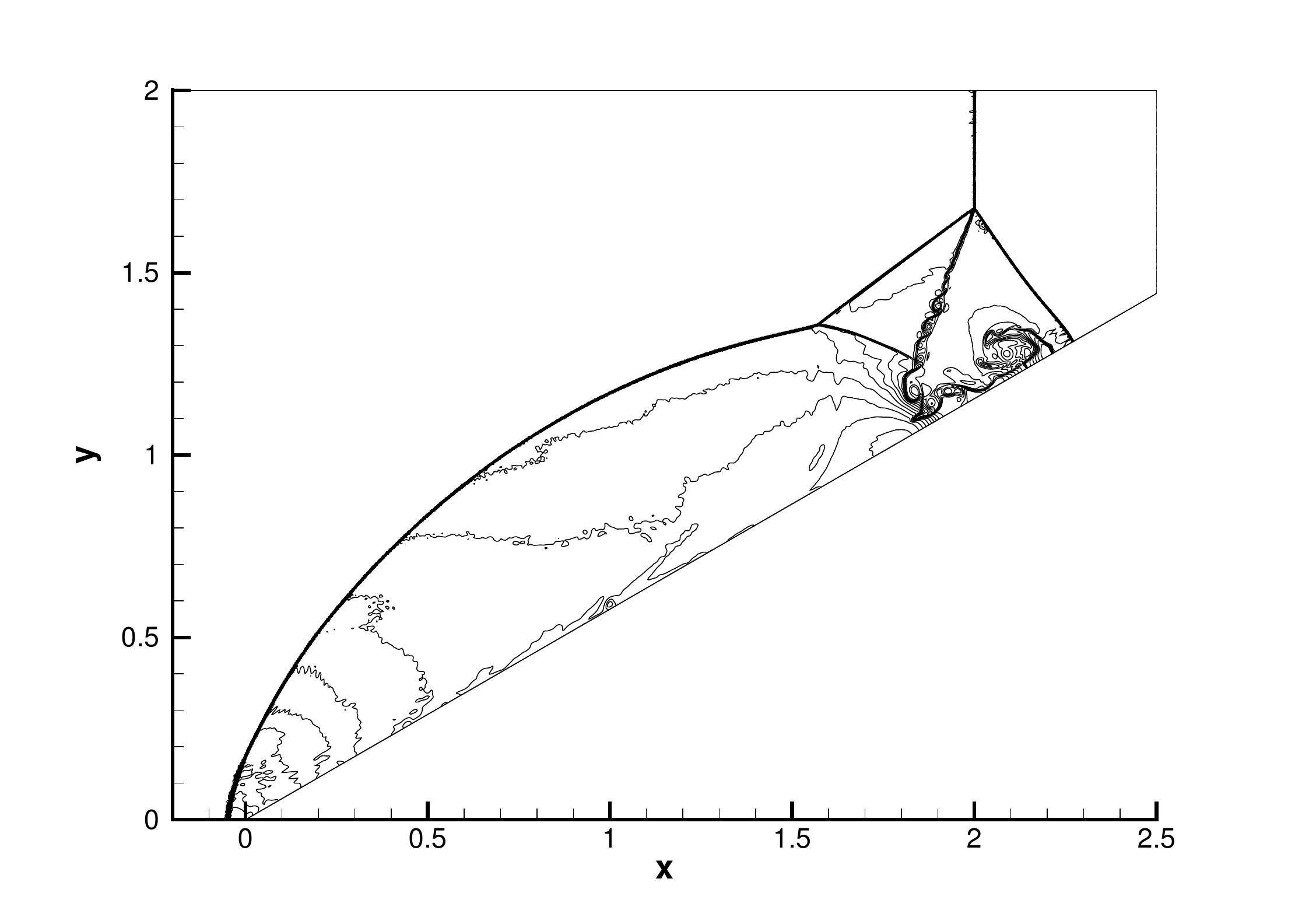}
	\includegraphics[height=0.37\textwidth]{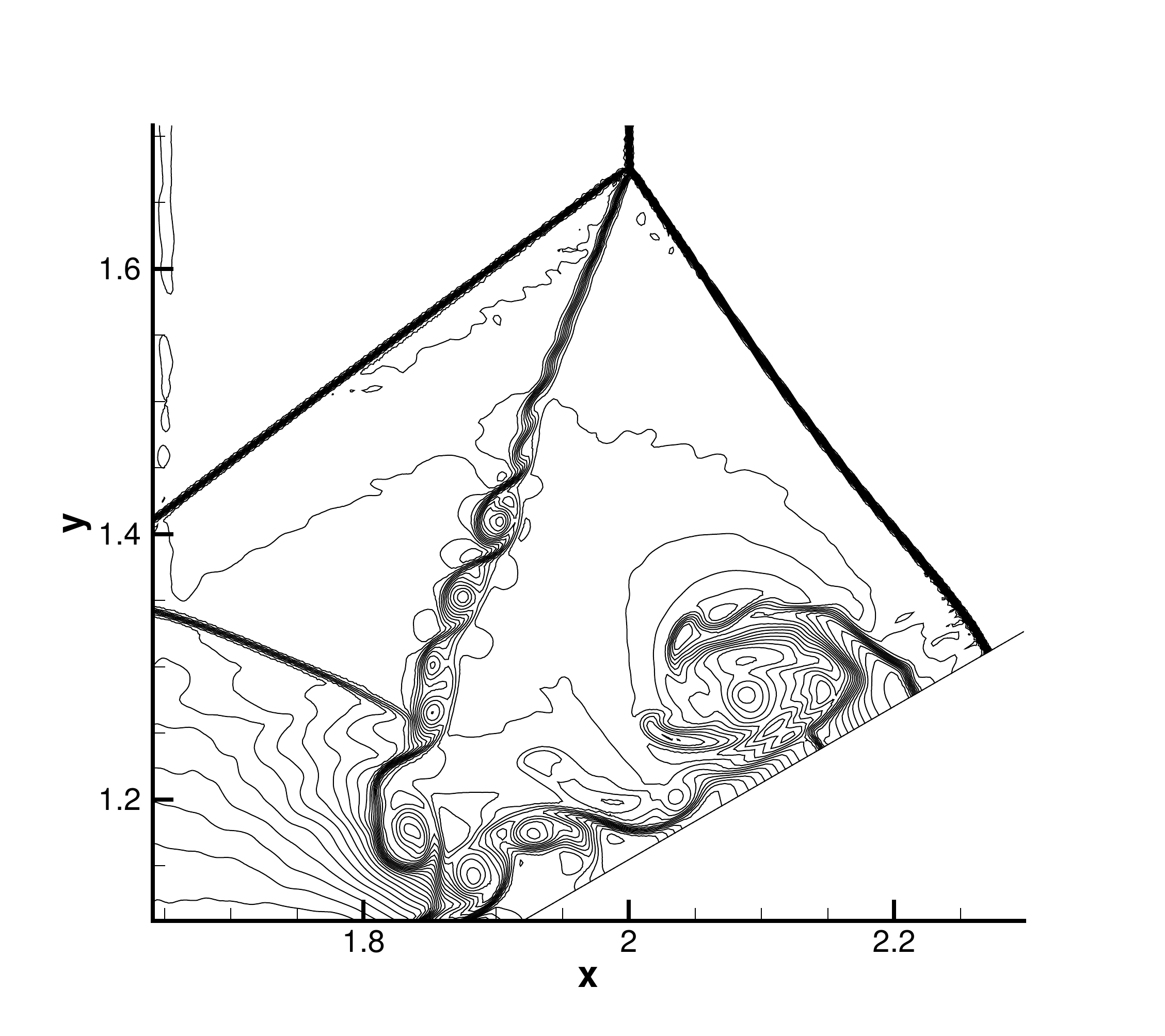}
	\caption{\label{double-mach-2}Double Mach reflection problem: density
		distributions and zoomed-in pictures around the Mach stem under non-uniform-mesh at $t=0.2$ with $h=1/160$, $1/320$.  $30$ contours are used.}
\end{figure}

We now consider the double Mach reflection problem \cite{woodward1984numerical} for the inviscid flow.
The computational domain is shown in Fig. \ref{double-mach-1}. Initially a
right-moving Mach $10$ shock is positioned at $(x,y)=(0, 0)$, and reflected by a $30 ^\circ$ wedge.  The initial pre-shock and
post-shock conditions are
\begin{align*}
(\rho, U, V, p)&=(8, 4.125\sqrt{3}, -4.125,
116.5),\\
(\rho, U, V, p)&=(1.4, 0, 0, 1).
\end{align*}
The reflecting boundary condition is used along the wedge.
The exact post-shock condition is imposed for the rest of bottom boundary.
The flow variables are set to follow  the motion of the shock at the top boundary.
The inflow boundary condition and the outflow boundary condition are set accordingly at the entrance and the exit.
The density distributions and the local enlargements with $h=1/160$ and $1/320$ at $t=0.2$  are shown in Fig. \ref{double-mach-2}.
The robustness of the compact GKS is validated, while the flow structure
around the slip line from the triple Mach point is resolved nicely by the compact scheme.

\subsection{Lid-driven cavity flow}

\begin{figure}[!htb]
	\centering
	\includegraphics[width=0.48\textwidth]{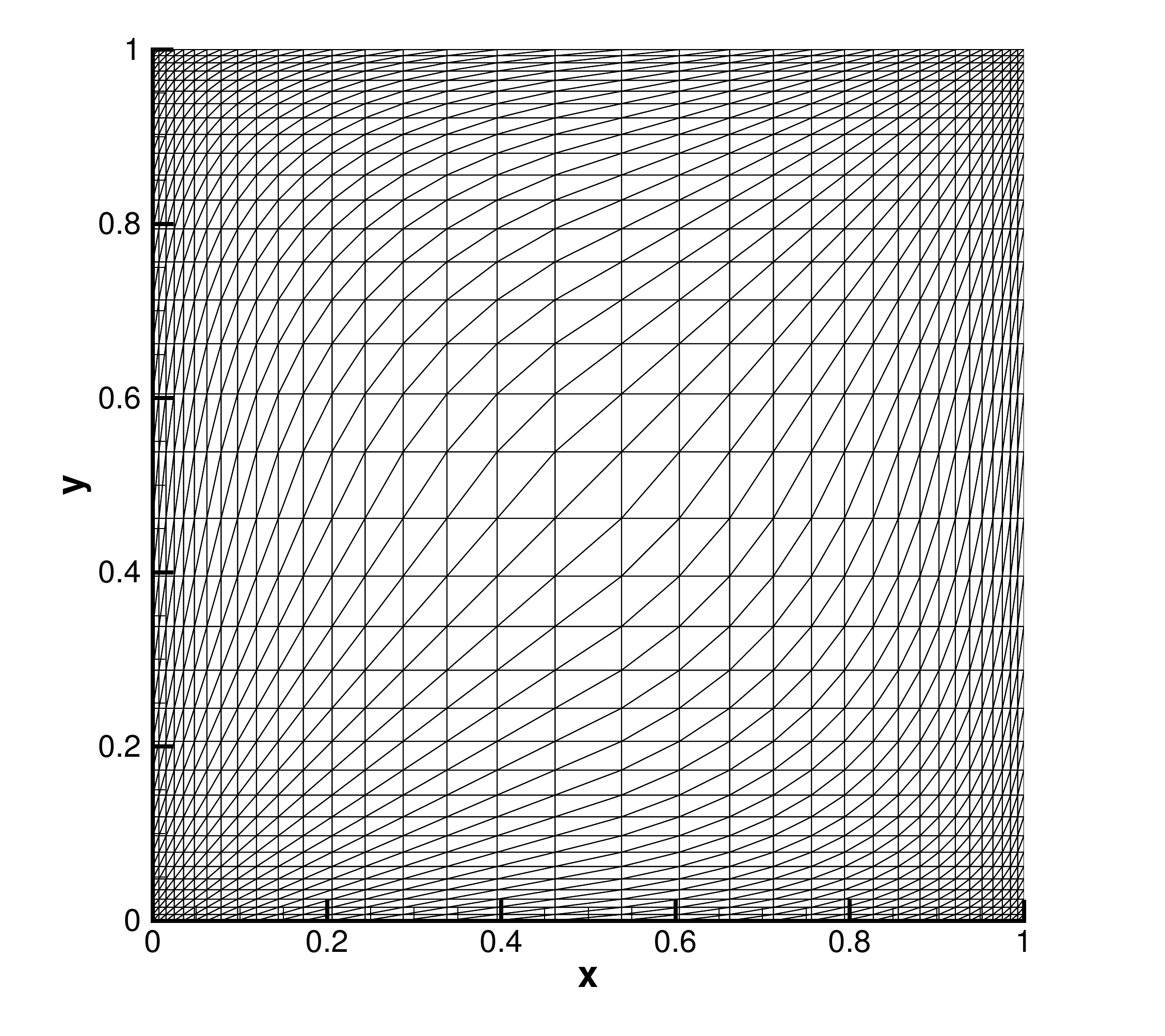}
	\includegraphics[width=0.48\textwidth]{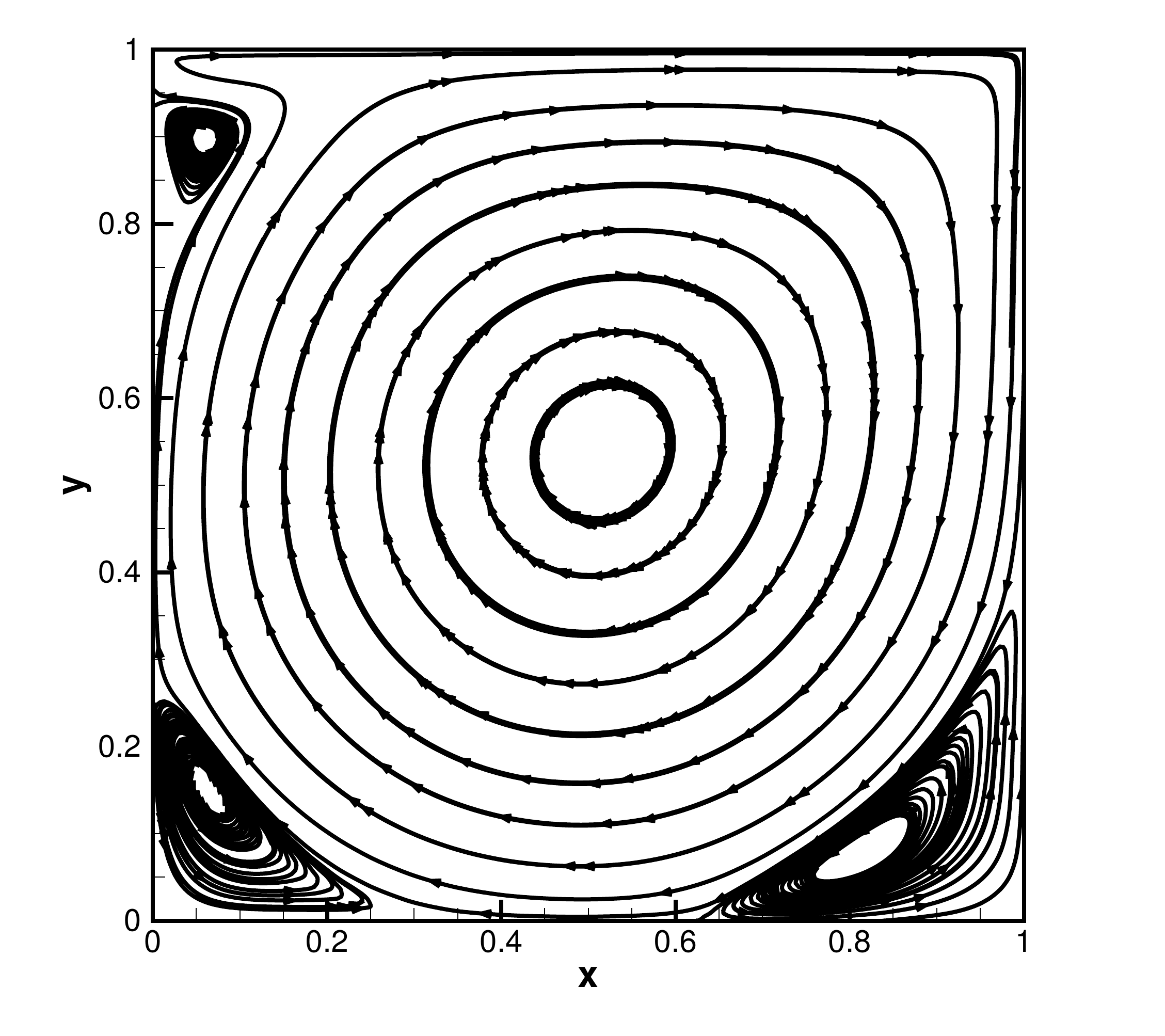}
	\caption{\label{2d-cavity} Lid-driven cavity flow. Left: mesh $35 \times 35 \times 2$ with $\Delta h \approx 0.007$ near the solid wall. Right: the stream line under $Re=3200$ case.}
\end{figure}

\begin{figure}[!htb]
    \centering
    \includegraphics[width=0.47\textwidth]{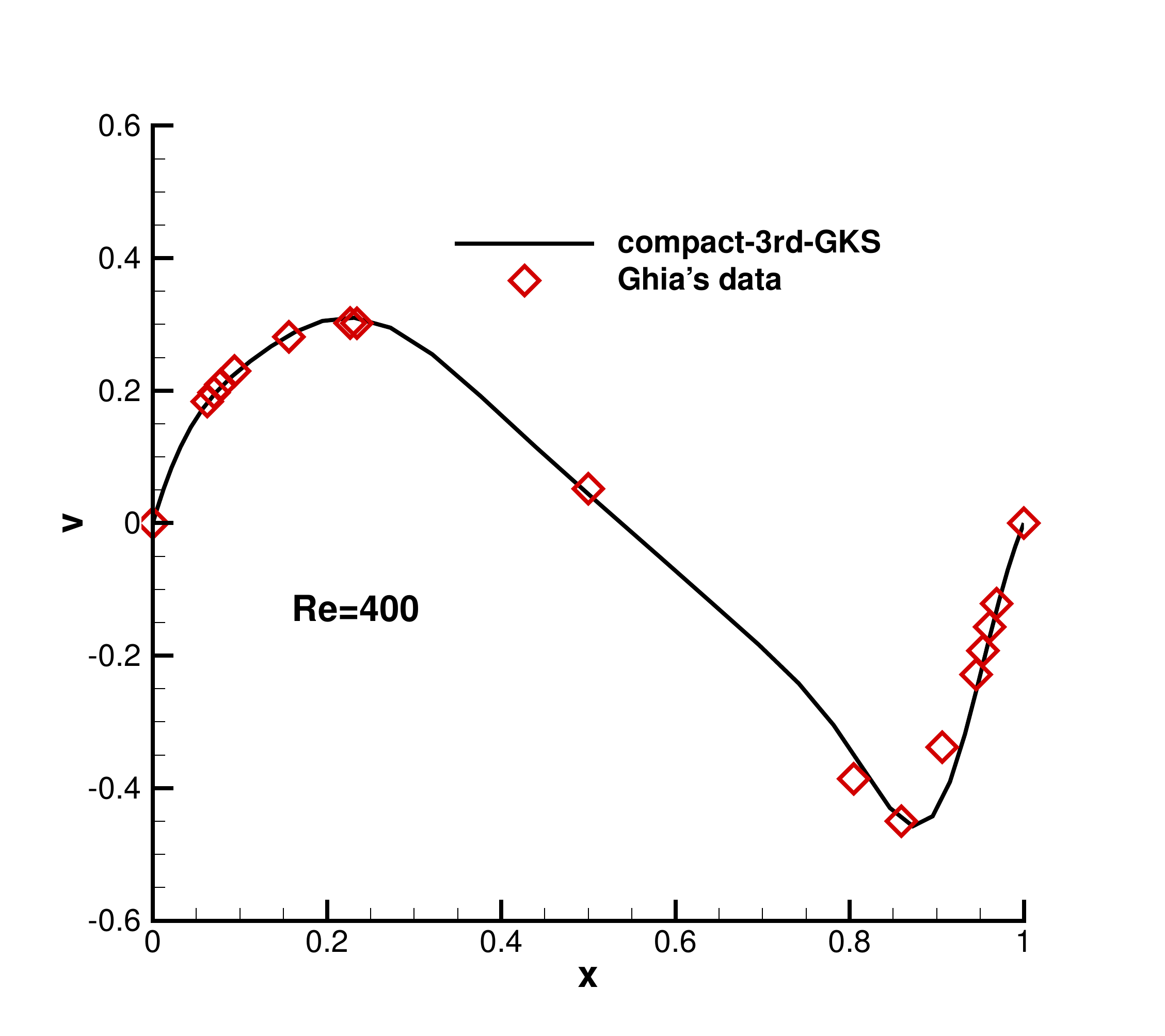}
    \includegraphics[width=0.47\textwidth]{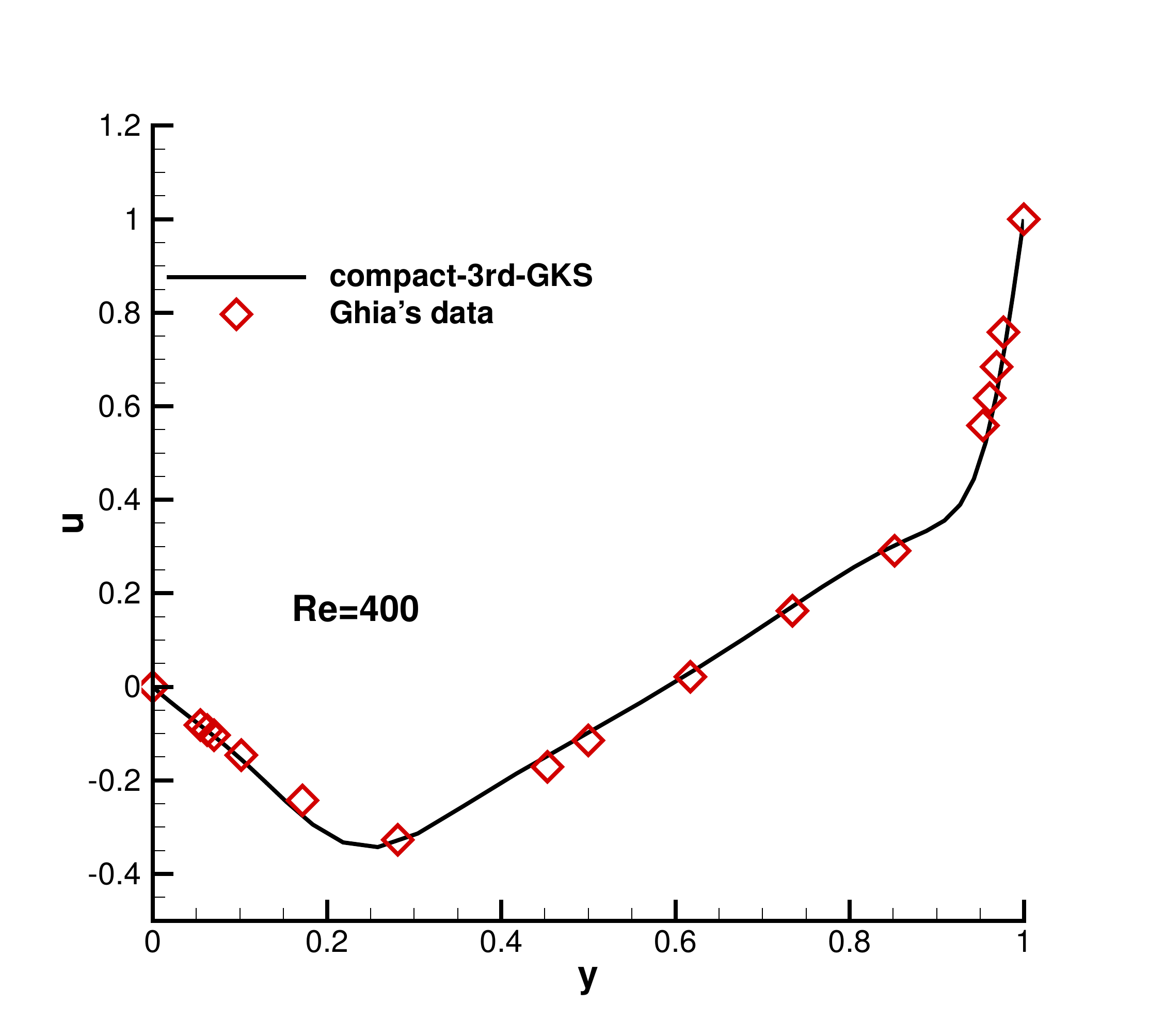}
    \includegraphics[width=0.47\textwidth]{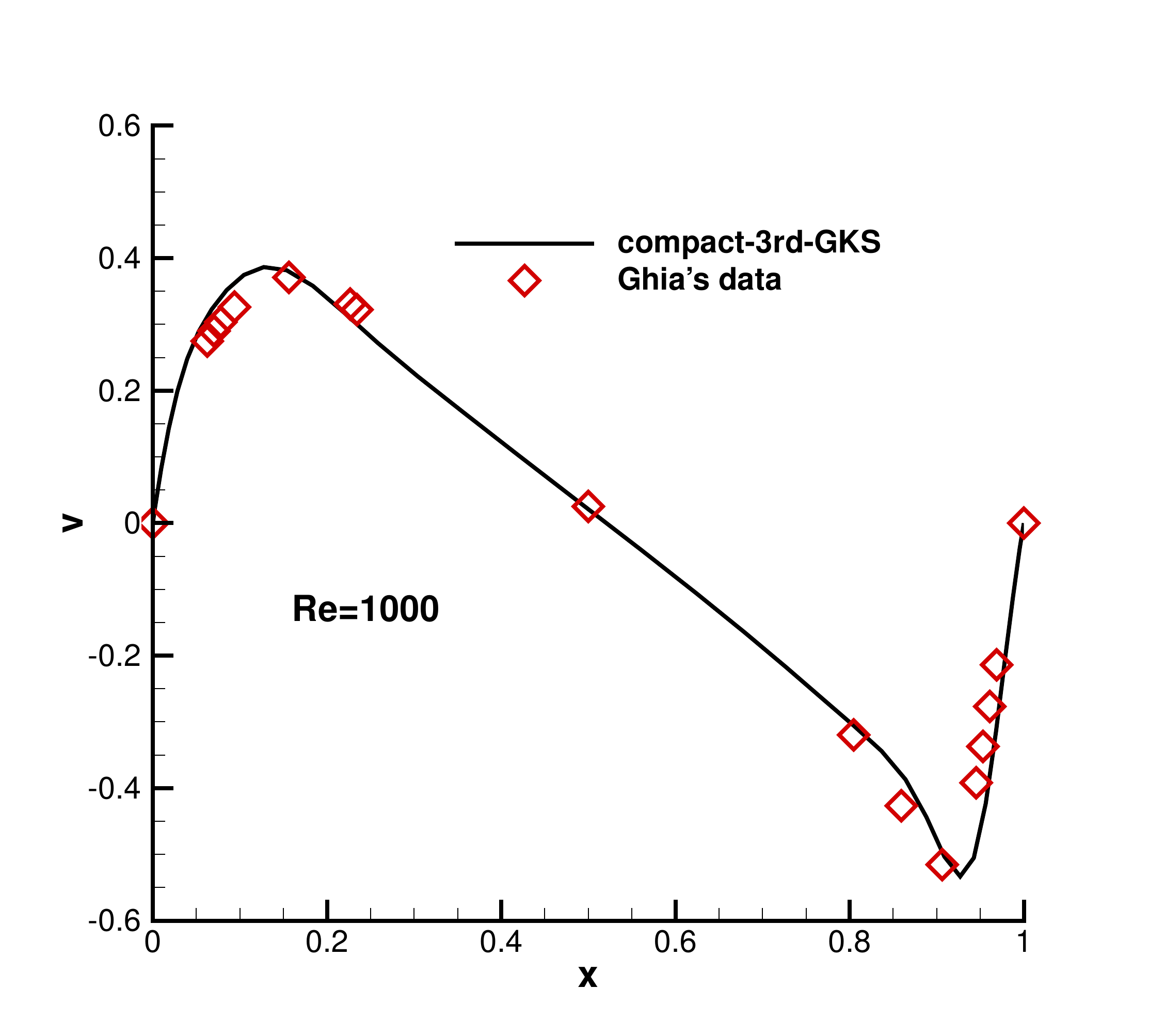}
    \includegraphics[width=0.47\textwidth]{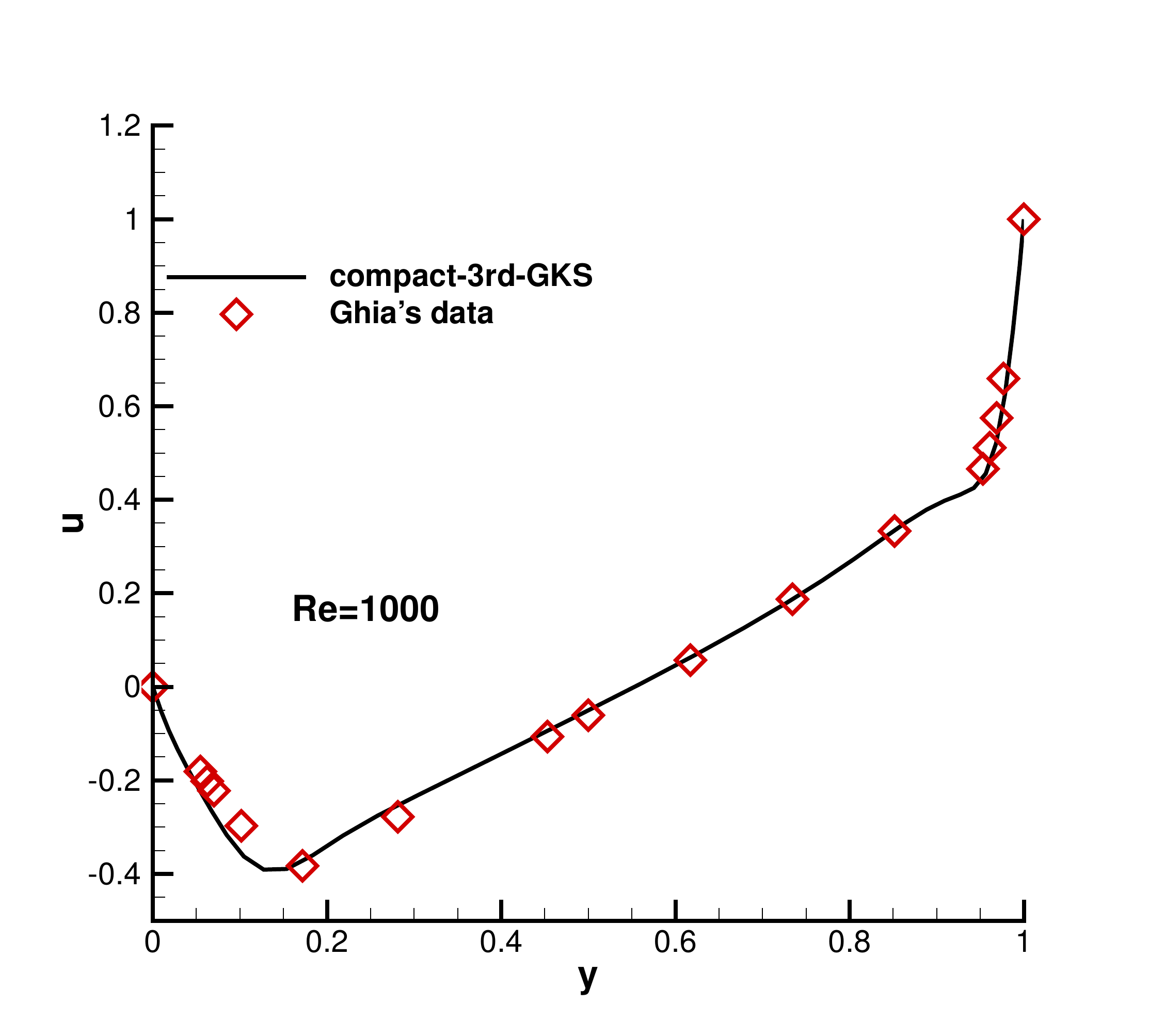}
    \includegraphics[width=0.47\textwidth]{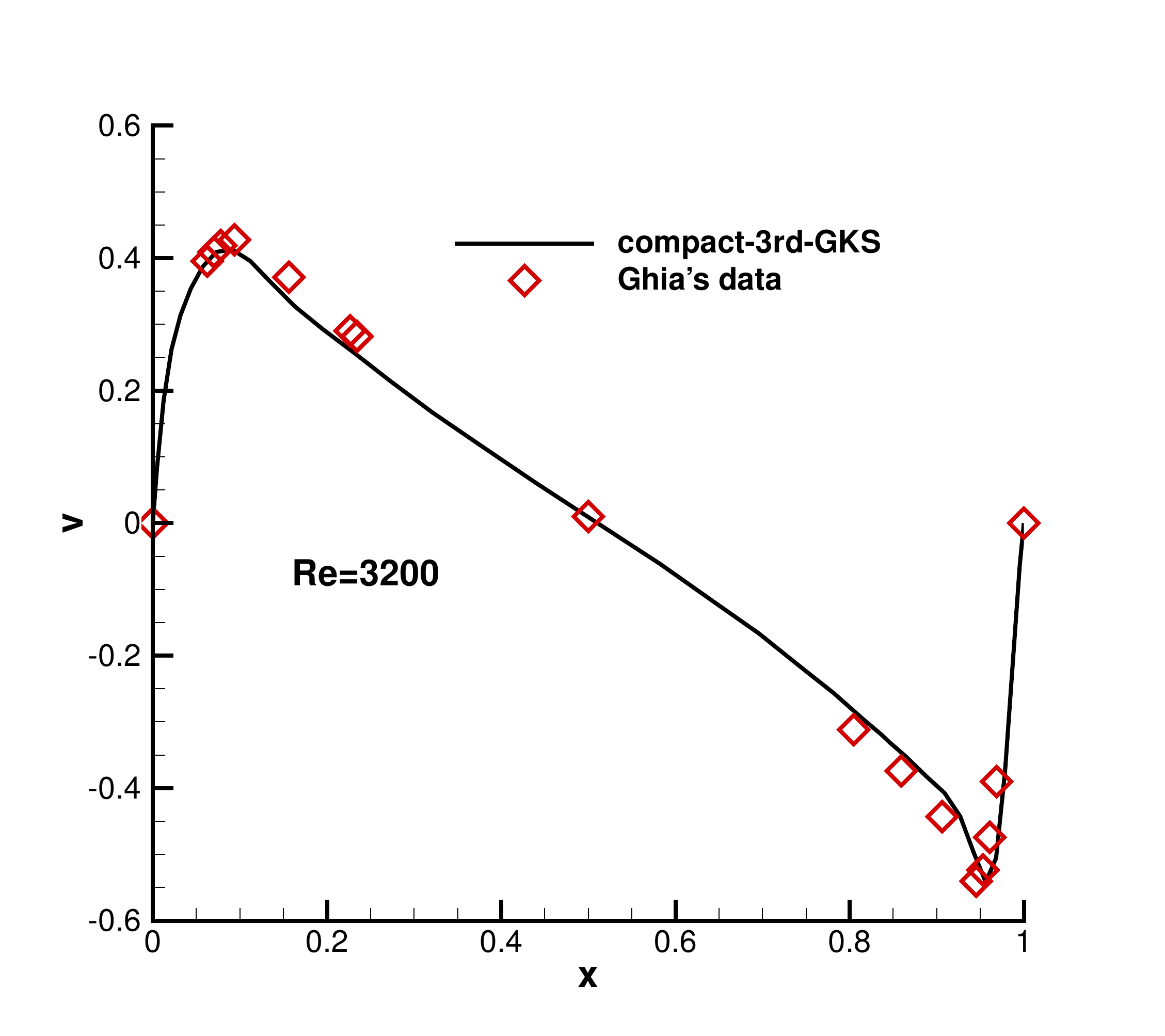}
    \includegraphics[width=0.47\textwidth]{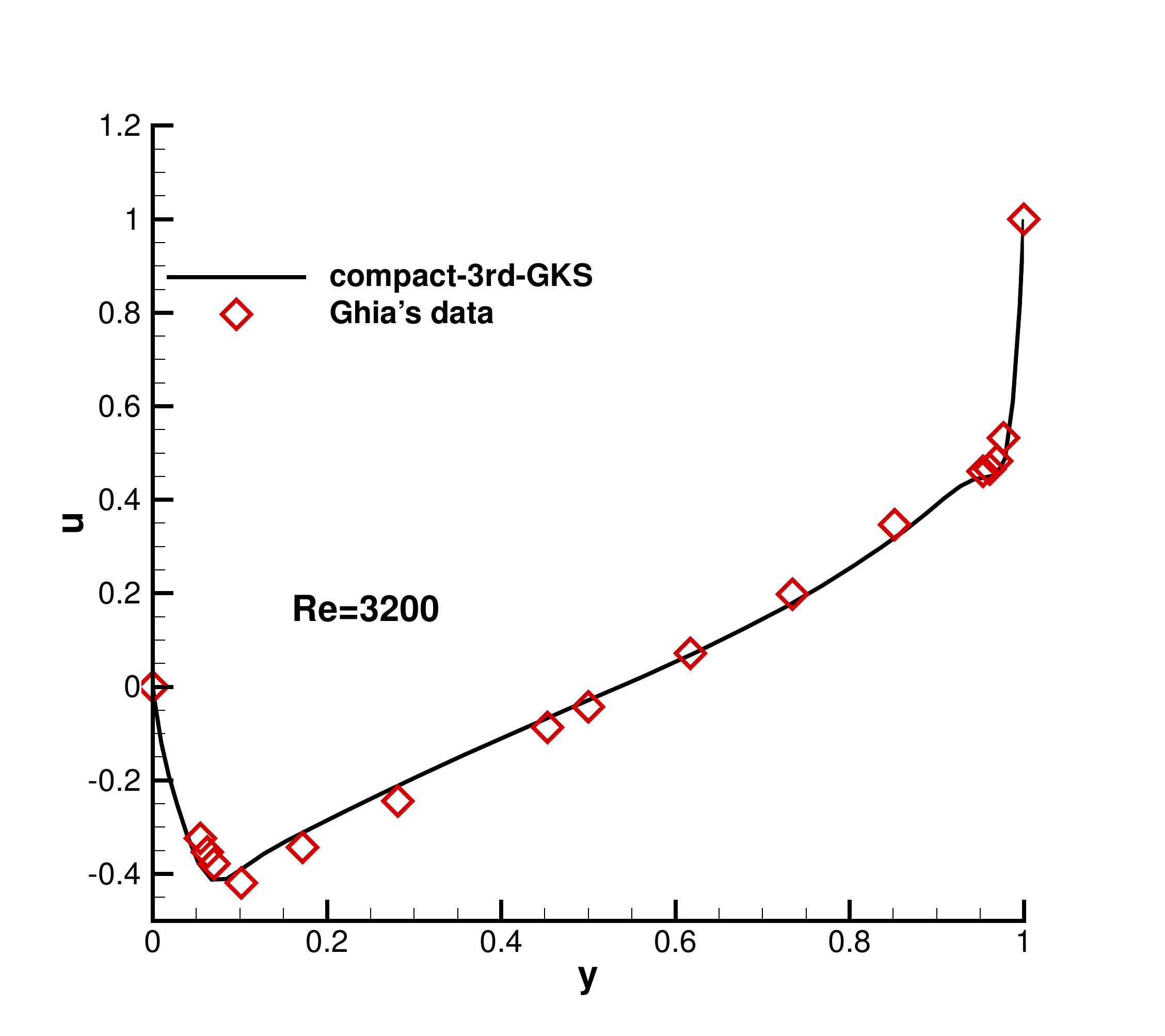}
    \caption{\label{cavity-line} Lid-driven cavity flow. V-velocities along the horizontal centerline and U-velocities along the vertical centerline with $Re=400,1000,3200$ (from top to bottom).}
\end{figure}

The lid-driven cavity problem is one of the most important benchmarks for validating incompressible or low-speed Navier-Stokes flow solvers.
The fluid is bounded in a unit square and is driven by a uniform translation of the top boundary. In this case, the flow is simulated with Mach number $Ma=0.15$ and all boundaries are isothermal and nonslip.
The computational domain is $[0, 1]\times[0,1]$.
As presented in Fig. \ref{2d-cavity}, the domain is covered by $35\times35 \times 2$ mesh points.
The stretching rate is $1.15$ with the minimum mesh size $h \approx 0.007$ near the wall boundary.
 Numerical simulations are obtained at three different Reynolds numbers, i.e., $Re=400$, $Re=1000$, and
$3200$. The streamlines for $Re=1000$ are shown in Fig. \ref{2d-cavity}. The $U$-velocity
along the center vertical line, and $V$-velocity along the center
horizontal line, are shown in Fig. \ref{cavity-line}. The benchmark data
\cite{ghia1982high} at the corresponding Reynolds numbers are also presented, and
the simulation results match well with these benchmark solutions.


\subsection{Laminar boundary layer}

\begin{figure}
    \centering
    \includegraphics[height=0.23\textwidth]{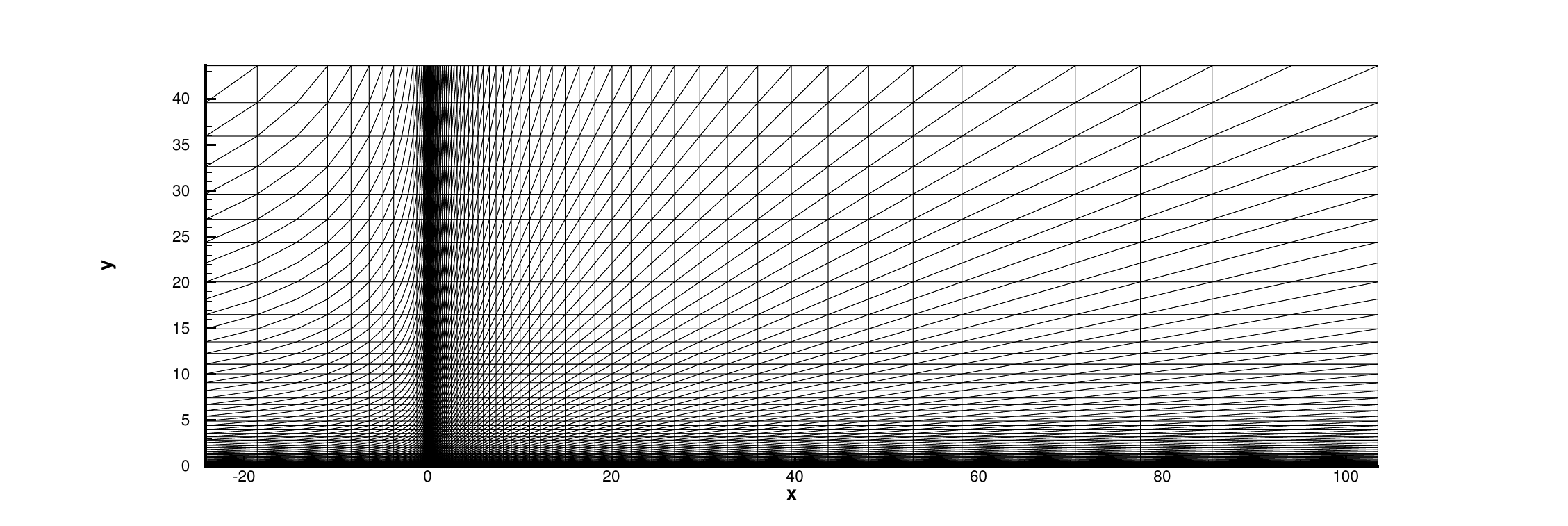}
    \includegraphics[height=0.23\textwidth]{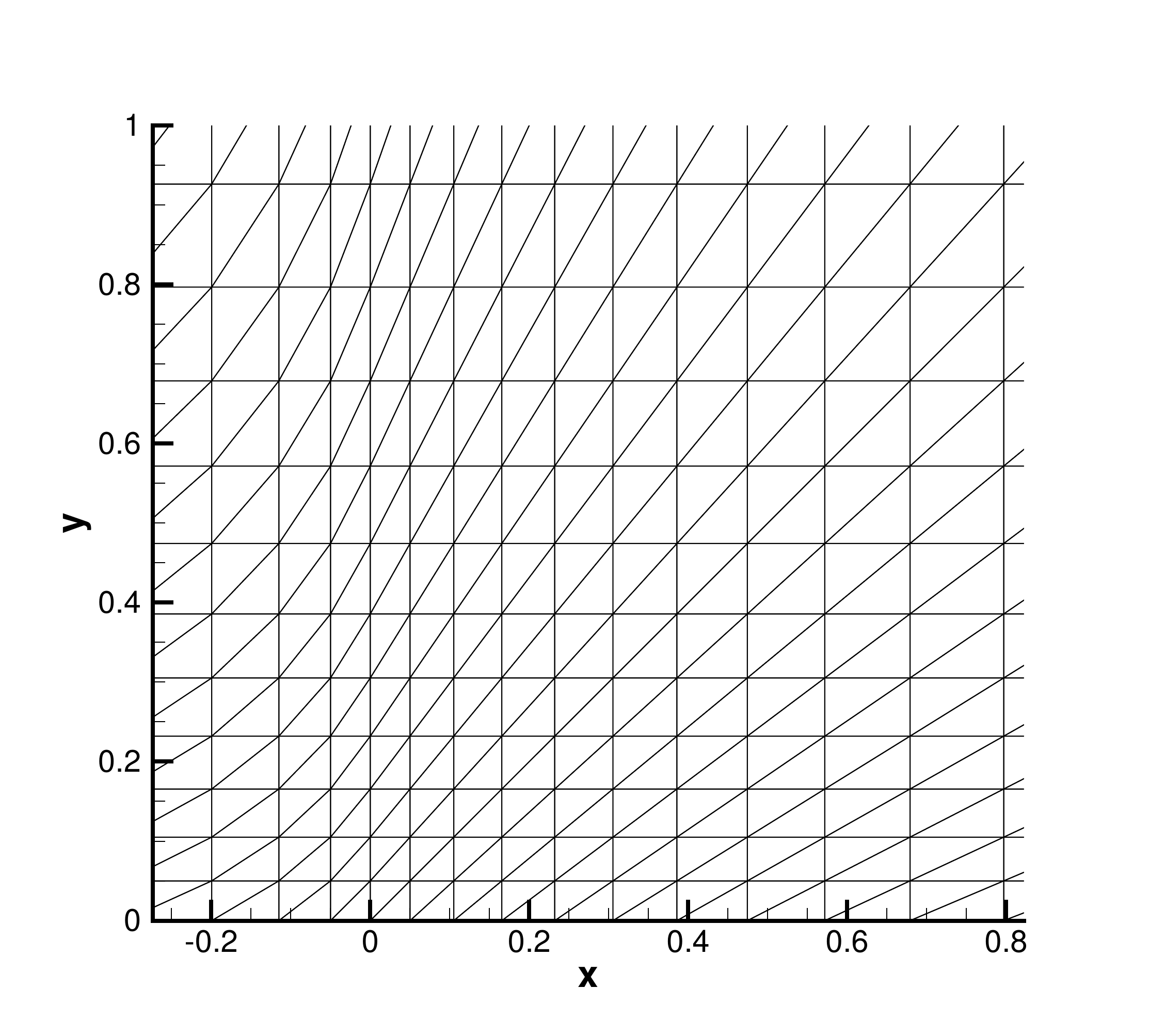}
    \caption{\label{boundary-layer-mesh}Computational domain for laminar boundary layer. $75 \times 47 \times 2$ mesh points are used with a wall thickness $h=0.05$ in the front of the flat plate.}
\end{figure}

\begin{figure}
    \centering
    \includegraphics[width=0.48\textwidth]{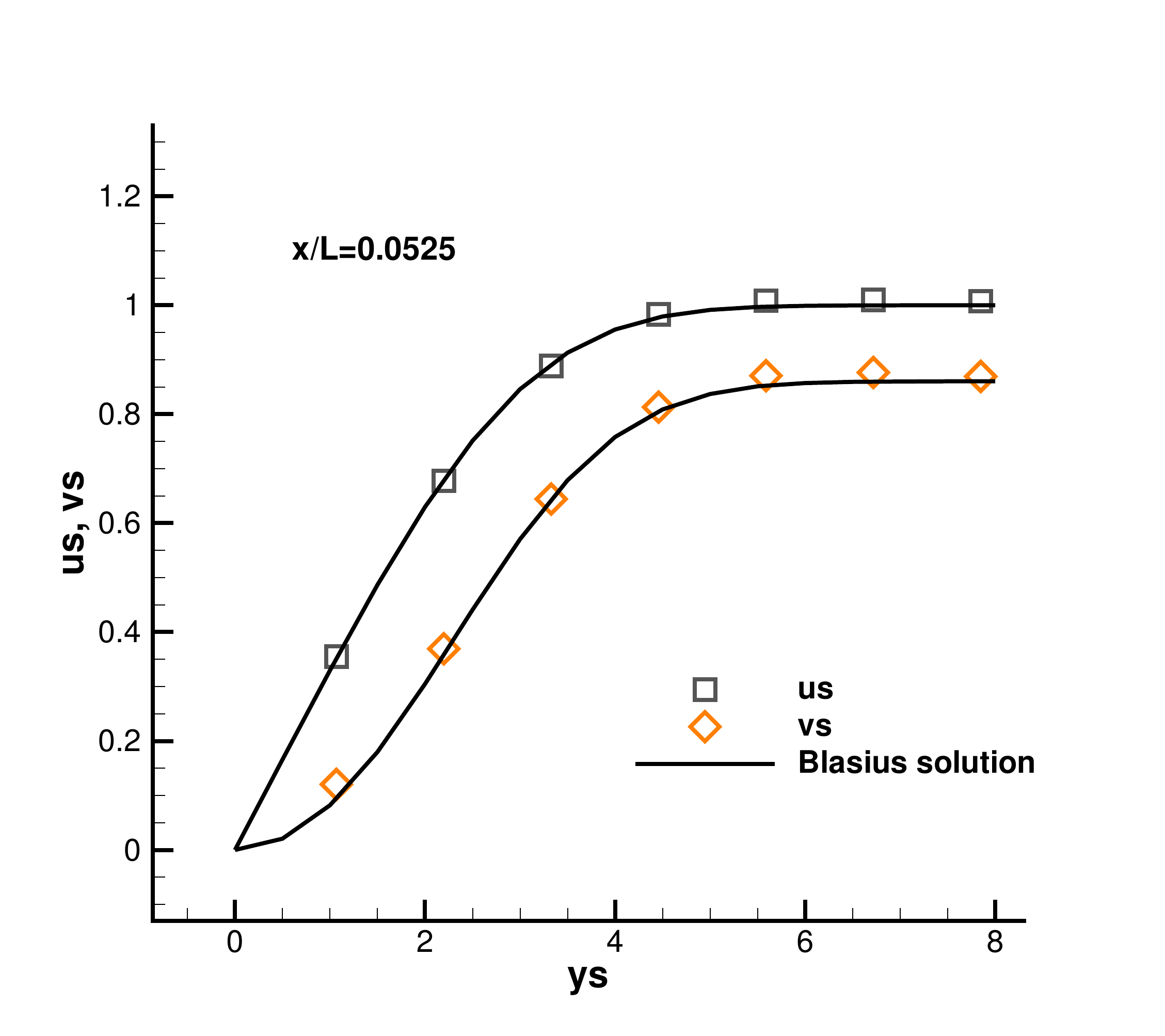}{a}
    \includegraphics[width=0.48\textwidth]{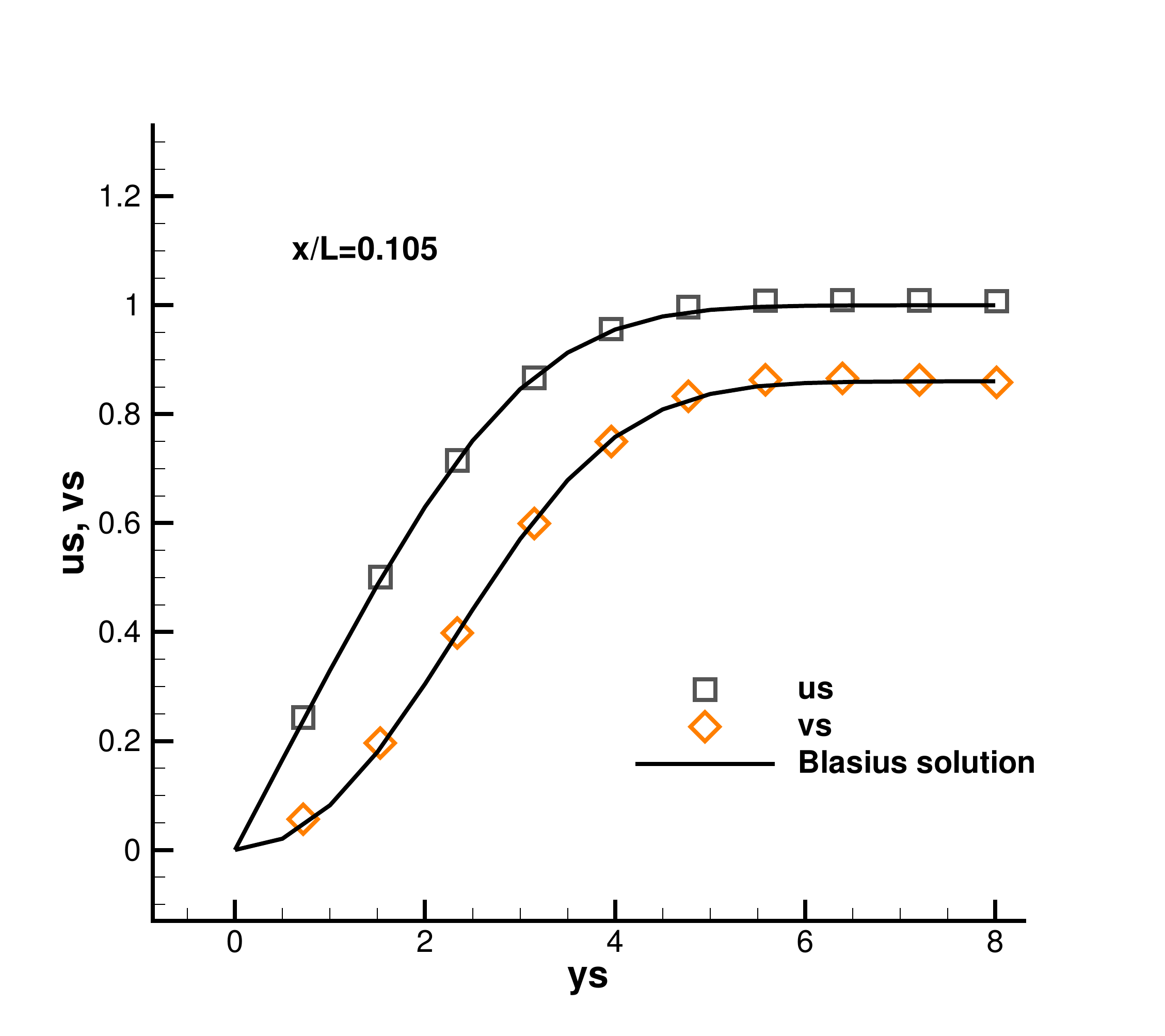}{b}
    \includegraphics[width=0.48\textwidth]{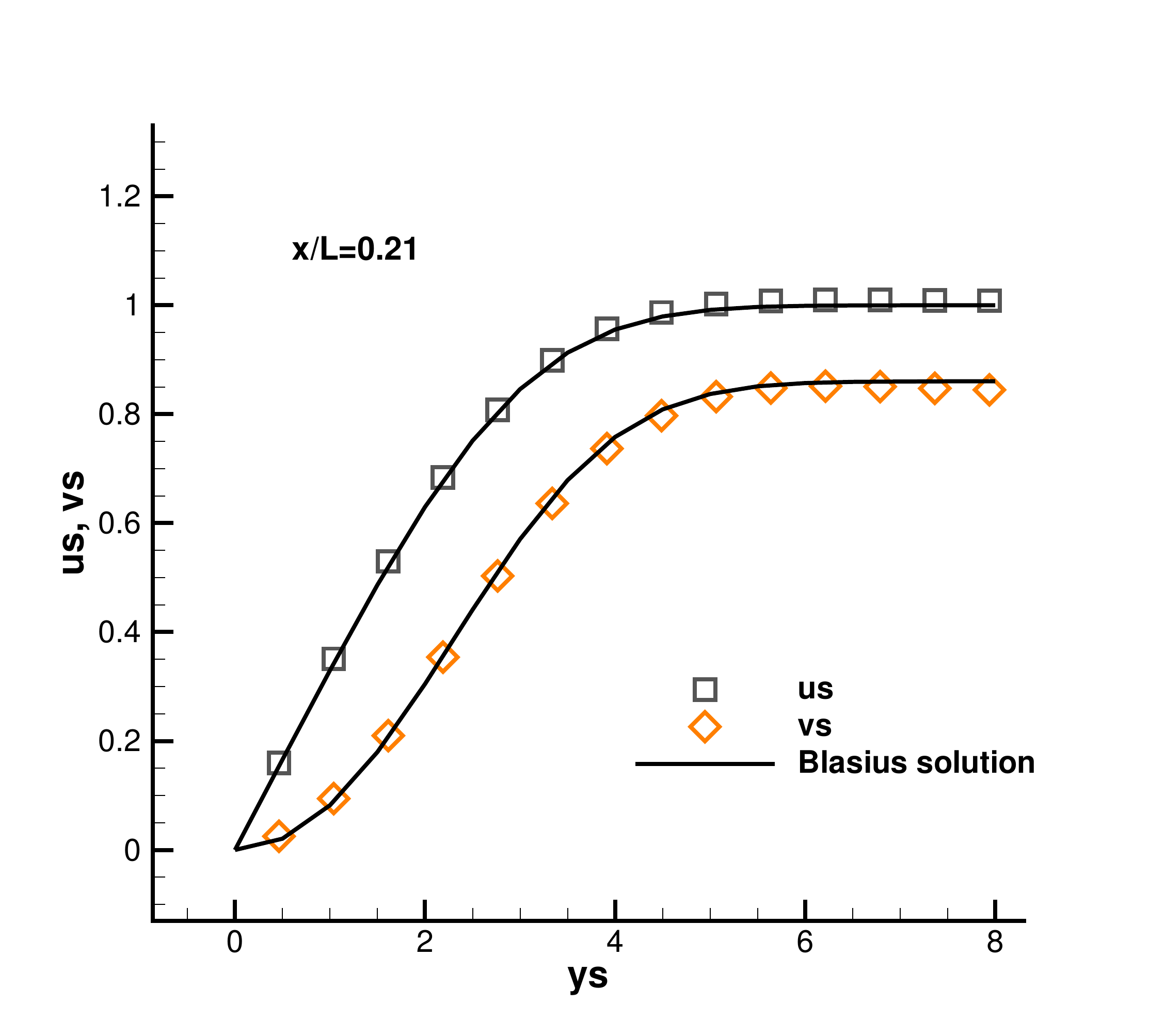}{c}
    \includegraphics[width=0.48\textwidth]{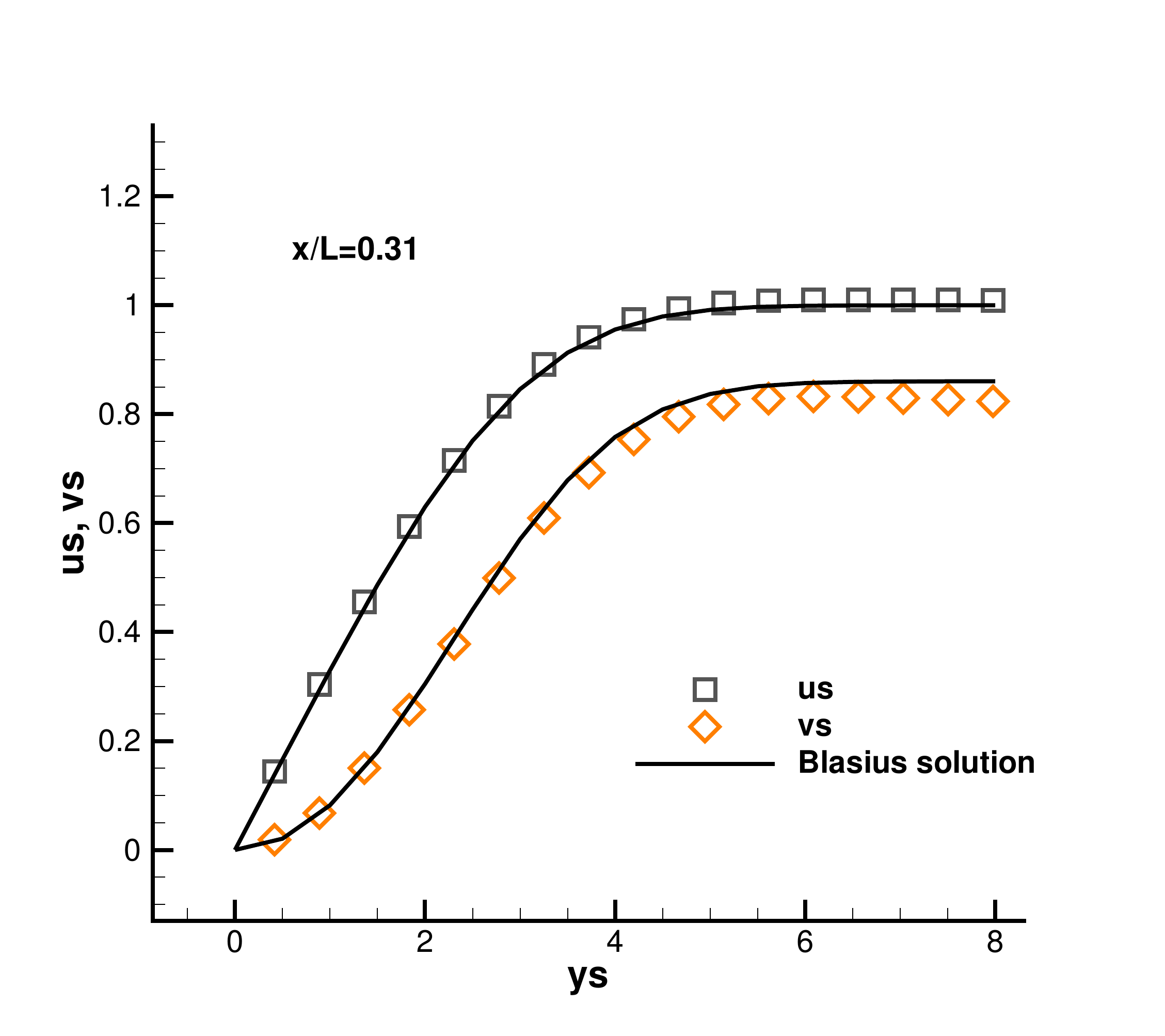}{d}
    \caption{\label{boundary-layer-vel} Laminar boundary layer: the non-dimensional velocity profile at $x/L=0.0525$ (a),
    $x/L=0.105$ (b), $x/L=0.21$ (c), and $x/L=0.31$ (d).}
\end{figure}

A low-speed laminar boundary layer with incoming Mach number $Ma=0.15$ is  simulated over a flat plate with Reynolds number $Re=U_{\infty}L/{\nu}=10^5$, where $L=100$ is the characteristic length.
The computational domain is shown in Fig. \ref{boundary-layer-mesh}, where the flat plate is placed at $x>0$ and $y=0$.
Total $75 \times 47 \times$ 2 mesh points are used in the domain with a refined cell size $h=0.05$ close to the boundary.
The mesh is generated from $(0,0)$ with a stretching ratio $1.1$ along the positive x-direction, $1.3$ along the negative x-direction,
and $1.1$ along the positive y-direction. There is $20$ mesh points in the front of the plate.
 An adiabatic non-slip boundary condition is imposed on the plate and a symmetric slip boundary condition is set
 at the bottom boundary in the front of the plate.
 The non-reflecting boundary condition based on the Riemann invariants is adopted for the other boundaries, where the free stream is set as $\rho_{\infty}=1, p_{\infty}=1/\gamma$.
The non-dimensional velocity U and V are given in Fig. \ref{boundary-layer-vel} at four selected locations. The numerical results match well with the analytical solutions even with a few mesh points at $x/L=0.0525$.

\subsection{Hypersonic flow around a cylinder }
In this case, both inviscid and viscous hypersonic flows impinging on a cylinder are tested to validate robustness of the current scheme.

\noindent{\sl{(a) Inviscid Cases}}

\begin{figure}[!htb]
    \centering
    \includegraphics[height=0.52\textwidth]{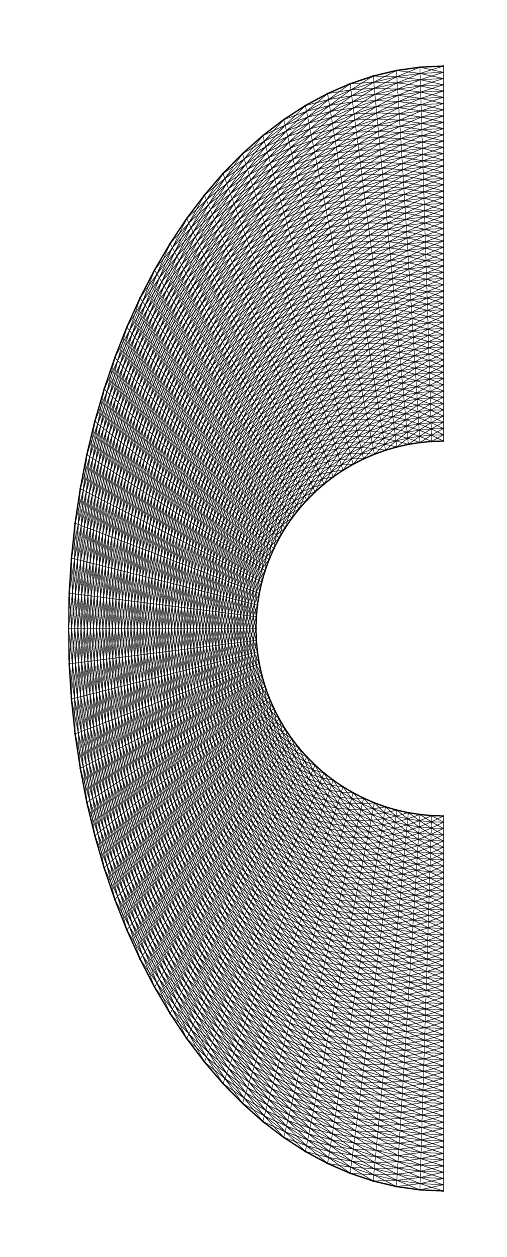}
    \includegraphics[height=0.52\textwidth]{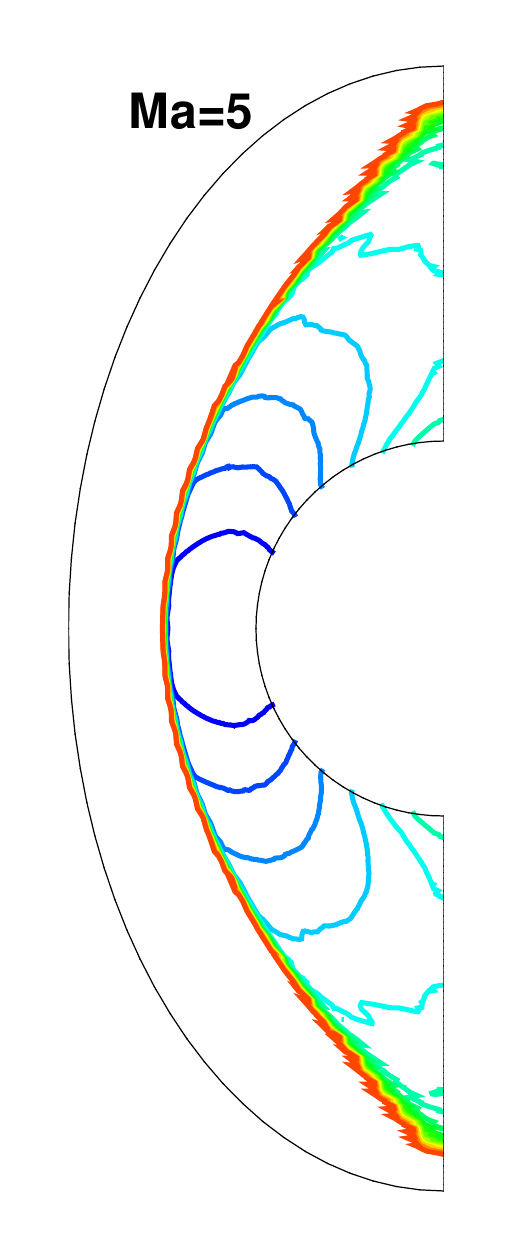}
    \includegraphics[height=0.52\textwidth]{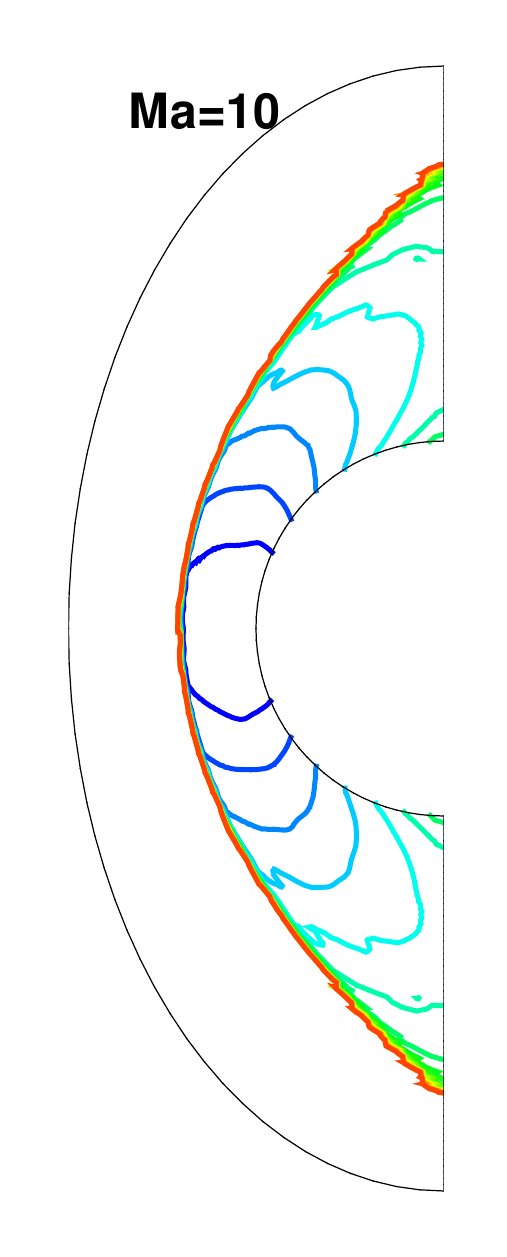}
    \includegraphics[height=0.52\textwidth]{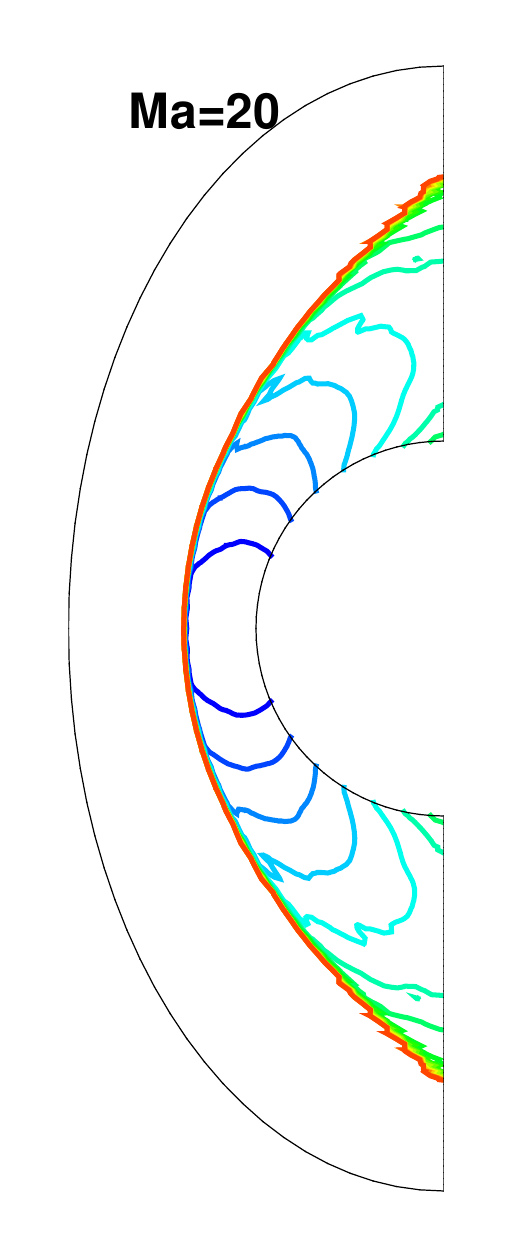}
    \caption{\label{mach-s} Hypersonic inviscid flow past a cylinder: Mach number distributions with Mach number $Ma=5, 10$, and $20$ under uniform mesh. CFL=$0.4$.}
\end{figure}

\begin{figure}[!htb]
    \centering
    \includegraphics[height=0.52\textwidth]{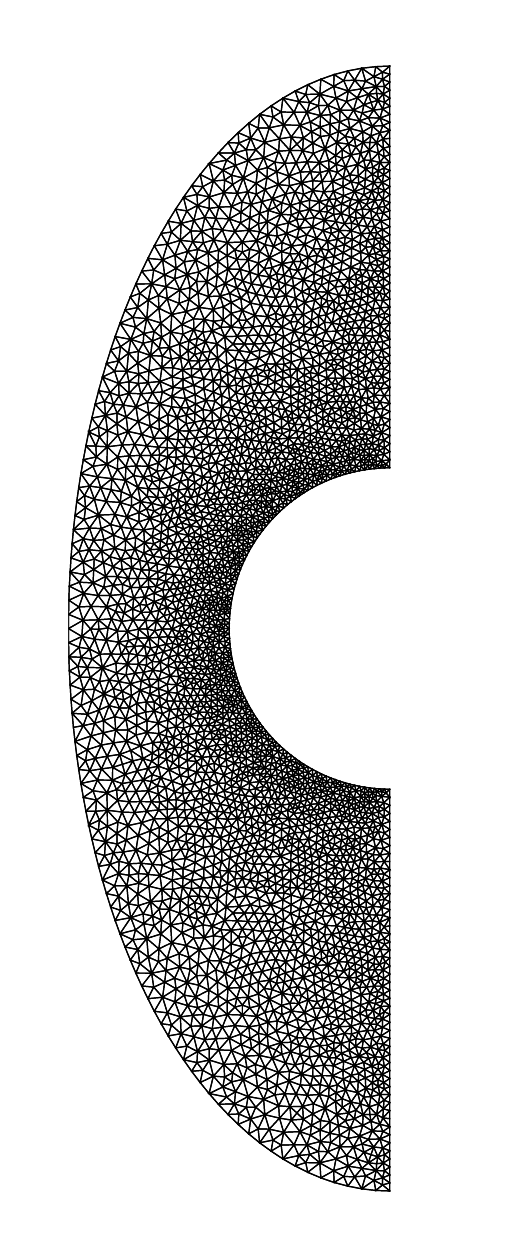}
    \includegraphics[height=0.52\textwidth]{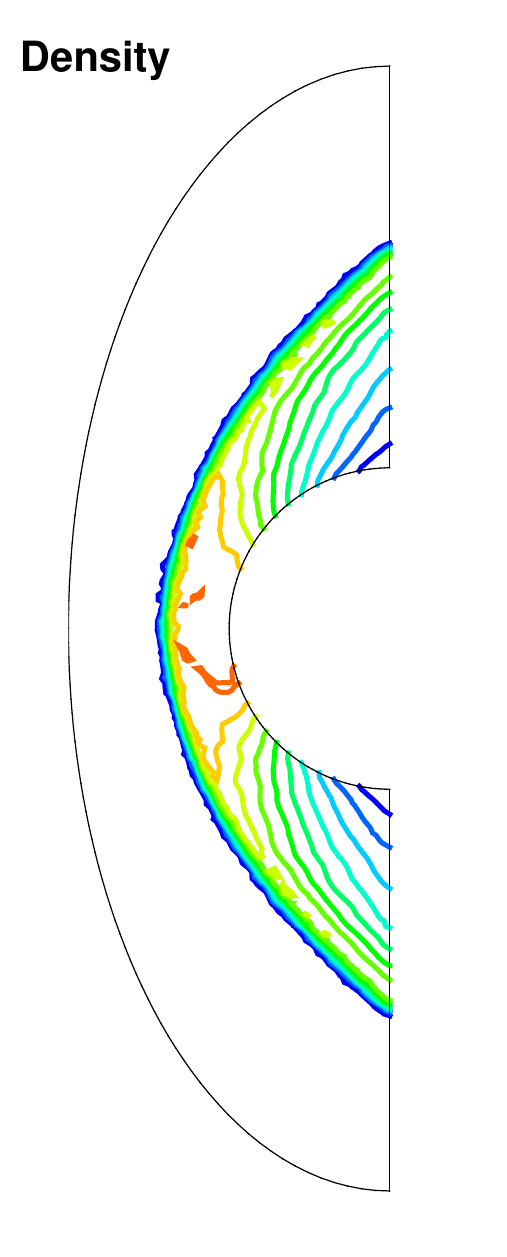}
    \includegraphics[height=0.52\textwidth]{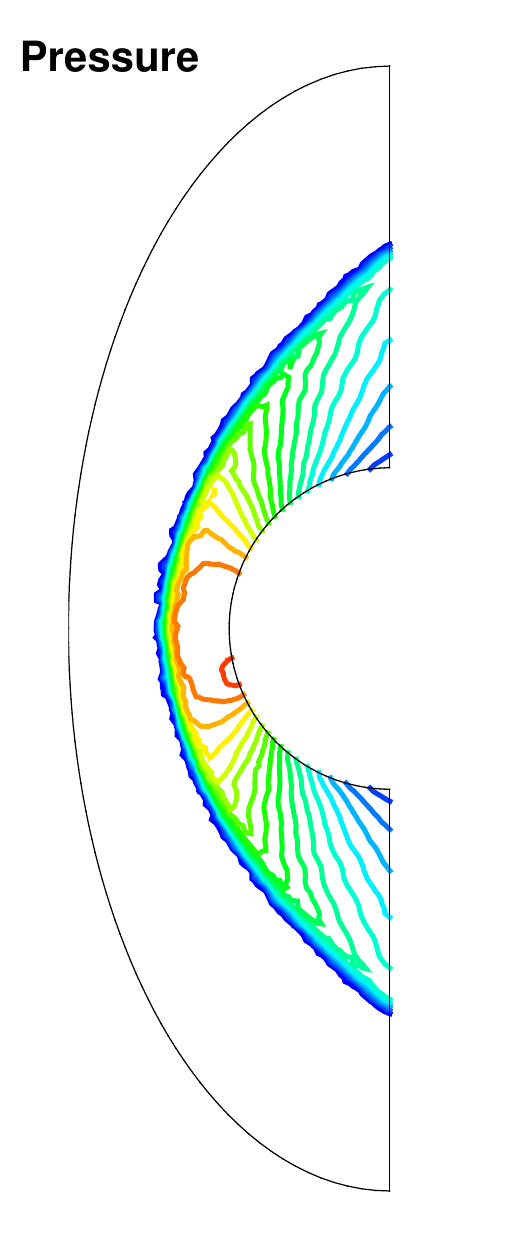}
    \includegraphics[height=0.52\textwidth]{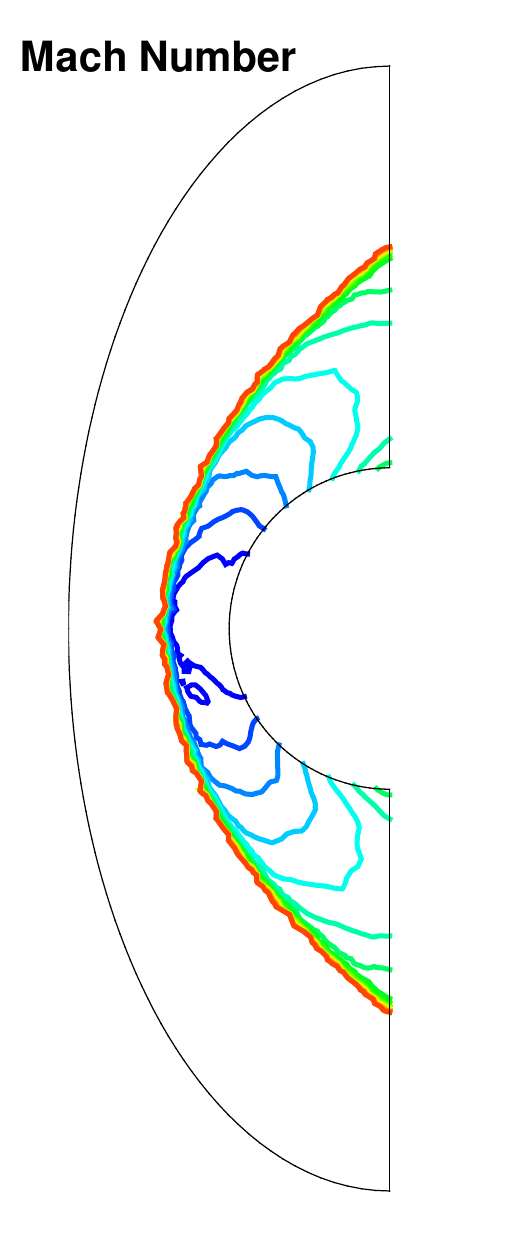}
    \caption{\label{mach-coarse} Hypersonic inviscid flow past a cylinder: mesh, density, pressure, and Mach number distributions by the compact 3rd-order GKS. Mach=$20$, CFL=$0.2$. }
\end{figure}

The first one is for the Euler solutions. The incoming flow has Mach numbers up to $20$ on the cylinder.
The reflective boundary condition is imposed on the wall of the cylinder while the right boundary is set as outflow boundary condition.
Firstly, a uniform quadrilateral mesh with $60$ cells along radial direction and $50$ cells along circumferential direction
is split diagonally in each cell for triangulation which is used in the computation.
The mesh size along radial direction is $0.03$. The uniform mesh and Mach number distributions are presented in Fig. \ref{mach-s}.
The results agree well with those performed under structured mesh by the original non-compact high order GKS \cite{Pan2016twostage}.
To further demonstrate the robustness of the compact scheme, an irregular mesh with near-wall thickness $h \approx 0.03$ is used
for this case as well. Under such a mesh at Mach number $20$, the limiter on the HWENO weights is triggered to avoid the appearance of negative temperature.
The mesh and  flow distributions are plotted in Fig. \ref{mach-coarse}.
In all cases with different triangular mesh, the current compact scheme
can capture strong shocks very well without carbuncle phenomenon. The robustness of the scheme is fully validated.

\noindent{\sl{(b) Viscous Case}}

\begin{figure}[!htb]
    \centering
    \includegraphics[height=0.6\textwidth]{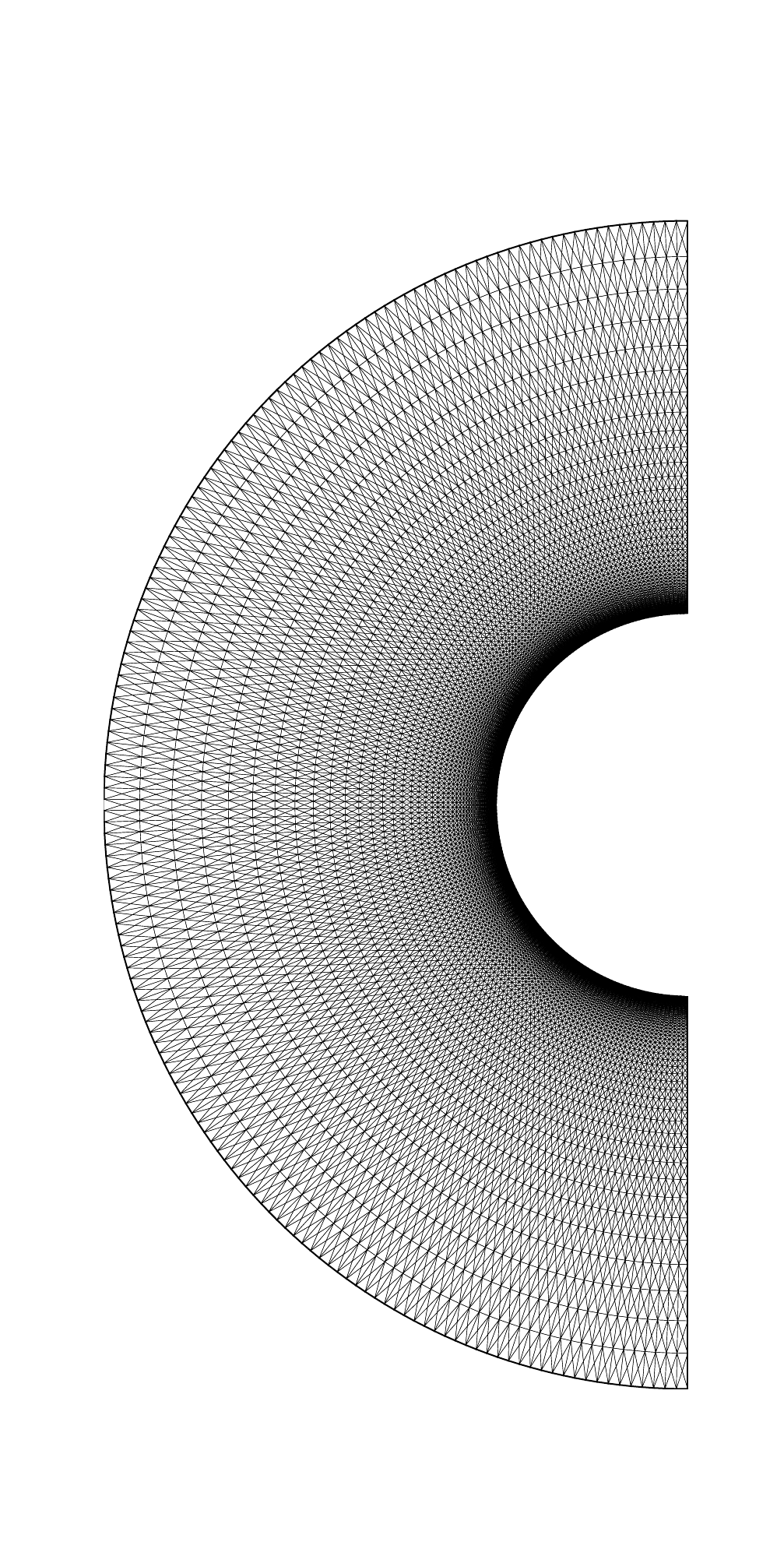}
    \includegraphics[height=0.6\textwidth]{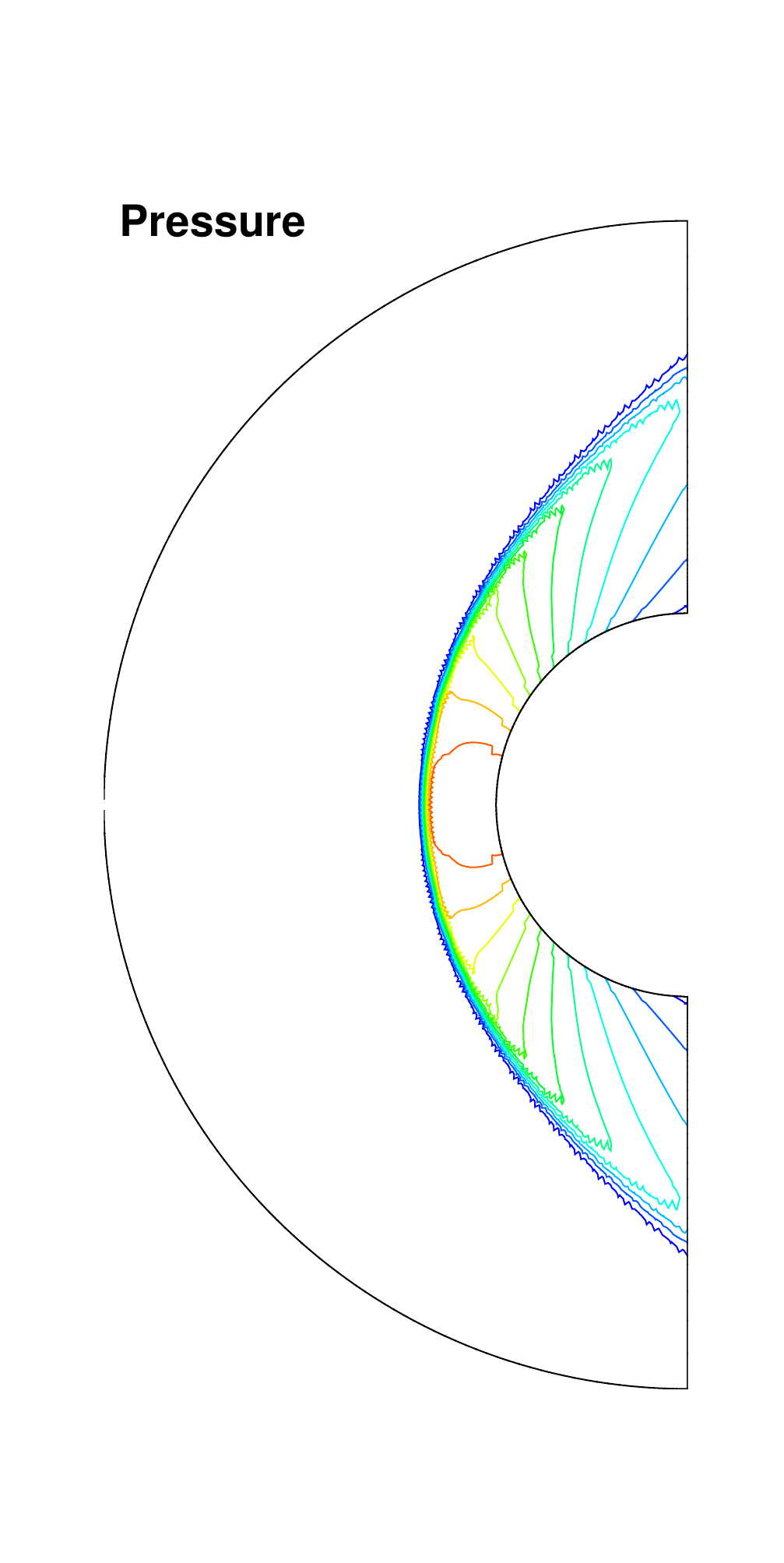}
    \includegraphics[height=0.6\textwidth]{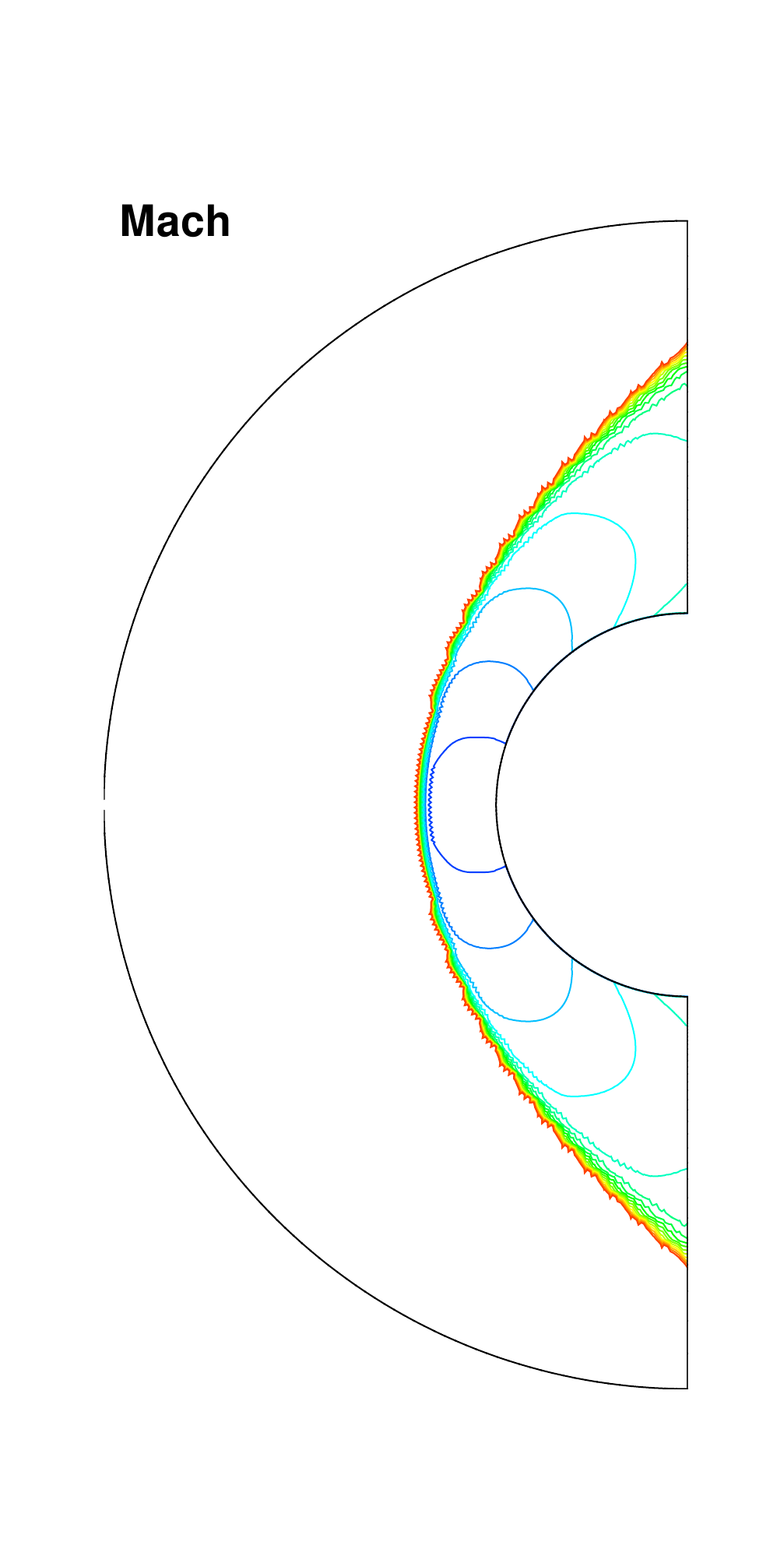}
    \caption{\label{vis-mach-s} A Mach number $8.03$ viscous flow past a cylinder with Mesh I: mesh, pressure, and Mach number distributions by the compact 3rd-order GKS. CFL=$0.1$.  The mesh size near the wall is $h \approx 10^{-4}$.}
\end{figure}

\begin{figure}[!htb]
    \centering
    \includegraphics[width=0.48\textwidth]{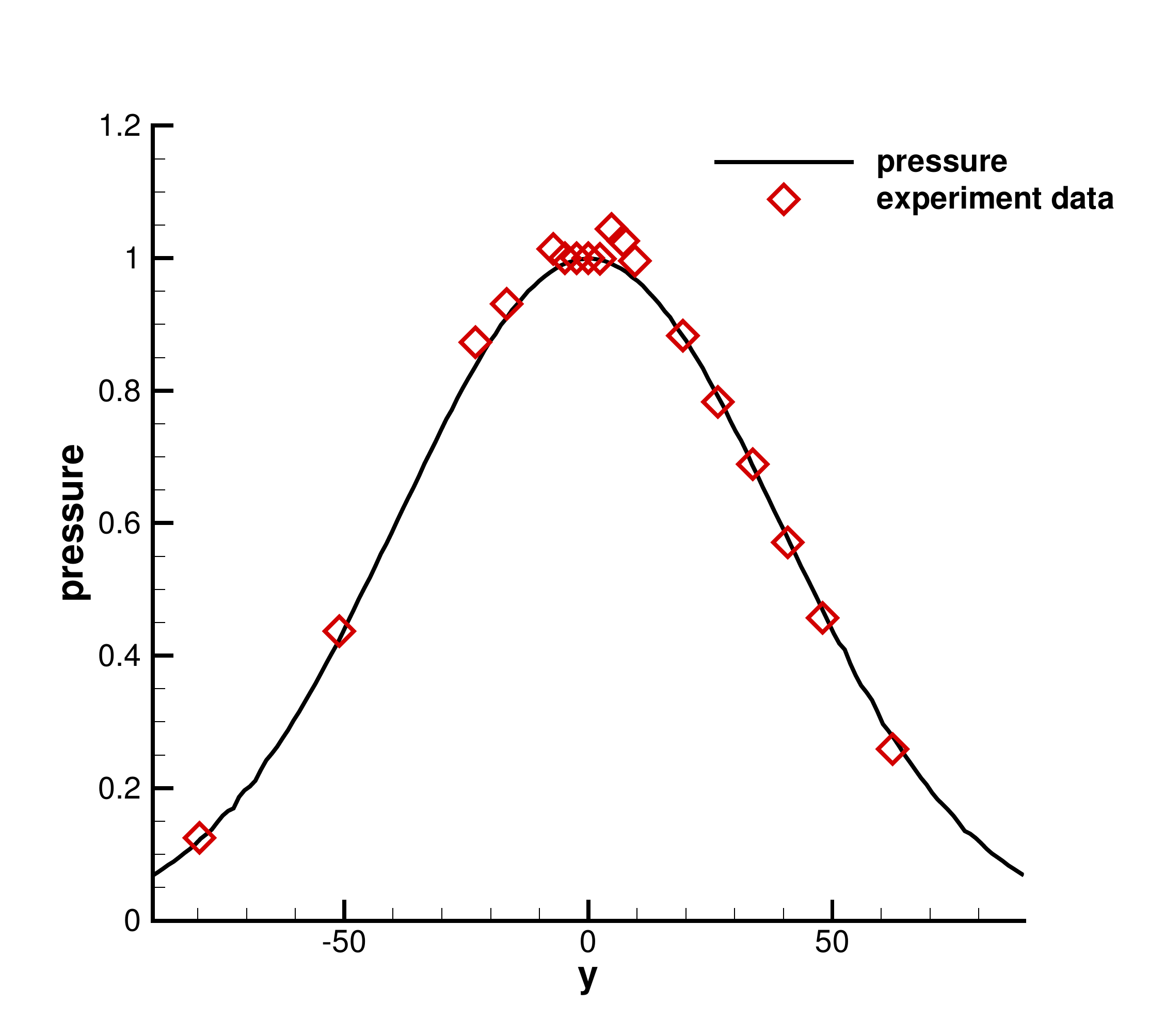}
    \includegraphics[width=0.48\textwidth]{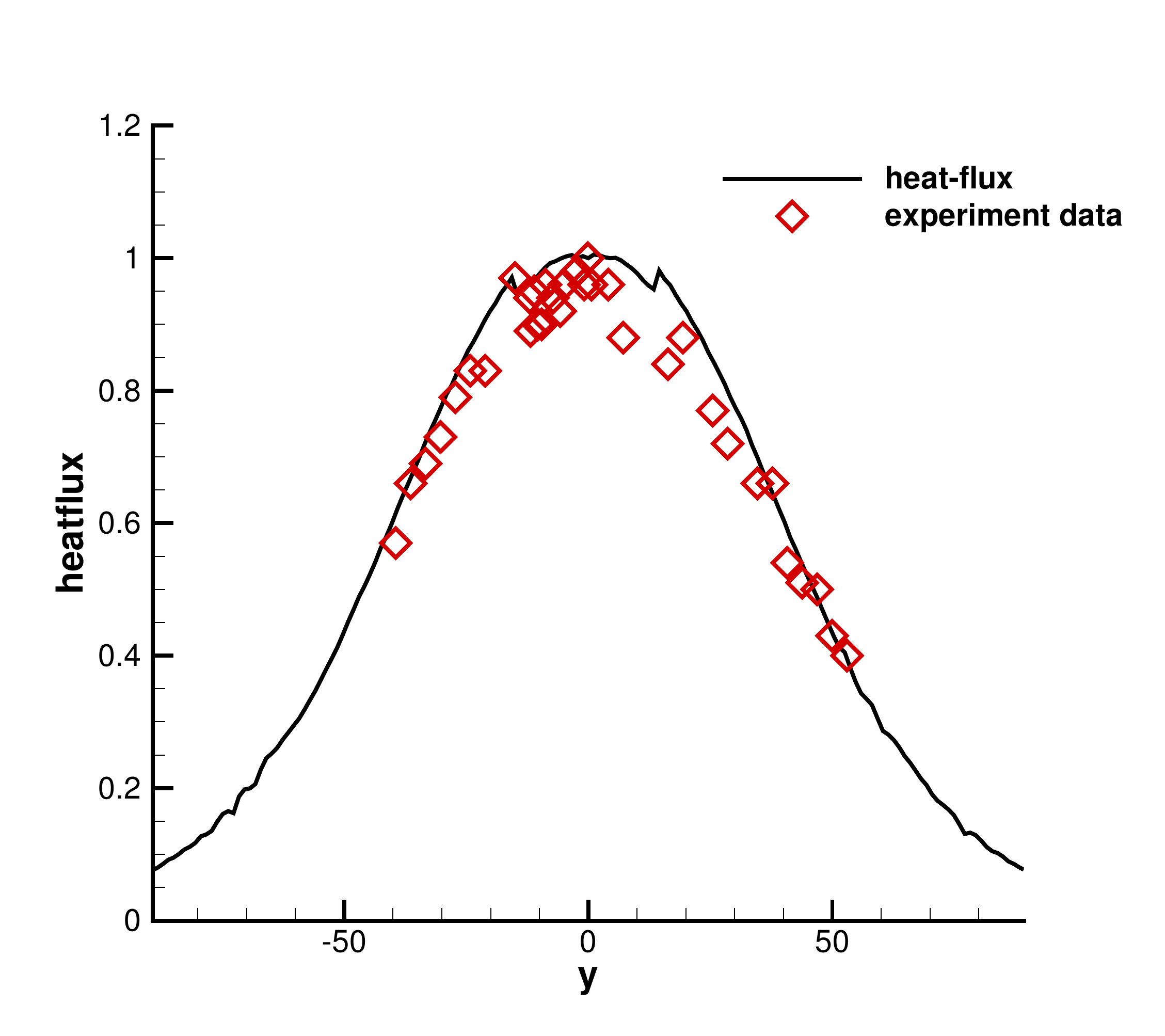}
    \caption{\label{vis-mach-s-line} A Mach number $8.03$ viscous flow past a cylinder with  Mesh I: non-dimensional pressure and heat flux distributions along the surface of cylinder which are compared with the experimental data in \cite{wieting1987experimental}.}
\end{figure}

\begin{figure}[!htb]
    \centering
    \includegraphics[height=0.5\textwidth]{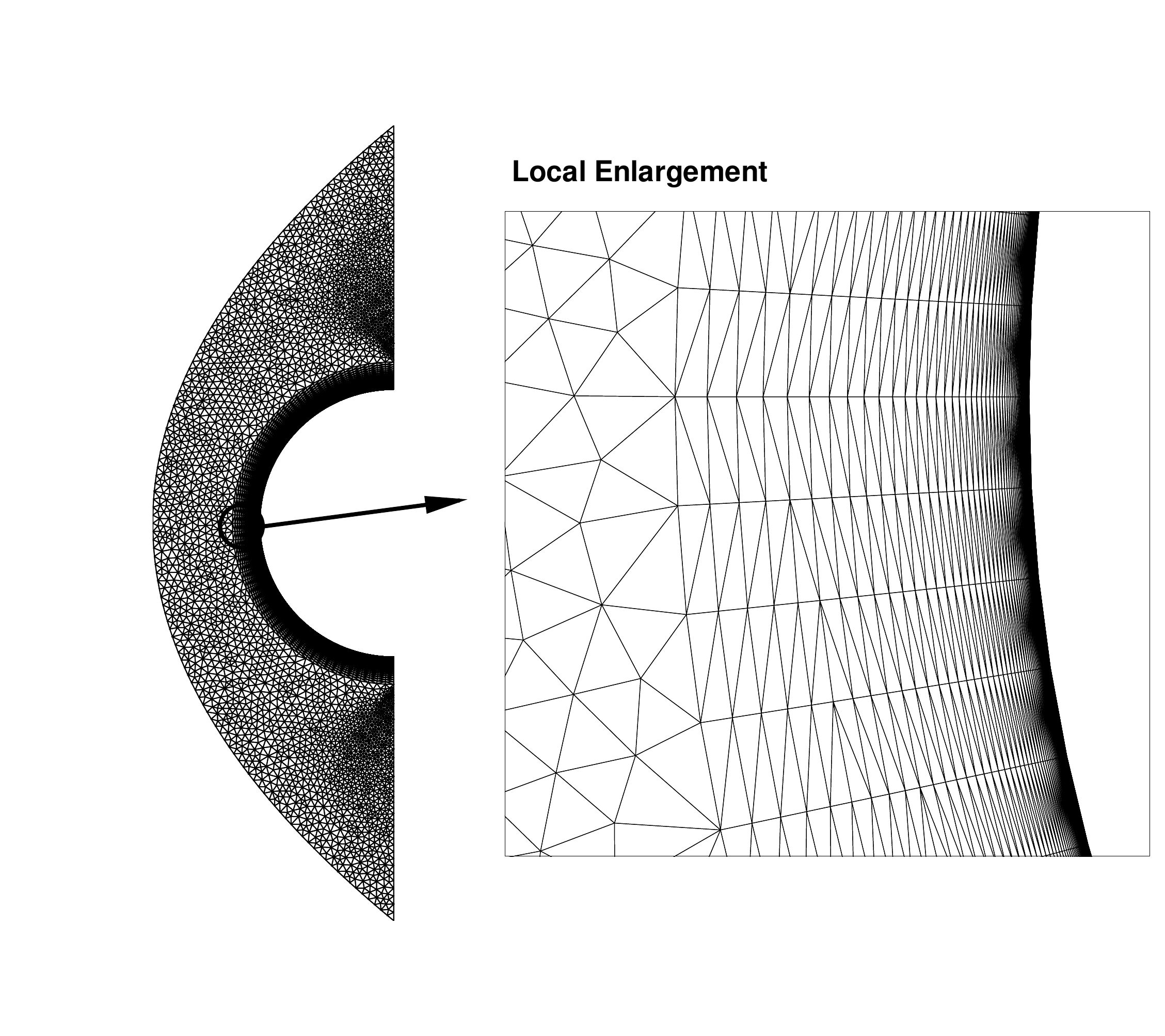}
    \includegraphics[height=0.5\textwidth]{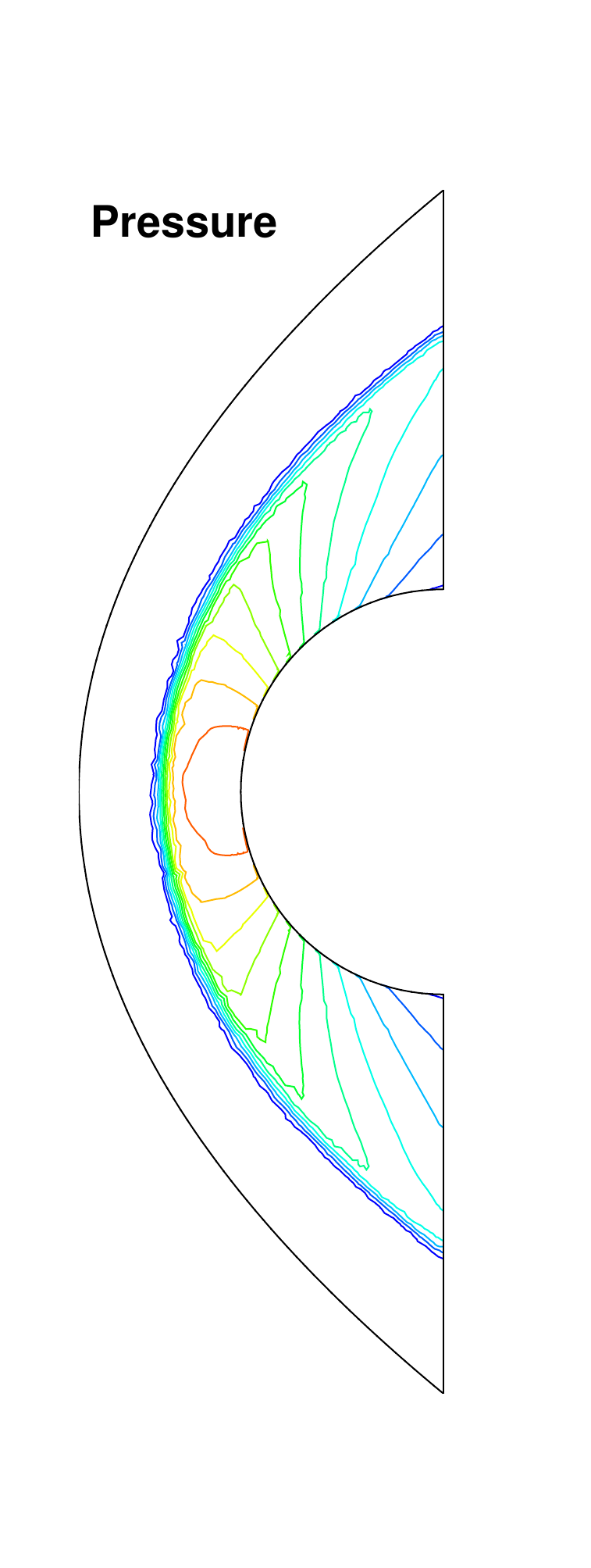}
    \includegraphics[height=0.5\textwidth]{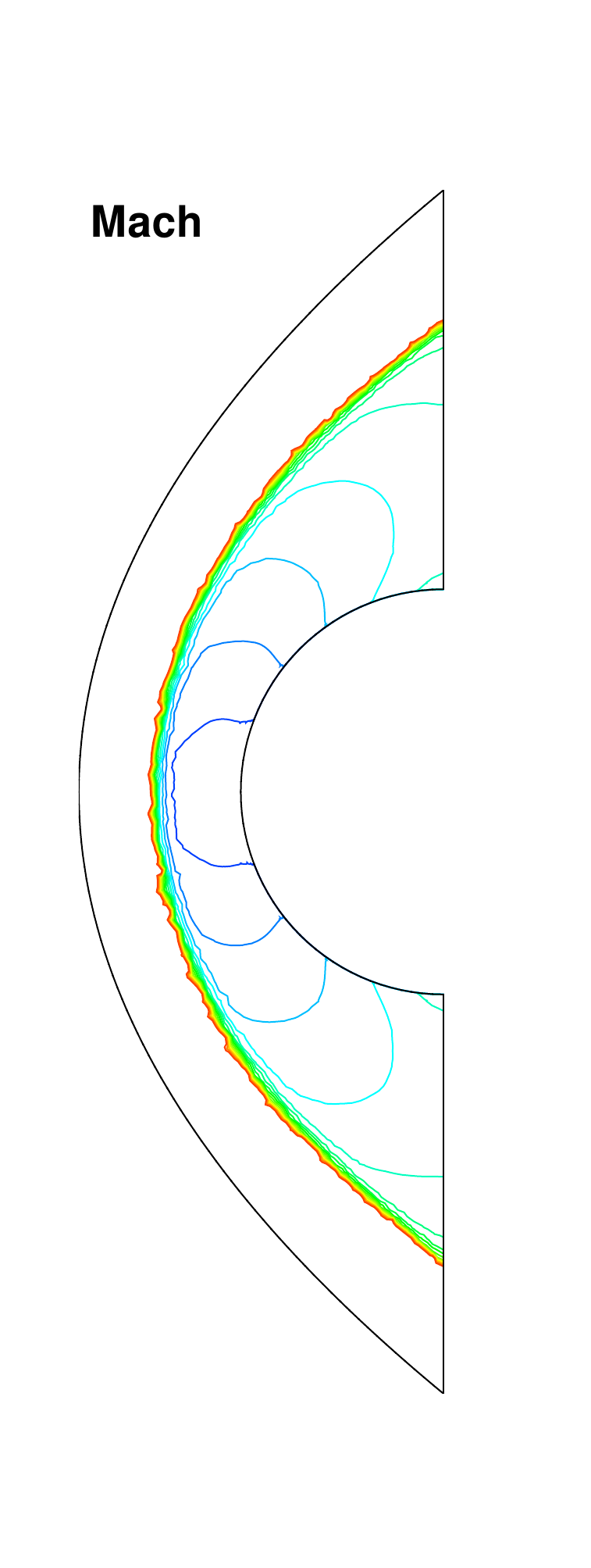}
    \caption{\label{vis-mach-o} A Mach number $8.03$ viscous flow past a cylinder with Mesh II: mesh, pressure, and Mach number distributions by the compact 3rd-order GKS. CFL=$0.1$.
        The mesh size near the wall is $h \approx 10^{-5}$.}
\end{figure}

\begin{figure}[!htb]
    \centering
    \includegraphics[width=0.48\textwidth]{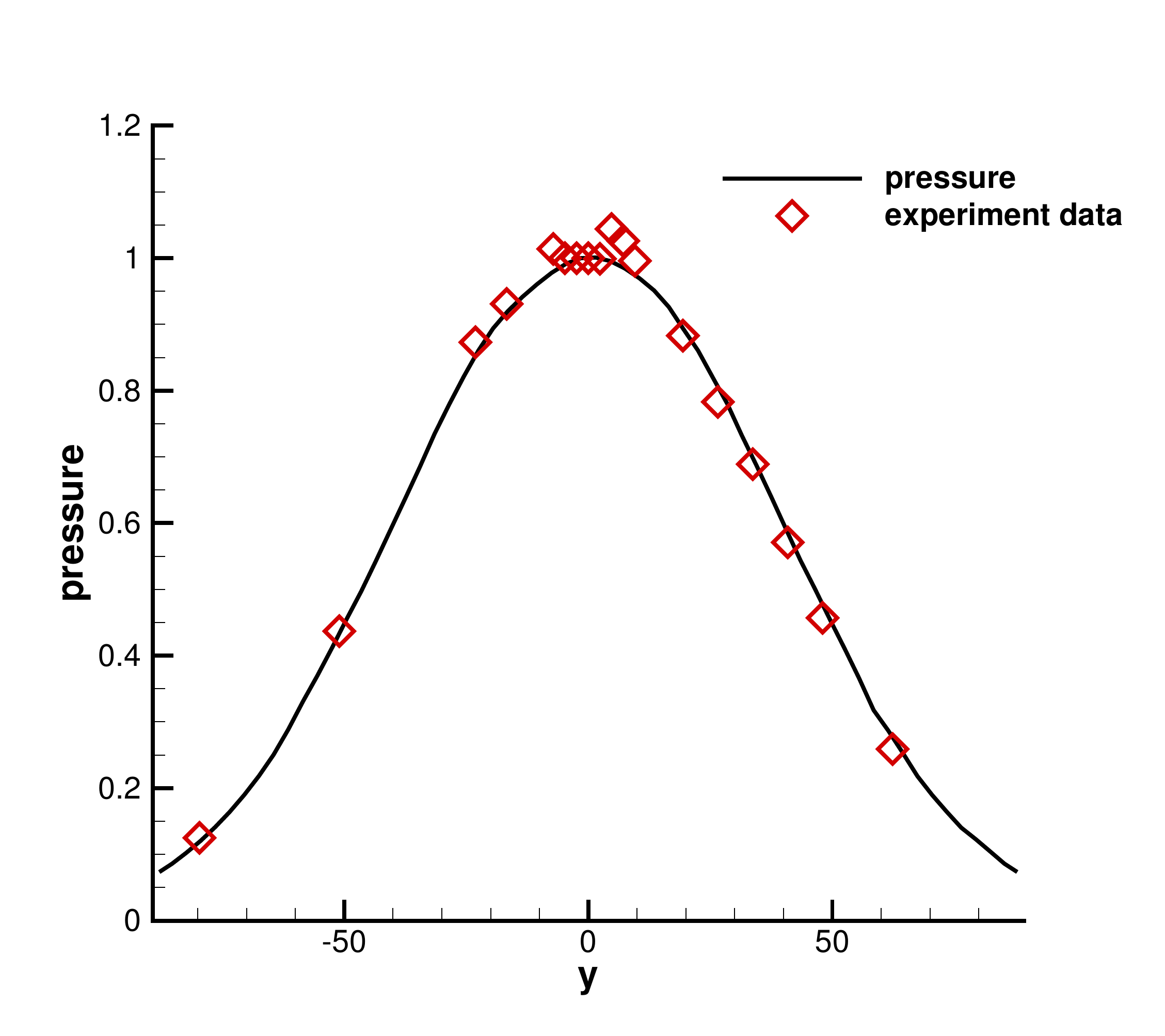}
    \includegraphics[width=0.48\textwidth]{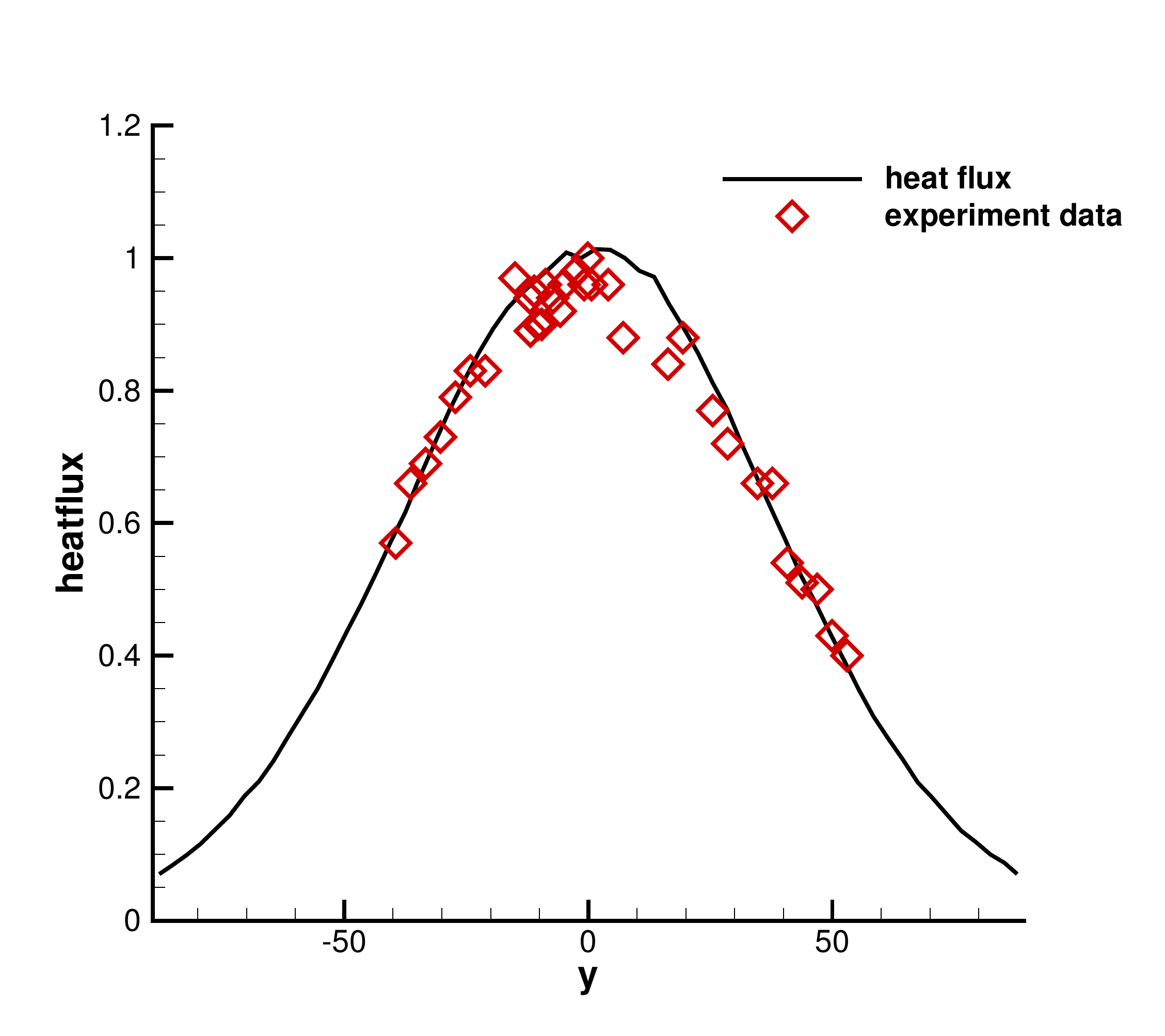}
    \caption{\label{vis-mach-o-line} A Mach number $8.03$ viscous flow past a cylinder with Mesh II: non-dimensional pressure and heat flux distributions along the surface of cylinder which are compared with the experimental data in \cite{wieting1987experimental}.}
\end{figure}

This test is taken from the experiment by Wieting \cite{wieting1987experimental}.
The non-slip isothermal boundary condition with wall temperature $T_w=294.44 K$ is imposed as the cylinder surface.
The far-field flow condition is given by $Ma_{\infty}=8.03, T_{\infty}=124.94 K$.
The Reynolds number is $Re=1.835 \times 10^5$.
Two meshes named as Mesh I and Mesh II are used for this test case.
As shown in Fig. \ref{vis-mach-s}, to resolve the boundary layer Mesh I is generated by simple triangulation
of a non-uniform quadrilateral mesh of $80 \times 161$ points with a near-wall thickness $h \approx 10^{-4}$ and a stretching ratio $1.1$.
As an explicit scheme, the CFL number is set as 0.1 due to the stiffness of viscous term.
To improve the efficiency, a primary flow field calculated by first order kinetic method \cite{GKS-lecture} is used as the initial field.
 The pressure and Mach number distributions are  given in Fig. \ref{vis-mach-s}. The non-dimensional pressure and heat flux along the cylindrical surface are extracted and shown in Fig. \ref{vis-mach-s-line}. Generally the numerical results agree well with the experiment data \cite{wieting1987experimental}. The heat flux is calculated by the Fourier's law through the temperature gradient on the wall.
 To show the capacity of mesh adaptability for the current scheme,
 Mesh II with high non-uniformity in the bow shock region is used, which is shown in Fig. \ref{vis-mach-o}.
 In the near-boundary region, the mesh size starts from $h \approx 10^{-5}$ and grows up to $80$ layers with a stretching ratio $1.1$.
 $61$ mesh points are used along the circumferential direction.
 The numerical results agree well with the experimental data as well with Mesh II, as shown in Fig. \ref{vis-mach-o-line}.

\section{Conclusion}

In this paper, a two-stage compact GKS has been developed on unstructured triangular mesh.
Different from the 1st-order Riemann solver, the GKS provides a time-accurate evolution solution of the gas distribution function
at a cell interface.
Besides providing numerical fluxes and their time derivatives, the explicit time evolution solution also updates the flow variables
at the cell interface,
which can be used to get the averaged gradients of flow variables inside each triangular mesh through the Green-Gauss theorem.
As a result, a high-order GKS can be constructed with compact reconstruction and one middle stage time stepping technique.
Even with the same stencils as other compact schemes, the current GKS distinguishes it from the original HWENO \cite{qiu2004hermite} and DG \cite{cockburn1989tvb} methods in the ways of gradients updates. In the DG formulation, the gradient inside each control volume is
directly evaluated at next time level through a weak formulation.
This main difference makes the CFL number used in GKS be much larger than that in the same order DG method.
With the implementation of a modified HWENO reconstruction, the current 3rd-order compact GKS has the same robustness as the 2nd-order shock capturing scheme. There is no trouble cell detection in all test cases in this paper.
Only under extreme condition, such as Mach $20$ flow passing through a cylinder with unfavorable irregular mesh,
a limiting technique on the HWENO reconstruction weights is triggered.
The proposed scheme shows good mesh adaptivity even for a highly irregular triangular mesh.
In the previous compact GKS methods, all pointwise values at cell interface Gaussian points are used to get a over-determined system
in the quadratic polynomial reconstruction. The use of cell averaged gradients in the current scheme reduces the
stiff connectivity in flow variables between cells in the previous approach. 
As a result, the current compact GKS with direct application of HWENO reconstruction 
becomes more robust and accurate than the previous method on unstructured mesh.
The extension of the current scheme to even higher-order accuracy is under investigation.

\section*{Acknowledgement}
The current research is supported by Hong Kong research grant council 16206617, and  National Science Foundation of China (11772281,11701038).


\vfill
\clearpage
\bibliographystyle{abbrv}
\bibliography{jixingbib}
\end{document}